\documentclass[fleqn,usenatbib]{mnras}

\usepackage[T1]{fontenc}
\DeclareRobustCommand{\VAN}[3]{#2}
\let\VANthebibliography\thebibliography
\def\thebibliography{\DeclareRobustCommand{\VAN}[3]{##3}\VANthebibliography}

\usepackage{graphicx}	
\usepackage{amsmath}	
\usepackage{amssymb}	
\usepackage{tikz}

\usepackage{xcolor}
\usepackage{enumitem}
\usepackage{newtxtext,newtxmath}

\newcommand{\MS}{\rm{M}_{\odot}}
\newcommand{\MH}{\rm{M}_{\rm 200c}}

\title[Disk survival through mergers of MW/M31 analogues]{The merger and assembly histories of Milky Way- and M31-like galaxies with TNG50: disk survival through mergers}

\author[Sotillo-Ramos et al.]{Diego Sotillo-Ramos,$^{1}$\thanks{E-mail:sotillo@mpia.de}
Annalisa Pillepich,$^{1}$ Martina Donnari,$^{1}$ Dylan Nelson,$^{2}$
Lukas Eisert,$^{1}$
\newauthor
Vicente Rodriguez-Gomez$^{3}$, Gandhali Joshi,$^{1}$ Mark Vogelsberger$^{4}$ and Lars Hernquist$^{5}$
\\
$^{1}$Max-Planck-Institut f{\"u}r Astronomie, K{\"o}nigstuhl 17, 69117 Heidelberg, Germany\\
$^{2}$Zentrum f{\"u}r Astronomie der Universit{\"a}t Heidelberg, ITA, Albert Ueberle Str. 2, D-69120 Heidelberg, Germany\\
$^{3}$Instituto de Radioastronom\'ia y Astrof\'isica, Universidad Nacional Aut\'onoma de M\'exico, Apdo. Postal 72-3, 58089 Morelia, Mexico\\
$^{4}$Kavli Institute for Astrophysics and Space Research, Massachusetts Institute of Technology, Cambridge, MA 02139, USA\\
$^{5}$Center for Astrophysics $|$ Harvard \& Smithsonian, 6P Garden St., Cambridge, MA 02138, USA
}

\begin{document}
\label{firstpage}
\pagerange{\pageref{firstpage}--\pageref{lastpage}}
\maketitle

\begin{abstract}

We analyze the merger and assembly histories of Milky Way (MW) and Andromeda (M31)-like galaxies to quantify how, and how often, disk galaxies of this mass can survive recent major mergers (stellar mass ratio $\ge$ 1:4).
For this, we use the cosmological magneto-hydrodynamical simulation TNG50 and identify 198 analog galaxies, selected based on their $z=0$ stellar mass ($10^{10.5-11.2}\,\MS$), disky stellar morphology and local environment. 
Firstly, major mergers are common: 85 per cent (168) of MW/M31-like galaxies in TNG50 have undergone at least one major merger across their lifetime. 
In fact, 31 galaxies (16 per cent) have undergone a recent major merger, i.e. in the last 5 Gyr. The gas available during the merger suffices to either induce starbursts at pericentric passages or to sustain prolonged star formation after coalescence: in roughly half of the cases, the pre-existing stellar disk is destroyed because of the merger but reforms thanks to star formation. Moreover, higher merger mass ratios are more likely to destroy the stellar disks.
In comparison to those with more ancient massive mergers, MW/M31-like galaxies with recent major mergers have, on average, somewhat thicker stellar disks, more massive and somewhat shallower stellar haloes, larger stellar ex-situ mass fractions, but similarly massive kinematically-defined bulges. All this is qualitatively consistent with the different observed properties of the Galaxy and Andromeda and with the constraints on their most recent major mergers, $8-11$ and ~2 Gyr ago, respectively. %
According to contemporary cosmological simulations, a recent quiet merger history is not a pre-requisite for obtaining a relatively-thin stellar disk at $z=0$.
\end{abstract}




\section{Introduction}
\label{sec:intro}
The \textit{Lambda} cold dark matter ($\Lambda$CDM) cosmological model successfully explains the evolution of the Universe and the formation of the large-scale cosmic web. Structures grow hierarchically, with more massive haloes growing by smooth accretion and by merging with smaller ones \citep{Genel2010}, in a process dominated by (cold) dark matter, down to galactic scales. 

In this context, mergers and interactions between galaxies are expected to be a  frequent phenomenon, more so the higher the redshift \citep{Fakhouri2008}.
In fact, galaxy mergers are known to play a central role in the mass growth of galaxies. Within the hierarchical growth of structure scenario, more massive haloes and galaxies undergo, on average, a larger number of mergers than lower-mass ones \citep{Fakhouri2010, RodGom2015}. Moreover, more massive galaxies have been shown, via cosmological hydrodynamical galaxy simulations, to be made of larger fractions of stellar material that is accreted from mergers and orbiting satellites \citep[e.g.][]{RodGom2016, Pillepich2018b}.

Following a suggestion originally by \citet[][]{Toomre1977}, galaxy mergers have also been shown to be able to drive morphological transformations in galaxies, i.e. to turn disky and rotationally-supported spiral galaxies into elliptical and dispersion-dominated ones. Already a few decades ago, it had been demonstrated via N-body simulations of isolated and idealized systems that the merger of two stellar disks generates descendants with spheroidal morphology \citep[e.g.][]{Barnes1988, Hernquist1991, Barnes1992a, Hernquist1993}.
Yet, more recently, it has also been found, with full-physics, cosmological hydrodynamical simulations of large samples of objects, that galaxies with similar stellar mass have undergone, on average, similar merger histories irrespective of their stellar morphology upon inspection \citep[e.g.][with Illustris and Horizon-AGN]{RodGom2017, Martin2018, Jackson2020}.

In the observed Universe, galaxies with disky stellar morphologies are ubiquitous, ranging across wide mass and size scales. In our most immediate vicinity, there are three representative examples: the Milky Way (MW),  Andromeda (M31), and Triangulum, all with stellar masses exceeding a few $10^{10}\,\MS$. Beyond the Local Group, at the MW mass scale, stellar disk morphology dominates, i.e. roughly two thirds of the observed galaxies exhibit stellar disk morphology: this is the case not only in the low-redshift Universe \citep[e.g. results with the SDSS and GAMA surveys,][respectively]{Park2007,Kelvin2014} but also up to $z\sim1.2$  \citep[e.g. results with VIMOS-VLT data by][]{Ilbert2006}. 

Recent observations have enabled estimates of the merger histories of both the MW and M31. It is now possible to measure the trajectories of satellite galaxies surrounding the Galaxy and Andromeda, and, for the case of the MW, also the positions and velocities of individual stars \citep[e.g. with RAVE, APOGEE, GAIA,][]{Kunder2017, Majewski2017, Lindegren2016}, both in the disk and the stellar halo, and in the satellite galaxies, together with ages and metallicities for many of these stars \citep[e.g. with LAMOST and GALAH,][]{Deng2012, DeSilva2015}. A series of stellar streams have been identified in the stellar halo that are associated with merger events.

Recently, \citet{Helmi2018} and \citet{Belokurov2018} have used data from the \textit{Gaia} mission \citep{GaiaDR22018} and inferred that the Galaxy has undergone a major merger event sometime at $z=1-2$ \citep[see also][]{Bonaca2020, XiangRix2022}:
the merging galaxy has been dubbed Gaia-Sausage-Enceladus (GSE) and its mass at the time of the collision has been estimated to be $\sim5-6\times10^8\, \MS$ in stars \citep{Naidu2021}. The GSE merger has been associated theoretically with the formation of the inner stellar halo and the thick disk \citep[][]{Grand2020}. Dynamical and chemical evidence of additional events in the MW's past has also been found, i.e. from the stellar remnants of past merging galaxies: these include Sequoia, accreted at a comparable epoch as GSE \citep{Myeong2019}; Thamnos \citep{Koppelman2019}; Helmi Streams \citep{Helmi1999}, that originated from a dwarf galaxy accreted approximately 5-8 Gyr ago \citep[][]{Koppelman2019}, and the long-known Sagittarius Stream \citep{Ivezic2000}. The latter is still clearly identifiable also in configuration space and is being stripped from the Sagittarius dwarf of $\sim5\times10^8\,\MS$ in stars, which is in the process of merging with the MW and whose infall time has been estimated to have occurred $\sim$8 Gyr ago \citep[][]{Dierickx2017}. These, together with the ongoing mergers with the Large and Small Magellanic Clouds (with stellar masses of $\sim2.7\times10^9\,\MS$ and $\sim3.1\times10^8\,\MS$, respectively \citep[][]{vanderMarel2006}, and thus with a current stellar mass ratios of $\sim$1:10 and $\sim$1:100) constitute the six most massive accretion events across the known history of the MW, with the GSE being the largest in estimated mass ratio. 

For M31, observational constraints are harder to obtain. However, for example, \citealt[][]{DSouza2018} recently suggested, by combining the results of cosmological models of galaxy formation and measurements of the ages and metallicities of halo stars, that Andromeda  underwent a major merger about 2 Gyr ago \citep[][]{DSouza2018} and that the observed satellite galaxy M32 is the remnant core of the secondary galaxy involved in the merger: the latter produced the giant stellar stream that is located in Andromeda's stellar halo, generated a recent burst of star formation, but did not destroy the stellar disk.

The merger histories of MW-mass haloes have also been studied from a theoretical perspective, and via cosmological simulations. For example, \citet{Stewart2008} used DM-only or gravity-only N-body $\Lambda$CDM simulations and quantified that 95 per cent of MW-mass haloes, i.e. gravitationally-collapsed objects of total mass of $\sim 10^{12}\MS$ h$^{-1}$ at $z=0$, have merged with at least one other halo more massive than 1:20 in the last 10 Gyr, this fraction reducing to 70 per cent for mergers with haloes more massive than 1:10. They concluded that halo-halo mergers involving (total) mass ratios of up to 1:5 must not destroy the stellar disks of the hosted galaxies if approximately two thirds of the observed galaxies are disky at $z=0$.

The reasoning mentioned above encapsulates and exemplifies a debate that has gone on for a few decades: namely, how to reconcile the fact that disk galaxies are found everywhere and at the same time the number of mergers that potentially would transform or destroy them are frequent. In the case of the Galaxy, \citet{Toth1992} claimed via analytical arguments that a thin stellar disk similar to the Milky Way's is only compatible with no minor nor major mergers in the last 10 Gyr, a conclusion similar to that derived via observational data as early as by \citet{Wyse2001} and \citet{Hammer2007}.

However, from a theoretical and numerical perspective, it has long been known \citep[e.g.][]{Hernquist1989, Barnes1996} that the formation and evolution of stellar disks -- also and especially in the presence of perturbation mechanisms such as major mergers -- cannot be addressed without accounting for the role and availability of gas. Numerical dissipational simulations of idealized mergers by e.g. \citealt[][]{Springel2005b, Naab2006, Robertson2006} showed that the remnants of {\it gas-rich} mergers can lead to the formation of a spiral and not an early-type galaxy. 

In the full cosmological context, i.e. with zoom-in cosmological N-body and hydrodynamical simulations of galaxies, the formation of thin stellar disks in MW-mass galaxies could be achieved only after solving for the ``angular momentum problem'' or ``overcooling catastrophe'' and after introducing appropriate (stellar) feedback and star formation recipes \citep{Guedes2011, Agertz2011} and refined numerical approaches  \citep{Keres2012, Sijacki2012, Vogelsberger2012}. A number of successes followed, with simulations capable of returning large, disk-dominated galaxies that resemble the Milky Way in many respects \citep[e.g.][]{Martig2012, Stinson2013, Marinacci2014}. Still, as of just some years ago, such zoom-in simulations of one or a few objects seemed to indicate that disk-dominated galaxies with stellar structures in accord with observations of the Galaxy could emerge, if, and only if, there is no major merger at $z\lesssim1$ \citep{RixBovy2013}.

The advent of large-volume cosmological hydrodynamical galaxy simulations \citep[][]{Vogelsberger2020} such as Illustris \citep{Vogelsberger2014, Vogelsberger2014b, Genel2014, Nelson2015} and EAGLE \citep{Schaye2015, Crain2015} allows us to reconsider the formation of disk galaxies from major mergers in a quantitative manner, thanks to thousand-strong galaxy samples, spanning wide mass ranges and without prior constraints on the past assembly histories or environments. For example, with Horizon-AGN \citep{Dubois2014} and Illustris galaxies, \citet{Martin2018} and \citet{Peschken2020} have confirmed that if enough gas is available during a merger event, new stars can form and can reform a stellar disk even if the latter gets destroyed in the collision. A similar scenario has been previously explored using zoom-in simulations by \citet[][]{Sparre2017}, who considered four $\sim$MW-mass (at $z=0$) galaxies that were disky and underwent one major merger at $z\sim0.5$. Focusing on very massive disk galaxies from Horizon-AGN ($M_{\ast} \geq 10^{11.4}\, \MS$), \citealt{Jackson2020} showed that, in the majority of cases, pre-existing stellar disks  either survive major mergers or reform shortly after a merger \citep[a pathway in qualitative agreement with observations, e.g.][]{Rothberg2004, McDermid2006} and, in fewer cases, stellar disks form after a spheroidal galaxy is established \citep{Hau2008,Kannappan2009}.

However, so far no quantitative analysis of the merger history and disk survival through mergers has been put forward with a focus on simulated analogues of the Galaxy and Andromeda. On the one hand, zoom-in simulation campaigns have begun to build up impressive samples of Milky Way-like galaxies: however,  these are typically not unbiased in formation history/isolation \citep[e.g. Auriga,][]{Grand2017} or are selected in relatively narrow ranges of total halo mass \citep[e.g. Artemis,][]{Font2020}. On the other hand, large-volume simulation projects have lacked the resolution to capture stellar disks as thin as a few hundred parsecs. Large, unbiased samples at relatively high-resolution are now possible with the cosmological magnetohydrodynamical simulation TNG50 \citep{Nelson2019b, Pillepich2019}, the highest-resolution run of the IllustrisTNG project (\url{www.tng-project.org}). There we can identify approximately two hundred galaxies at $z=0$ with global properties similar to the MW's and M31's and with stellar disk heights as small as $100-200$ parsecs. In addition to gravity and hydrodynamics, these simulations account for the effects of stellar and AGN feedback, magnetic field evolution, gas cooling and heating, etc. 

In this paper, we hence describe the TNG50 cosmological simulation, the method to create the merger trees, and the selection criteria for the MW/M31 analogues in Section \ref{sec:methods}. We quantify the merger and assembly histories of TNG50 MW/M31-like galaxies in Section \ref{sec:statistics}. In Section \ref{sec:disc_survival} we delve into the evolutionary pathways and physical mechanisms at play in the case of MW/M31 analogues that underwent at least one major merger in recent epochs, e.g. in the last 5 Gyr.
We compare the $z=0$ structural and global properties of these galaxies with the rest of the TNG50 MW/M31-like objects in Section \ref{sec:properties} and summarize and discuss our findings in Section \ref{sec:conclusions}.

\section{Methods: simulation and sample selection}
\label{sec:methods}

\subsection{The TNG50 simulation}
\label{sec:simul}

In this paper, we focus on Milky Way and Andromeda-like (MW/M31-like) galaxies realized within the TNG50 cosmological simulation. TNG50 \citep{Nelson2019b, Pillepich2019} is the highest-resolution run of the IllustrisTNG project \citep{Naiman2018, Marinacci2018, Pillepich2018b, Nelson2018, Springel2018}. It consists of a periodic cubic volume with side length of $\approx 51.7$ comoving Mpc, contains 2160$^3$ dark matter (DM) particles and the same initial number of gas cells. The mass of the DM particles is uniform, $m_\rmn{DM} = 4.5 \times 10^5$ M$_{\odot}$, and the average mass of the gas cells (and stellar particles) is $m_\rmn{baryon} = 8.5 \times 10^4$ M$_{\odot}$. The code \textsc{Arepo} \citep[][]{Springel2010}, which makes use of an unstructured moving mesh that is based on the Voronoi tessellation of the space and that is automatically adaptive in resolution, solves the gravity and magnetohydrodynamical equations in an expanding universe from $z\approx127$ to $z=0$. Additionally, a series of physical processes are included to follow galaxy formation: these and their implementation are described in detail in the methods papers by \citealt{Pillepich2018, Weinberger2017}. Crucially and differently from most high-resolution simulations of Milky Way-like galaxies \citep[e.g. zooms without SMBH feedback by][]{Wetzel2016, Buck2020, Renaud2021}, the IllustrisTNG model not only includes star formation and its ensuing expected feedback and metal enrichment, but also magnetic fields and the seeding, growth, and feedback from supermassive black holes (SMBHs). All simulations in the IllustrisTNG series adopt values for the cosmological parameters from \cite{Planck2016}.

\subsection{Halo identification and histories of simulated galaxies}
\label{sec:merger_trees}

Throughout our analysis, the identification of haloes and subhaloes in the simulated domain is performed with the Friends-of-Friends \citep[FoF;][]{Davis1985} and \textsc{Subfind} \citep[][]{Springel2001} algorithms, respectively. The latter identifies all particles and resolution elements that are measured to be gravitationally-bound to a self-gravitating structure.   
We define as galaxies those subhaloes with non-vanishing stellar mass. The ``central'' galaxy of a FoF halo is typically the most massive one, whereas the remaining subhaloes within a FoF halo are called satellites. Throughout the analysis, we consider only subhaloes with cosmological origin, i.e. non-spurious subhaloes \citep{Nelson2019b}, and galaxies with at least 50 stellar particles.

\subsubsection*{Merger histories and merger summary statistics}
Information for all resolution elements, subhaloes, and haloes is stored for one hundred snapshots, starting at $z=20$ \citep[see][for details]{Nelson2019b}. At recent epochs, the time spacing between consecutive snapshots is approximately 150 million years. The (sub)haloes at different snapshots are linked to describe their merger histories (or merger trees) using the \textsc{Sublink} \citep{RodGom2015} and \textsc{LHaloTree} \citep{Springel2005a} algorithms. Both return similar results. In this work, we use \textsc{SublinkGal}, which connects subhaloes across cosmic epochs based on their star particles and star-forming gas cells, instead of DM particles.

According to \textsc{SublinkGal} \citep[described in detail][]{RodGom2015}, the main progenitor of a galaxy is the one with the most massive history behind it: we will refer to its temporal evolution as the ``main-progenitor history'' (or ``main-progenitor branch'').

In the following, we describe the merger history of a galaxy by analyzing the time of its mergers and the {\it stellar-mass} ratio between the two involved galaxies, i.e. between the progenitors. Namely: 

\begin{itemize}
\item {\it Time of a merger:} As per \cite{RodGom2015}, the time of a merger ($t_\rmn{merger}$) is given by the earliest snapshot when both progenitors are identified by \textsc{Subfind} as part of the same unique subhalo. Such a merger time represents the time of coalescence or just after it, i.e. at the next available snapshot.\\

\item {\it Time of the merger-ratio estimate:} The stellar mass ratio (the smallest of the ratios between the secondary and the primary) is measured at the time when the secondary reached its maximum stellar mass ($t_\rmn{max}$). This can be considered as the time when a merger starts.\\

\item {\it Merger duration:} The duration of a merger is the time elapsed between $t_\rmn{max}$ and the time of coalescence, i.e. the time of a merger. This is therefore similar to the infall time, i.e. the time spent by the satellite in orbit within the host halo.
\end{itemize}

Mergers are classified into three types: major, whereby the stellar-mass ratio is 1:4 or larger;  minor, with ratios between 1:10 and 1:4, and mini (below 1:10). 
In the case of several mergers occurring at the same time, i.e. between the same pair of snapshots for the same galaxy, we account for all of them, separately, and their stellar mass ratios are recorded separately for each binary event.

In comparison to the default functioning of the \textsc{SublinkGal} code, we apply additional measures to avoid counting as mergers the interactions with spurious subhaloes or with subhaloes that are resolved with too few stellar particles, i.e. to account for the selections mentioned above. The former may be objects without a cosmological origin (e.g. fragments of disks or other types of galactic sub structures) that \textsc{Subfind} identifies as subhaloes. They are identified with the \textit{SubhaloFlag} \citep{Nelson2019a} and are here excluded from the merger statistics. Galaxies whose main progenitor history is very brief (less than three snapshots before the time of coalescence) are also excluded.   
Finally, we also neglect all mergers between galaxies with fewer than 50 star particles ($\sim 5\times10^{6} \rm{M}_{\odot}$), across cosmic epochs. Thinking about the merger history of massive galaxies like MW/M31 analogues at $z=0$, this allows us to avoid counting as major those mergers whose $t_\rmn{max}$ may occur very early on in the cosmic history of a galaxy, when primary and secondary objects might have had comparable masses, but whose actual coalescence occurs several billion years later and at times when the two progenitors actually differ by many orders of magnitude in mass.

The three conditions above allow us to obtain a clean and complete catalog of merger events, particularly at lower redshifts. On the other hand, these conditions also imply that our analysis is neither sensitive nor complete for the merger history of MW/M31-like main progenitors prior to $z\sim5$. 

\begin{figure*}
	\includegraphics[width=\columnwidth]{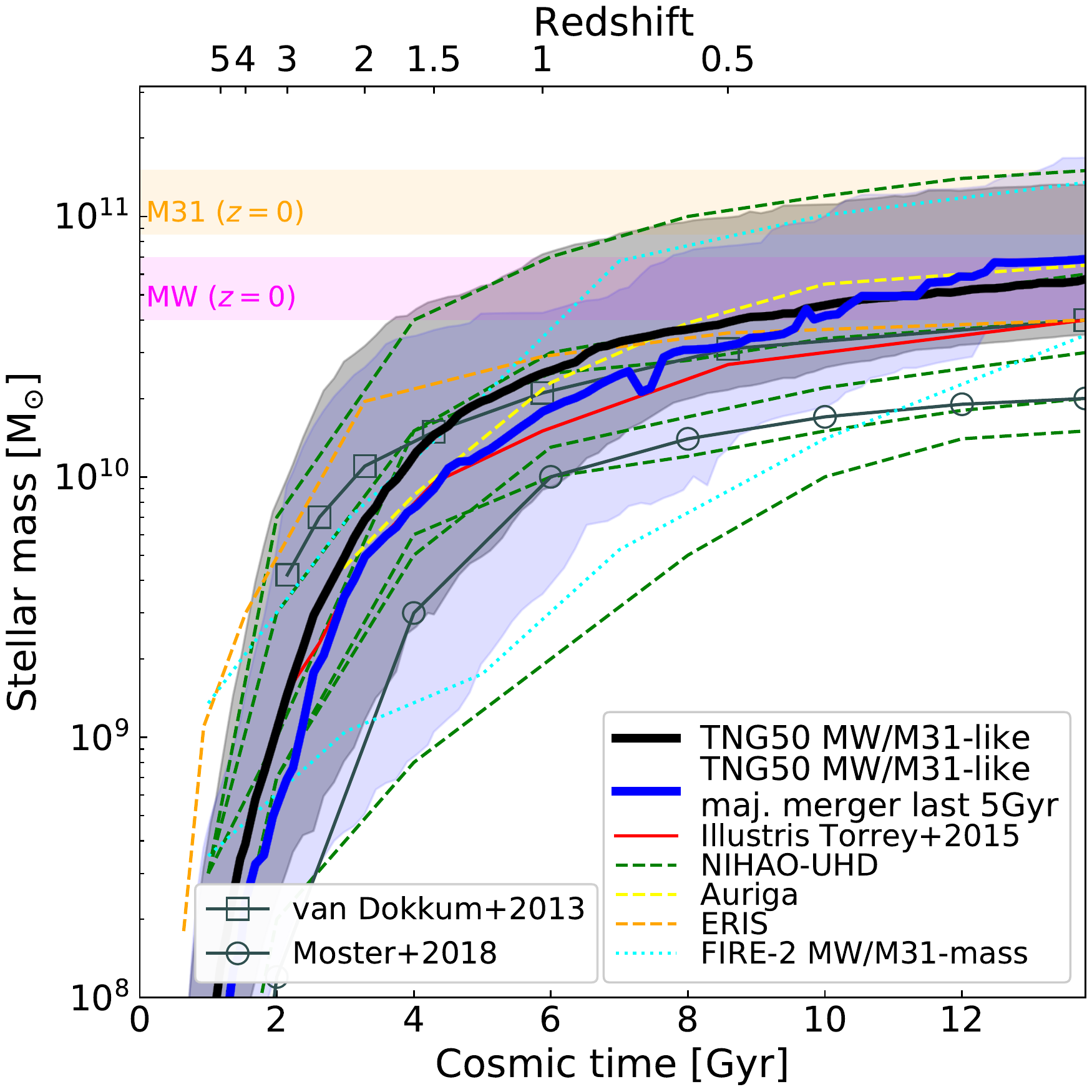}
	\includegraphics[width=\columnwidth]{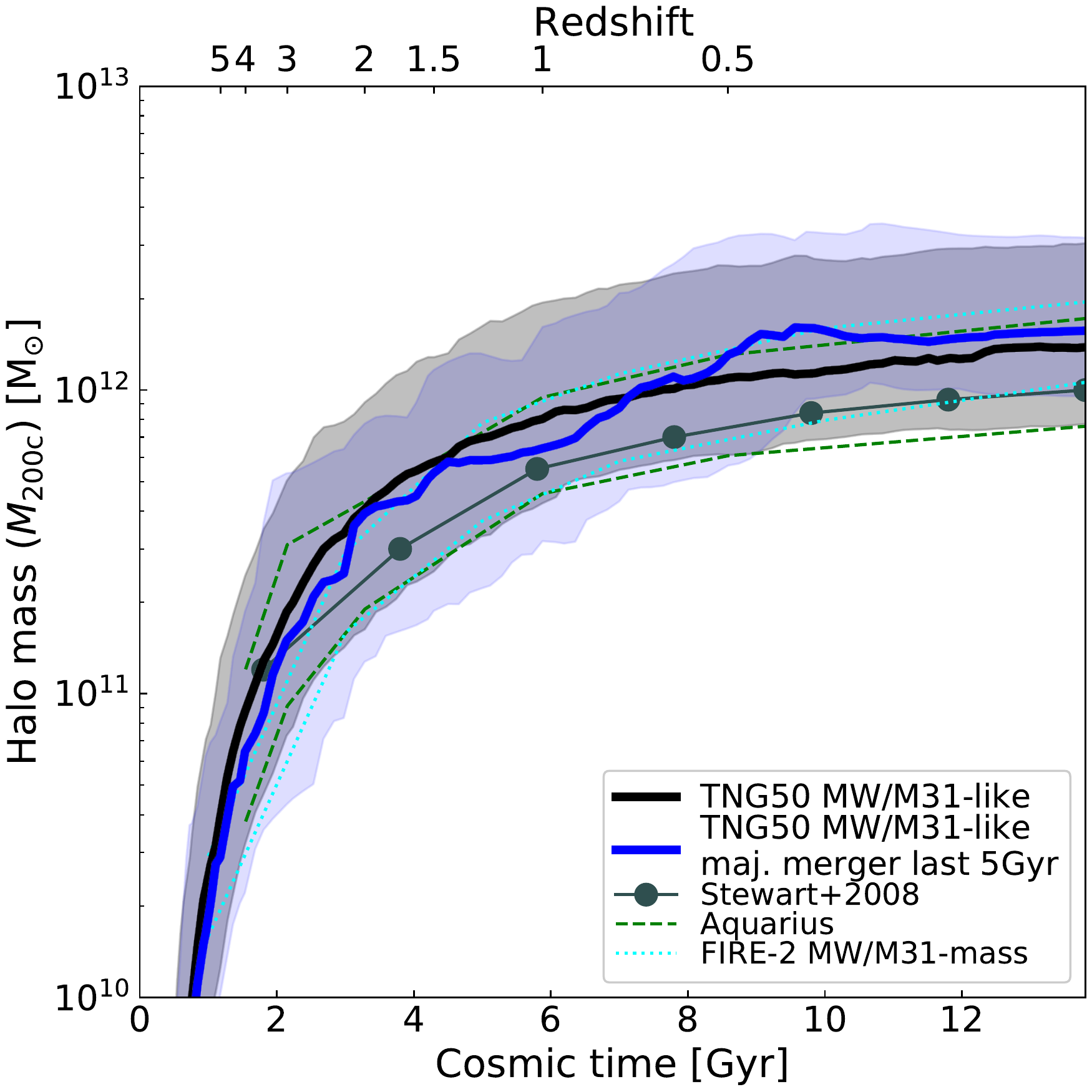}
    \caption{Mass assembly histories of MW/M31 analogues in TNG50, for galaxy stellar mass (left) and total host mass (right). We show the complete TNG50 MW/M31 sample in black and a sub-sample of galaxies with recent major merger in blue (see next Sections): median tracks at fixed cosmic time are marked as thick curves, whereas shaded areas denote interquantiles (10th-90th percentiles). Additional curves represent results from other simulations (colored lines) or observationally-derived constraints  (grey lines with empty markers). For TNG50 and the additional simulations, the subhalo's total stellar mass is plotted. For $z=0$ MW/M31-like galaxies, 10 per cent of their stellar mass is assembled by $z=2$ and 50 per cent by $z=1$. At $z=2$ the scatter in galaxy stellar mass extends for two orders of magnitude ($1.8\times10^8-2.2\times10^{10}\,\rmn{M}_{\odot}$, 10th-90th percentiles), in contrast to a $z=0$ selection ranging 0.7 dex. 
    In terms of total host mass, $z=0$ MW/M31-like galaxies have assembled 50 per cent of their halo mass by $z=2$.
    We show annotations with observational estimates for the stellar mass at $z=0$ of the MW \citep[in magenta, as an envelope of the values from][]{Flynn2006,Licquia2015, Bland-Hawthorn2016} and 
    M31 \citep[orange, from][]{Geehan2006, Barmby2006, Tamm2012, Sick2014}.
    }
    \label{fig:assemblyHistory}
\end{figure*}

\subsubsection*{In-situ vs. ex-situ stars}
The cosmological simulations allow us to also follow the evolutionary tracks of individual stellar particles, so that we can identify the galaxy where each formed and the location at subsequent times. Based on the merger trees and for any galaxy at $z=0$, we categorize its stars as in situ if they formed in a progenitor that belongs to the main-progenitor branch of the galaxy, and ex situ otherwise, according to the definitions and catalogs described in \citet[][]{RodGom2016}. Ex-situ stars are hence stellar particles stripped and consequently accreted into a galaxy from its mergers and orbiting satellites.

\subsection{Galaxy and stellar-particle properties: circular orbits and diskyness or D/T ratio}
\label{sec:galStarProperties}

Throughout the paper, the following conventions are intended unless otherwise stated.\\

\textit{Galaxy stellar mass}: sum of the mass of all gravitationally-bound star particles.\\

\textit{Total host mass}: the total mass of the FoF group in which a galaxy is found, $M_\rmn{200c,Host}$, measured inside a sphere whose mean density is 200 times the critical density of the Universe (at the considered time).\\

\textit{Stellar half-mass radius} of a galaxy ($R_\rmn{1/2}$): the radius of the sphere, centered at the particle with the minimum gravitational potential energy in the galaxy, that contains half of the stellar mass.\\

\textit{Star formation rate (total, SFR)} is measured by adding the individual instantaneous SFRs of all gas cells that are gravitationally-bound to a galaxy at the time of inspection.\\

Other targeted galaxy properties are described in the course of the analysis.

For any given galaxy, the orbital properties of its stars are described via the circularity parameter, according to the definition by \citet{Scannapieco2009}: $\epsilon=j_\rmn{z}/j_\rmn{circ}$, where $j_\rmn{z}$ is the specific angular momentum of the star in the direction perpendicular to the galactic disk, and $j_\rmn{circ}$ is the specific angular momentum of a star at the same radius, but on a perfectly circular orbit. Namely, $j_\rmn{circ}=rv_\rmn{circ}$, with $v_\rmn{circ}=\sqrt{GM(\leq r)/r}$ being the circular velocity of the galaxy at the considered radius. The up vector or vertical axis of a galaxy is chosen as the direction of the total angular momentum of all stars within two half-mass radii. Stellar orbits with $\epsilon\gtrsim 0.7$ are considered circular, i.e. in rotational motion.

We define the diskyness or disk-to-total (D/T) ratio of a galaxy as the fractional mass of stars in circular orbits, i.e. with $\epsilon > 0.7$, minus the fraction of stars with $\epsilon < -0.7$, hence removing the contributions of bulge and stellar halo and assuming that the latter are symmetric around $\epsilon=0$. The stellar mass in circular orbits and the total stellar mass are evaluated, for the purposes of D/T, within an aperture of five stellar half-mass radii.

The exact choice of separating circular and cold orbits from the rest with the  threshold $\epsilon > 0.7$ is not crucial: \citet[][]{Aumer2013} found that the D/T ratio obtained from our circularity definition is roughly equivalent to the values obtained with another usual definition of the circularities, based on the ratio of the specific angular momentum of the star and the maximum specific angular momentum at the specific binding energy of the star, using in both cases a similar circularity threshold.

\subsection{Sample selection: MW/M31-like galaxies in TNG50}
\label{sec:sample_selection}

From the TNG50 box, which at $z=0$ samples hundreds of massive galaxies, we select the best analogues of the Milky Way and Andromeda based on their $z=0$ properties, i.e. without imposing conditions on their evolution or any restrictions on their morphological changes with time.

In practice, we use the selection presented and motivated in  \citet{PillepichInPrep} and already used by \cite{Engler2021, Pillepich2021a} and others. To be selected as “MW- or M31-like”, a TNG50 galaxy must meet, at $z=0$, the following three conditions, based on galaxy stellar mass, stellar diskyness, and isolation:
\begin{enumerate}[label=\Alph*)]
    \item Galaxy stellar mass: log$_{10}$(M$_{\ast}$/M$_{\odot}$) within [10.5, 11.2], with stellar mass measured within a 3D circular aperture of 30 kpc.\\
    
    \item Diskyness: either of the following conditions: 
    \begin{enumerate}
        \item Stellar minor-to-major axis ratio \citep[][]{Pillepich2019}, $c/a$, $ < 0.45$, whereby $c$ and $a$ are respectively the minor and major axis of the ellipsoidal distribution of the stellar mass between one and two times the stellar half-mass radius. 
        
        \item Disky by visual inspection of 3-band images, both face-on and edge-on.
    \end{enumerate}
    
    \item Isolation: no galaxy with galaxy stellar mass $\geq 10^{10.5}$ M$_{\odot}$ within 500 kpc distance and total host mass $M_\rmn{200c,Host}$ < 10$^{13}$ M$_{\odot}$.
\end{enumerate}

There are 198 galaxies in the TNG50 simulation that match such conditions.

The stellar mass constraint is an envelope of the most accurate available estimates of the stellar mass of both the MW and M31 \citep[e.g.][]{Bland-Hawthorn2016, Boardman2020}.
The diskyness criteria aim at including all simulated galaxies with a global stellar disk morphology and with spiral arms. Whereas typically, disky galaxies are defined as those with minor-to-major axis ratio $<0.33$ (and middle-to-major axis ratio $>0.66$; see e.g. Fig. 8 of \citet{Pillepich2019} and reference therein), the connection of this metric to other measures of ``diskyness'' (e.g. kinematic or photometric D/T ratios, scale length to scale height ratio, etc.) can be very varied, depending also on where, within a galaxy’s body, the measurements are taken. Here we adopt a looser upper limit on the minor-to-major axis ratio to accommodate for the ratio of e.g. the stellar disk height to the stellar disk length of the Galaxy, particularly of its geometrically-thick component. The latter measures $\sim 0.45$ (in comparison to $\sim0.11$ for the geometrically-thin disk) when adopting the best values for the structural properties of the Milky Way disk from \citet[][thin and thick disk lengths = $2.6\pm0.5$ and $2\pm0.2$ kpc, respectively, and thin and thick disk heights at the Sun location = $300\pm50$ and $900\pm180$ pc]{Bland-Hawthorn2016}.
The isolation criterion avoids the presence of a galaxy, with mass equal to the lower MW estimate or larger, at a distance closer than 500 kpc -- the distance between the Galaxy and Andromeda is $\sim$ 770 kpc \citep[][]{McConnachie2005, Riess2012}.
Additionally, the requirement on total host mass is meant to exclude galaxies located within a very massive galaxy group or cluster, as we know this is not the environment of the Galaxy and Andromeda. However, it is more relaxed than requiring a galaxy to be the central of its halo, while allowing certain galactic environments such as those similar to the Local Group, whereby the Galaxy and Andromeda may be sharing the same dark matter halo.

As the galaxies are selected from a volume-limited sample, a larger number of TNG50 objects have stellar mass closer to the MW's than to the more massive M31's. In fact, among the 198 TNG50 MW/M31 analogues, 130 galaxies have masses more compatible with the MW's ($\le 10^{10.9}\,\MS$), and will be dubbed `MW-mass' and 68 are instead more representative of Andromeda and will be dubbed `M31-mass'. 

No constraints on the past history are imposed, in contrast to what has been typically done with cosmological zoom-in simulations of Milky Way analogues -- e.g. those by \citet{Governato2004}, \citet{Okamoto2005}, Eris \citep{Guedes2011}, \citet{AgertzKravtsov2015}, GARROTXA \citep{RocaFabrega2016}, VINTERGATAN \citep[][]{Renaud2021}, AGORA \citep{RocaFabrega2021} --, usually by limiting the last major merger to redshift $z \geq 1$. Furthermore, no a-priori limitations are placed for e.g. the range of current star formation rates, stellar disk structural properties, and SMBH masses \citep[see][for an overview]{PillepichInPrep}. 

\section{The past histories and merger statistics of MW/M31-like galaxies, according to TNG50}
\label{sec:statistics}

\subsection{Mass assembly}
We quantify the assembly history of TNG50 MW/M31-like galaxies by plotting their mass across cosmic epochs, i.e. along their individual main progenitor branches.
The evolution of the galaxy stellar mass (left) and of the total host mass (within $R_\rmn{200c}$, right) of all MW/M31 analogues are shown in Fig.~\ref{fig:assemblyHistory}. For TNG50 and the additional simulations the total (i.e. gravitationally-bound) stellar mass is plotted. Thick curves denote the median evolutionary tracks across the galaxy populations at fixed redshifts, whereas the shaded areas indicate the 10th-90th percentiles. It should be noted that, although the medians appear smooth, individual assembly histories may exhibit assembly histories that are very different from the median curves. 

The selection by galaxy stellar mass at $z=0$ -- in the range of $3.2\times10^{10}-1.6\times10^{11}\,\rmn{M}_{\odot}$, within 30 kpc aperture as per Section~\ref{sec:sample_selection}, left panel -- translates into a $z=0$ host mass range of $\MH = 8.3\times10^{11}-2.5\times10^{12}\,\rmn{M}_{\odot}$ 
at $z=0$, i.e. 0.48 dex, within 10th-90th percentiles (right panel): this is a consequence of the effective stellar-to-halo mass relation resulting in TNG50 \citep{Engler2021}. On average, the host haloes of MW/M31-like galaxies grow by a factor of $\sim7$ between $z=3$ and today, while the average galaxy stellar mass of their progenitors at $z=3$ was a factor of $\sim40$ lower than today.

\begin{figure*}
	\includegraphics[width=1.1\columnwidth]{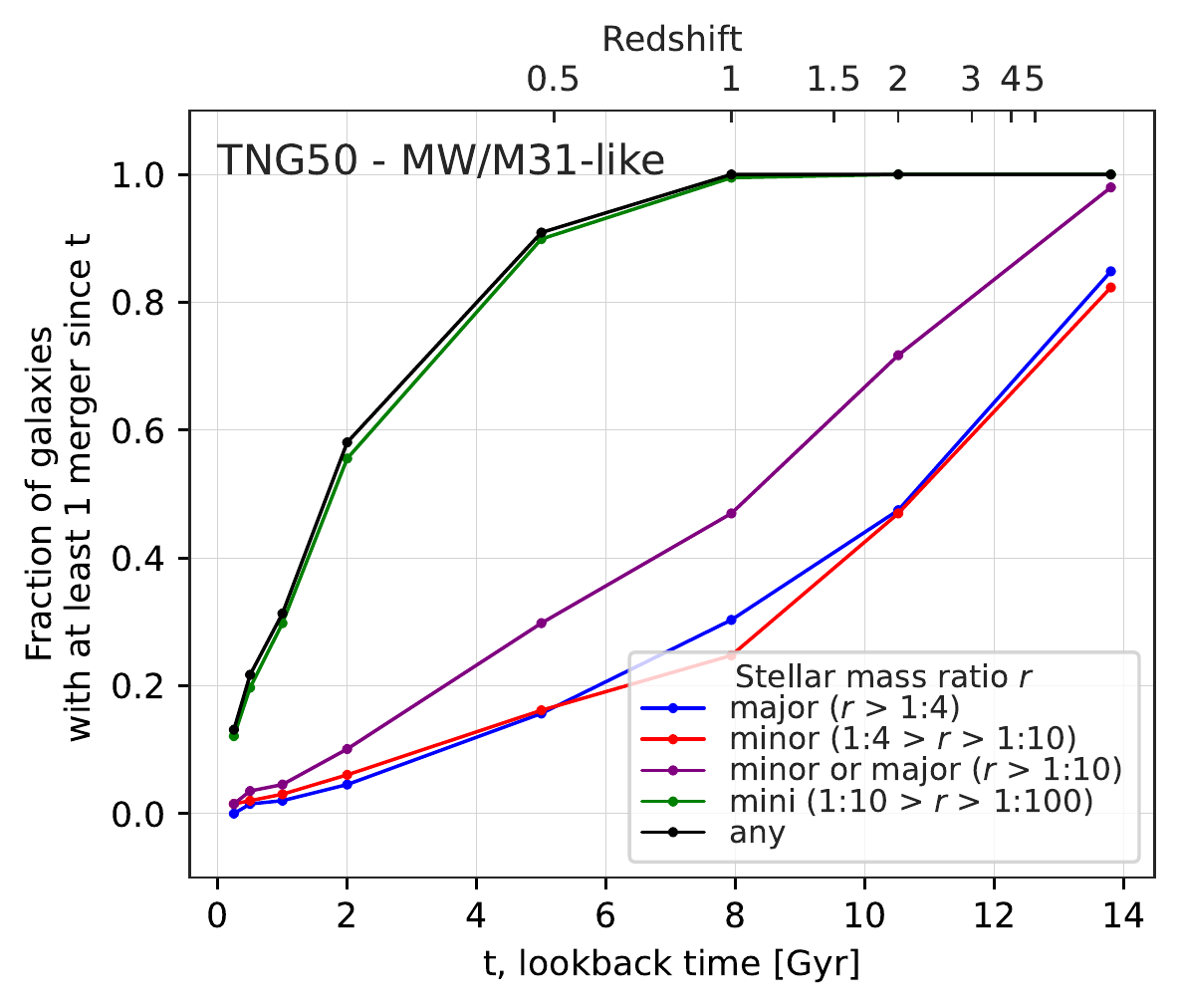}\\
    \includegraphics[width=.343\textwidth,trim={0 0 2.65cm 0},clip]{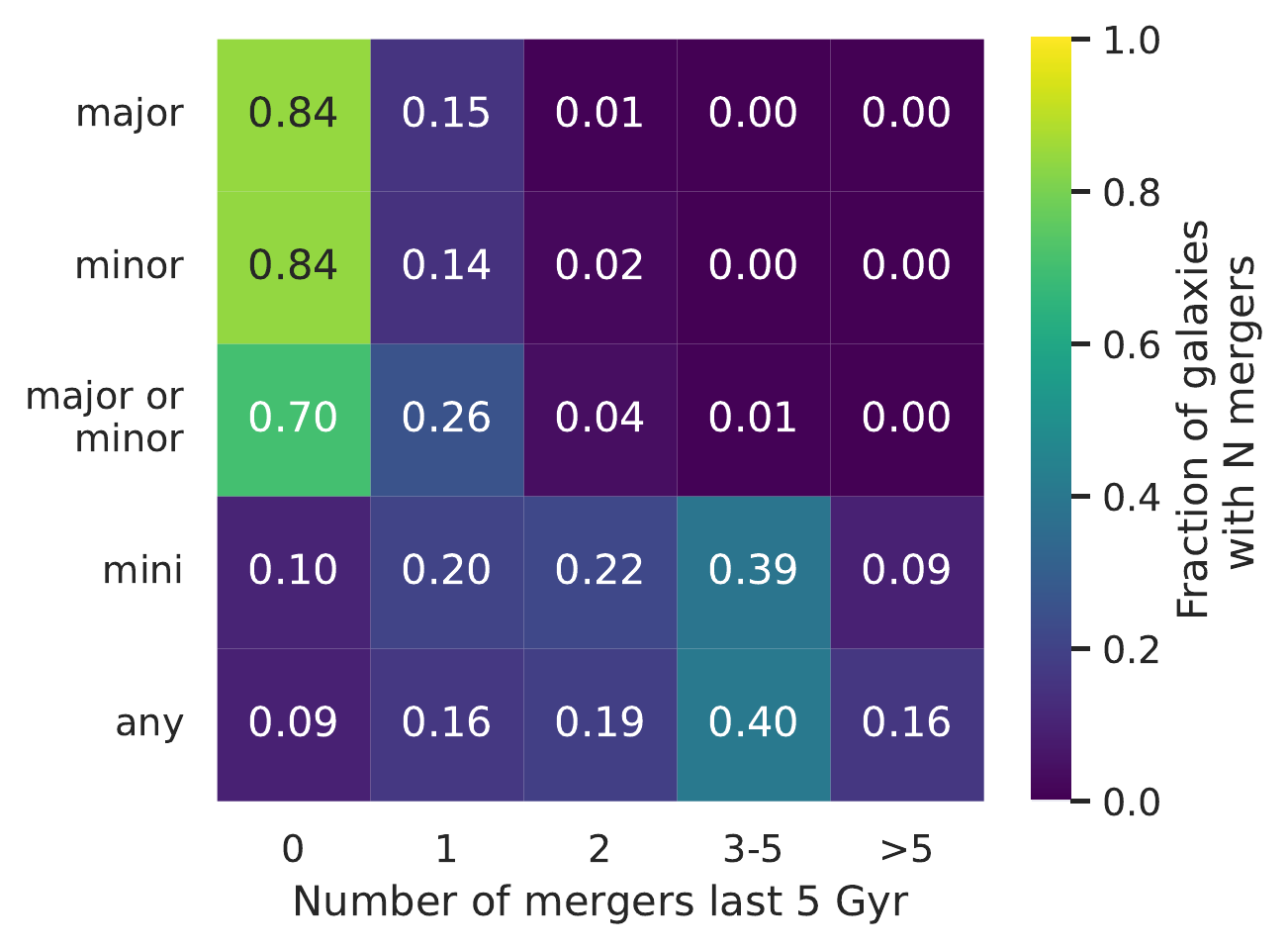}
    \includegraphics[width=.28\textwidth,trim={1.87cm 0 2.65cm 0},clip]{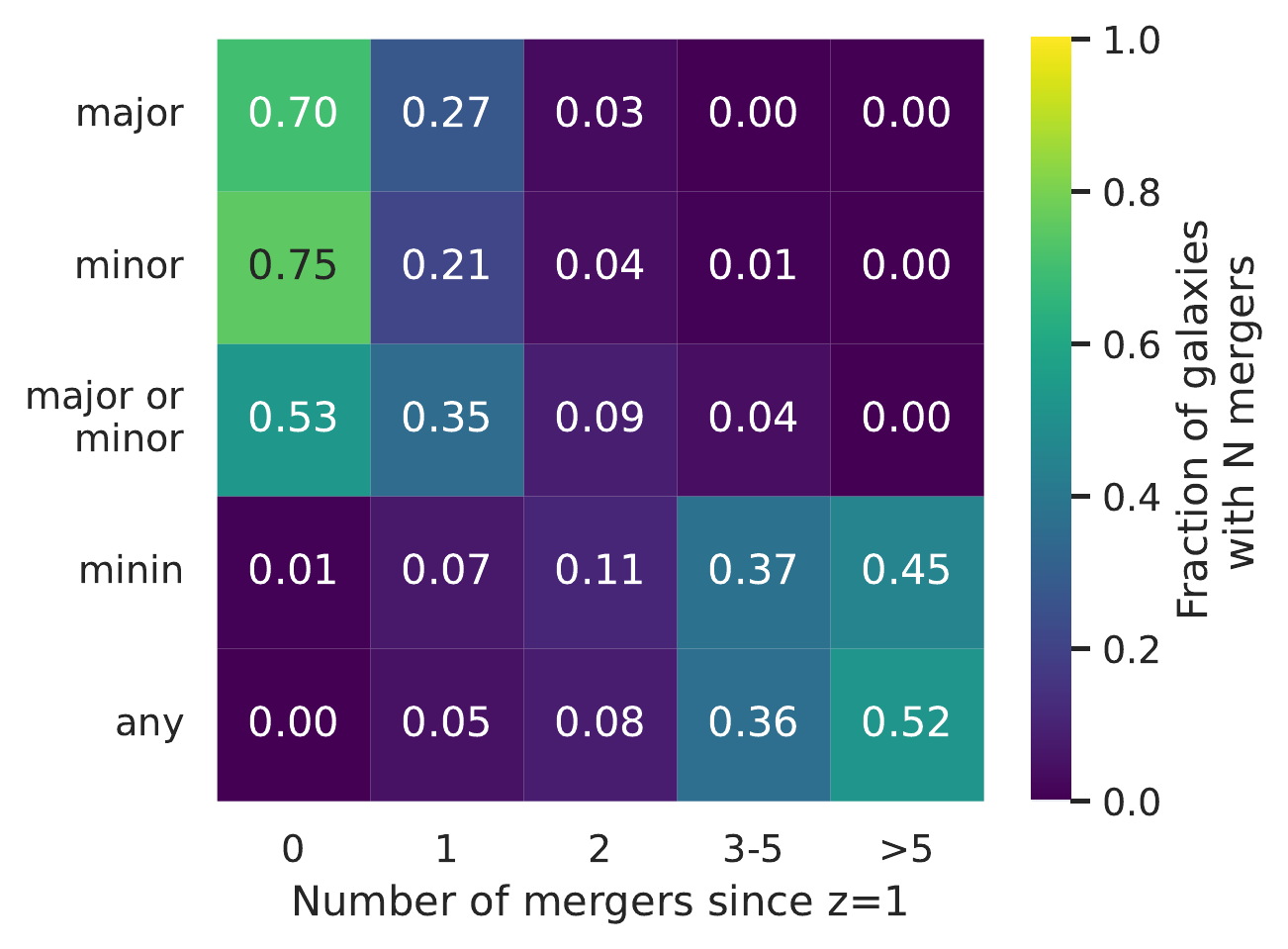}
    \includegraphics[width=.365\textwidth,trim={1.87cm 0 0 0},clip]{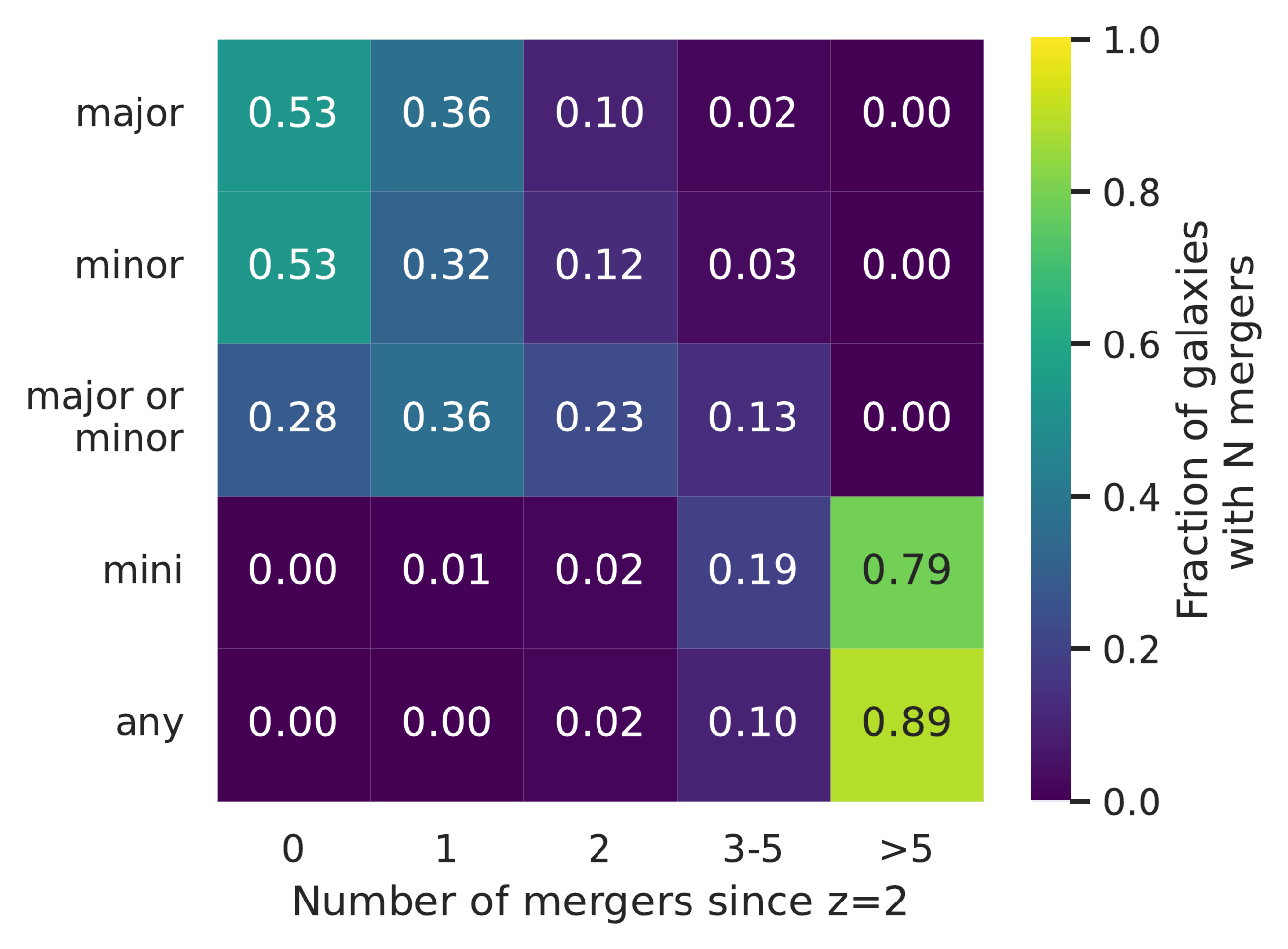}
    \caption{
    Number of mergers (minor, major, both, mini and any) in different periods of time for MW/M31-like galaxies in TNG50. In the tables, different rows correspond to different stellar mass ratios.
    \textit{Top panel:} fraction of galaxies that have undergone at least one merger in different time periods (lookback time measured from $z=0$) and different stellar mass ratios: major (red), minor (blue), minor or major (purple), mini (green) and any (black). Dots represent the exact periods for which the measurements are available.
    \textit{Bottom panels}: fraction of galaxies undergoing different types of mergers for three defined time periods. \textit{Left}: last 5 Gyr ($z\sim0.5$). \textit{Middle}: since $z=1$ (lookback time $\sim7.98\,\rm{Gyr}$). \textit{Right}: since $z=2$ (lookback time $\sim$ 10.51 Gyr). The color denotes the fraction of galaxies undergoing a certain number of mergers. These fractions are not cumulative.
    18 per cent of the MW/M31-like galaxies have recent (i.e. over the last 5 Gyr) major mergers. 34 per cent have a major merger since $z=1$. 
    }
    \label{fig:mwlike_mergersvstime_tables_plot}
\end{figure*}

The progenitors of MW/M31-like galaxies span a host mass range (within 10th-90th percentiles) at $z=2$ similar to that at $z=0$, namely $0.5-0.6$ dex: on the other hand, the progenitors of MW/M31-like galaxies at e.g. $z=2$ span almost three orders of magnitude in stellar mass across all sampled galaxies: $1.8\times10^8-2.2\times10^{10}\,\rmn{M}_{\odot}$ within the 10th-90th percentiles, i.e. 2 dex. The pathways leading to MW/M31-like galaxies at $z=0$ are rich and varied.

These theoretical predictions are of the essence to connect galaxy populations across cosmic time for the purposes of contrasting e.g. MW-like galaxies with their expected progenitors at high redshifts. For comparison, in Fig.~\ref{fig:assemblyHistory} we show the corresponding results from previous cosmological simulations of MW-like galaxies, albeit differently selected: solid and dashed thin lines. The thirty zoom-in Auriga MW analogues \citep[yellow dashed curve, for their median growth,][]{Grand2017}, where galaxies are selected as isolated haloes at $z=0$ with $M_{200c}$ in the range $(1-2)\times10^{12}\,\MS$, have a very similar median stellar assembly history to those from TNG50.
For the Illustris simulation \citep[][]{Torrey2015}, where MW analogues are all galaxies with stellar mass within the range $(4-5)\times10^{10}\,\MS$, the median stellar mass is lower at all redshifts (red solid curve).
The six NIHAO-UHD MW-like galaxies \citep[green dashed curves,][]{Buck2020} span a similarly wide range in mass growth across all cosmic epochs as TNG50. On the other hand, the Eris galaxy \citep[orange dashed curve,][]{Guedes2011} exhibits a much earlier mass growth \citep{Pillepich2015}, similarly as to the FIRE-2 MW-mass zoom-ins by \cite{Garrison-Kimmel2018}, of whom we show the growth curves for the least and most massive at $z=0$ (dotted cyan).

\begin{figure*}
\includegraphics[width=18cm]{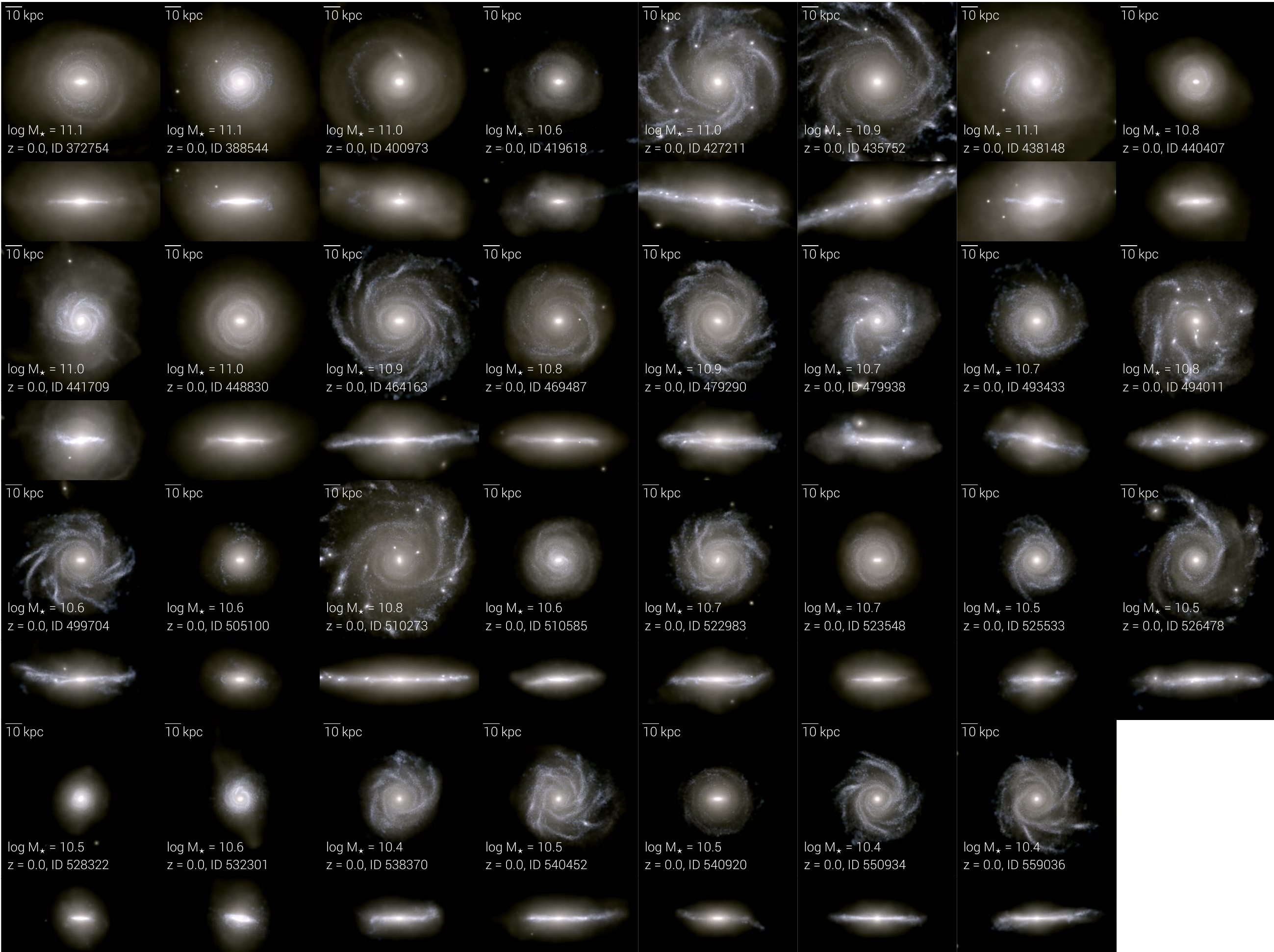}
    \caption{Stellar-light composite images of the 31 MW/M31-like galaxies from TNG50 at $z=0$ that have undergone a recent major merger (i.e. in the last 5 Gyr). Every disk galaxy is shown in face-on and edge-on projections. Each panel spans horizontally 80 kpc, and each galaxy is identified by its stellar mass and Subhalo ID number in the TNG50 $z=0$ catalog. 
    These galaxies exhibit a great diversity in stellar sizes and in terms of stellar structures within the disks (see text for comments on galaxies with Subhalo IDs 400973 and 41961): for example, there are cases with marked spiral arms and others with central bars.
    }
    \label{fig:images}
\end{figure*}

In Fig.~\ref{fig:assemblyHistory}, assembly histories obtained by assuming that galaxies preserve their number density in time \citep[e.g.][]{vanDokkum2013} are shown for contrast. As in the original Illustris simulation \citep[][]{Torrey2015, Torrey2015b}, also according to TNG50 and for MW/M31-like galaxies only, we find that the average stellar mass evolution inferred via a constant comoving number density assumption is systematically shallower than when tracking galaxies via their merger trees, with $z\sim3$ progenitors characterized by galaxy stellar masses a factor of a few larger than those obtained along the main progenitor branches of galaxies. Therefore, the mass evolution from constant comoving number density assumptions cannot be used to validate the results of zoom-in simulations \citep{Buck2020}.

\subsection{Merger statistics}

Fig.~\ref{fig:mwlike_mergersvstime_tables_plot} shows the merger statistics of MW/M31-like galaxies according to TNG50, selected as described in Section~\ref{sec:sample_selection}. To our knowledge, this is the first time that such statistics, based on galaxy stellar masses rather than halo masses, are quantified and are obtained from a cosmological hydrodynamical simulation where MW/M31-like galaxies are selected based on observable (rather than host halo) properties.

The upper panel shows what fraction of MW/M31 analogues undergo at least one merger over varying past periods of time (in lookback Gyr from $z=0$), for different stellar mass ratios: major, minor, major or minor, mini and any ratio. In the lower panels, we give the fractions of galaxies that experienced varying number of mergers, over three time periods: from left to right, in the last 5 Gyr (i.e since z $\sim$ 0.5), since $z=1$ (lookback time $\sim$ 7.98 Gyr), and since $z=2$ (lookback time $\sim$ 10.51 Gyr). The color code in the lower panels represents the fraction of MW/M31-like galaxies that underwent a certain number of mergers -- these fractions are not cumulative. In all panels, mergers are counted based on the times of coalescence, i.e. the time of the mergers as defined in
Section~\ref{sec:merger_trees}.
About thirty per cent of MW/M31-like galaxies have undergone at least one major or minor merger (i.e. at least one merger with stellar mass ratio larger than 0.1) over the last 5 billion years: this fraction increases to 48 (72) per cent since $z=1$ ($z=2$).

Interestingly, for MW/M31-like galaxies, on average, the frequency of past major mergers is similar, if not slightly larger, than the frequency of minor mergers. Not only are the fractions of galaxies undergoing at least one minor or one major merger similar (upper panels), but also comparable are the fractions of galaxies undergoing different numbers of minor and major mergers since a fixed period of time (lower panels).

Whereas it has often been assumed, also prior to the newest results with Gaia, that our Galaxy has had a very quiet merger history at least since $z\sim1$ with no major mergers since then (see Introduction), according to TNG50 and our selection, the fraction of  MW/M31-like galaxies that have merged with at least one other similarly-massive galaxy since $z\sim1$ is about 30 per cent. This fraction is slightly lower, 27 per cent (35 of 130 galaxies), if we only consider TNG50 analogues with stellar mass below $10^{10.9}\rmn{M}_{\odot}$, i.e. if we exclude Andromeda-mass galaxies.

From the lower panels of Fig.~\ref{fig:mwlike_mergersvstime_tables_plot}, it can also be seen that 9 per cent of MW/M31-like galaxies have not merged with any other galaxy, irrespective of its stellar mass, over the last 5 billion years. On the other hand, a handful (12 per cent) of the selected TNG50 galaxies have experienced multiple major mergers since $z=2$, as many as between 3 and 5. Yet, more than one merger event per galaxy (major or minor) since e.g. $z=1$ is infrequent\footnote{For future references and analyses, the $z=0$ Subhalo IDs of the TNG50 MW/M31-like galaxies with two or more major mergers since $z=1$ read: 400973, 435752, 441709, 479938, 526478, 538370; and in the last 5 Gyr: 441709.}.

It is important to notice that the numbers summarized in Fig.~\ref{fig:mwlike_mergersvstime_tables_plot} are based on a full-physics (i.e. not DM only) model for galaxy formation in the full cosmological context, they apply to a specific galaxy selection, and the merger mass ratios are characterized in terms of the stellar mass (and not DM mass) of the merging objects. Fig.~\ref{fig:mwlike_mergersvstime_tables_plot} can be considered as the stellar-mass based update of e.g. Figs. 5 and 6 of \citealt{Stewart2008}, which were based on N-body only simulations, and the TNG updates of the merger estimates for MW-mass haloes given for Illustris by \citealt{RodGom2015} and quoted in the Introduction. 


\begin{figure*}
\includegraphics[width=\columnwidth]{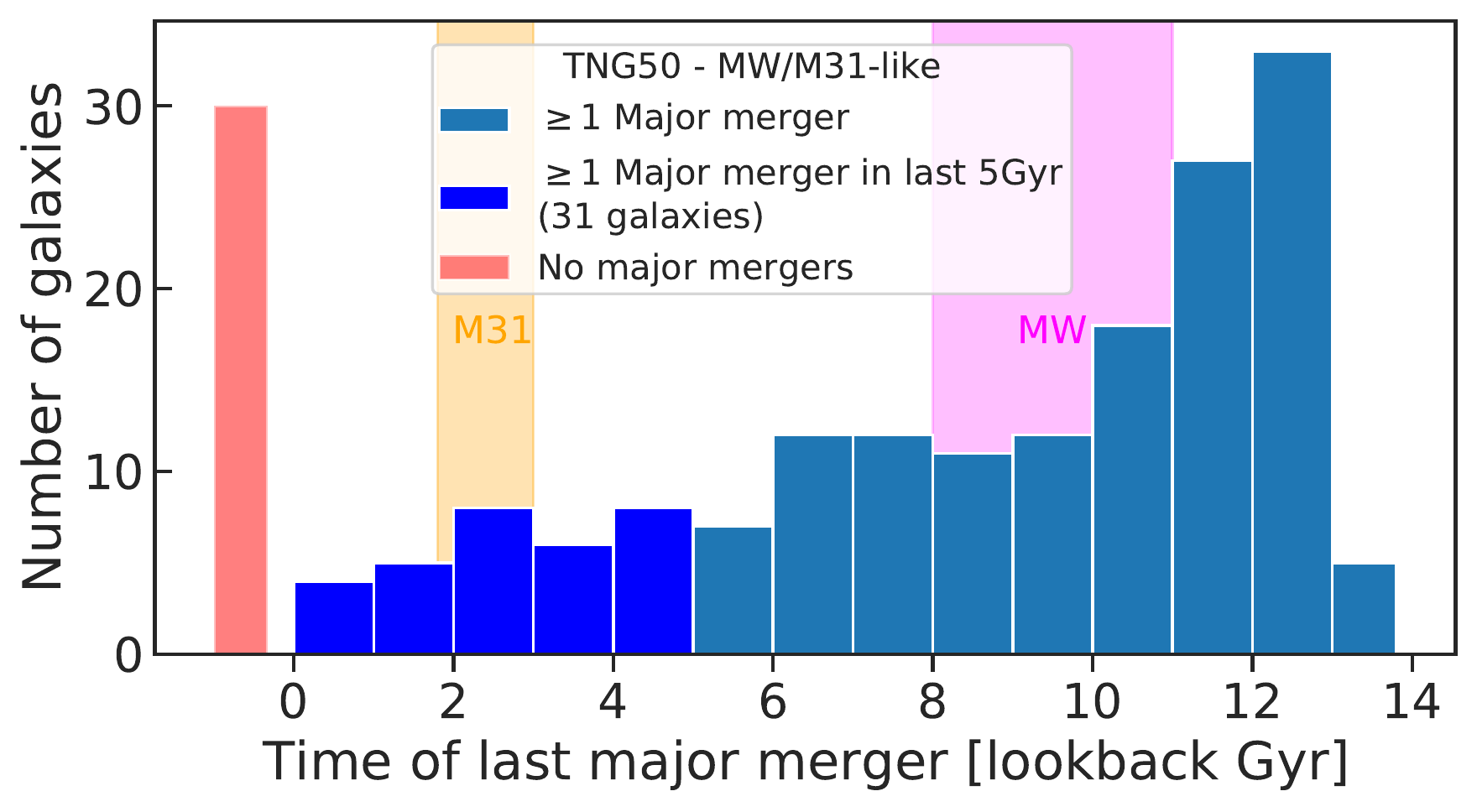}
\includegraphics[width=\columnwidth]{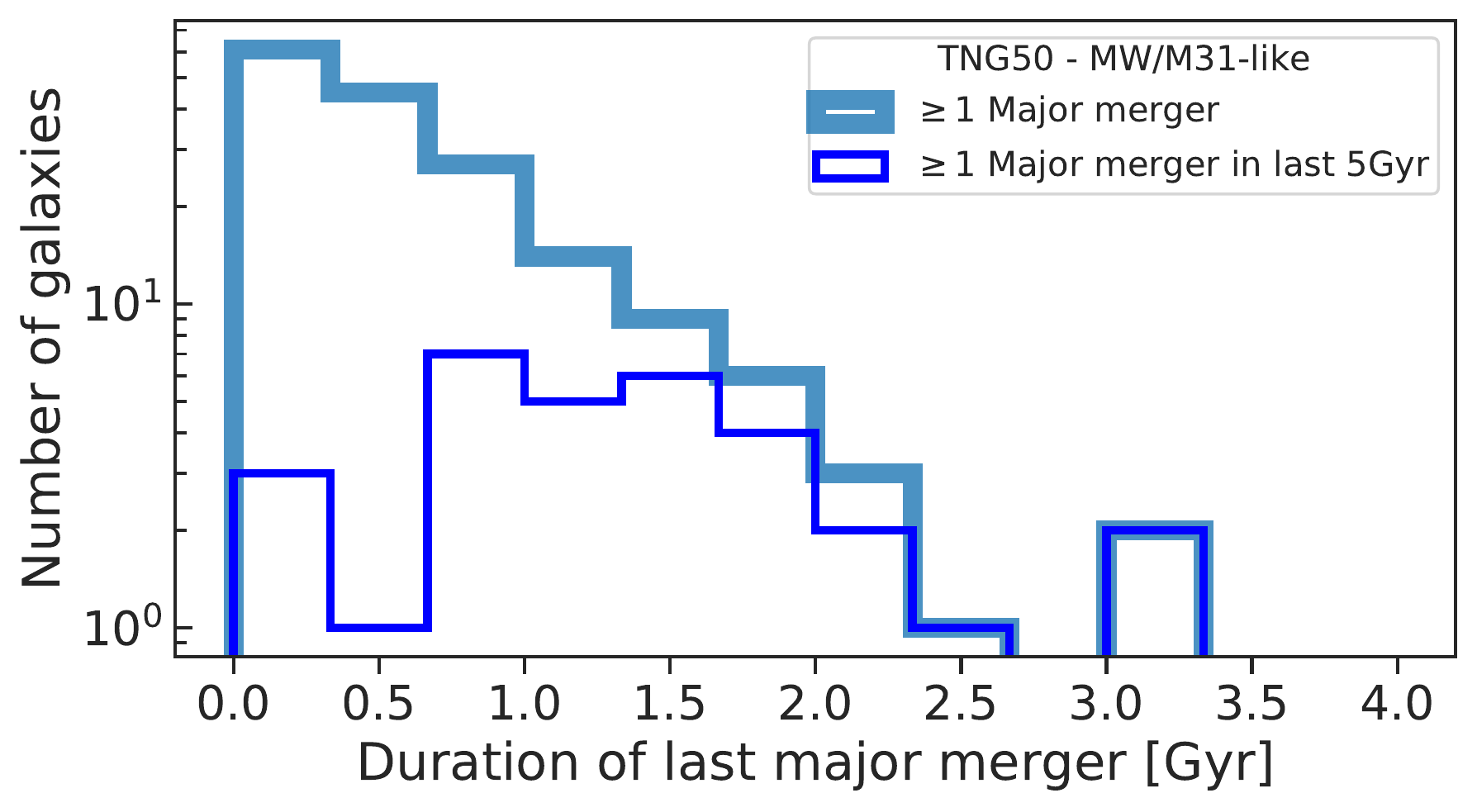}
\includegraphics[width=\columnwidth]{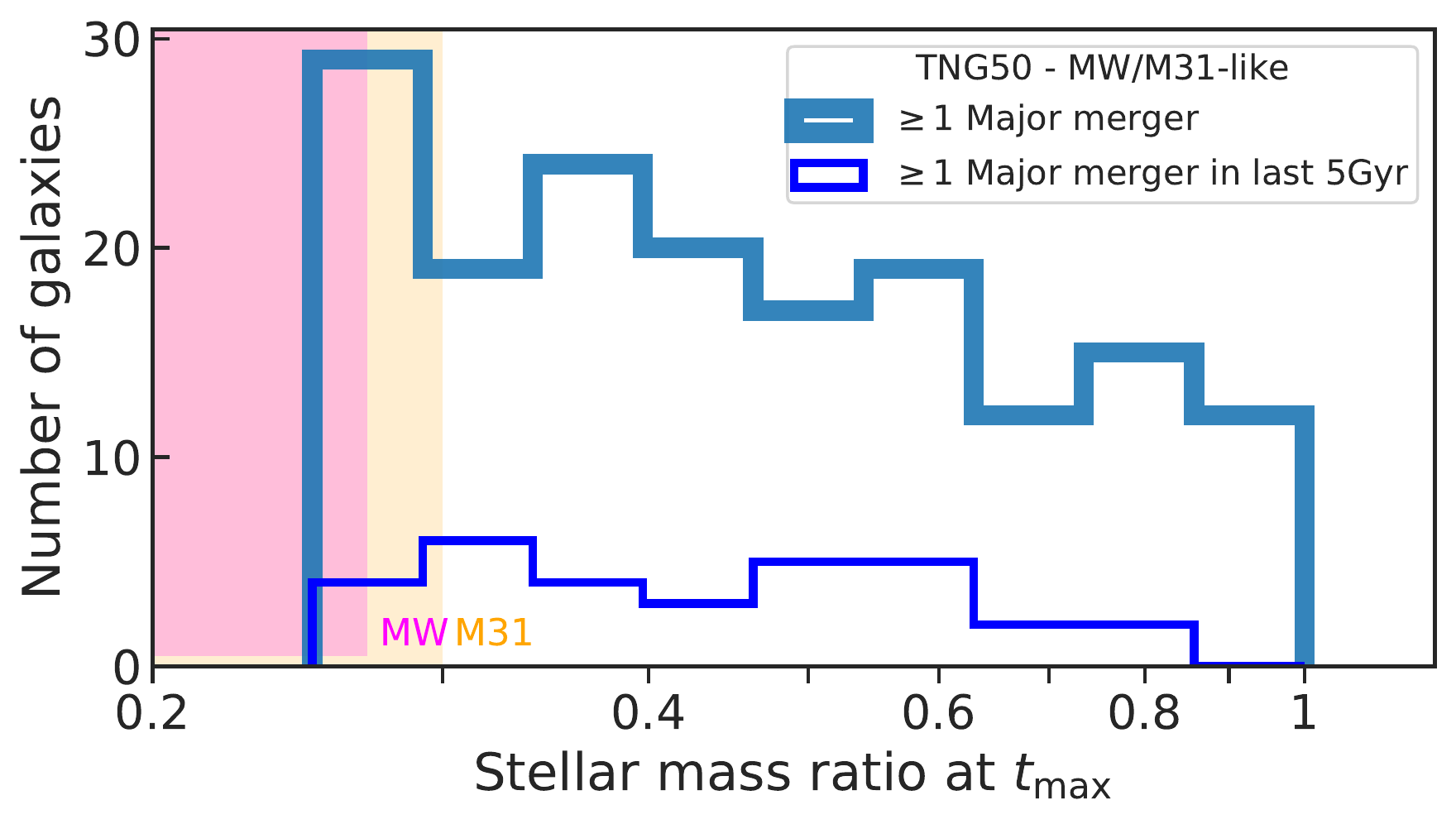}
\includegraphics[width=\columnwidth]{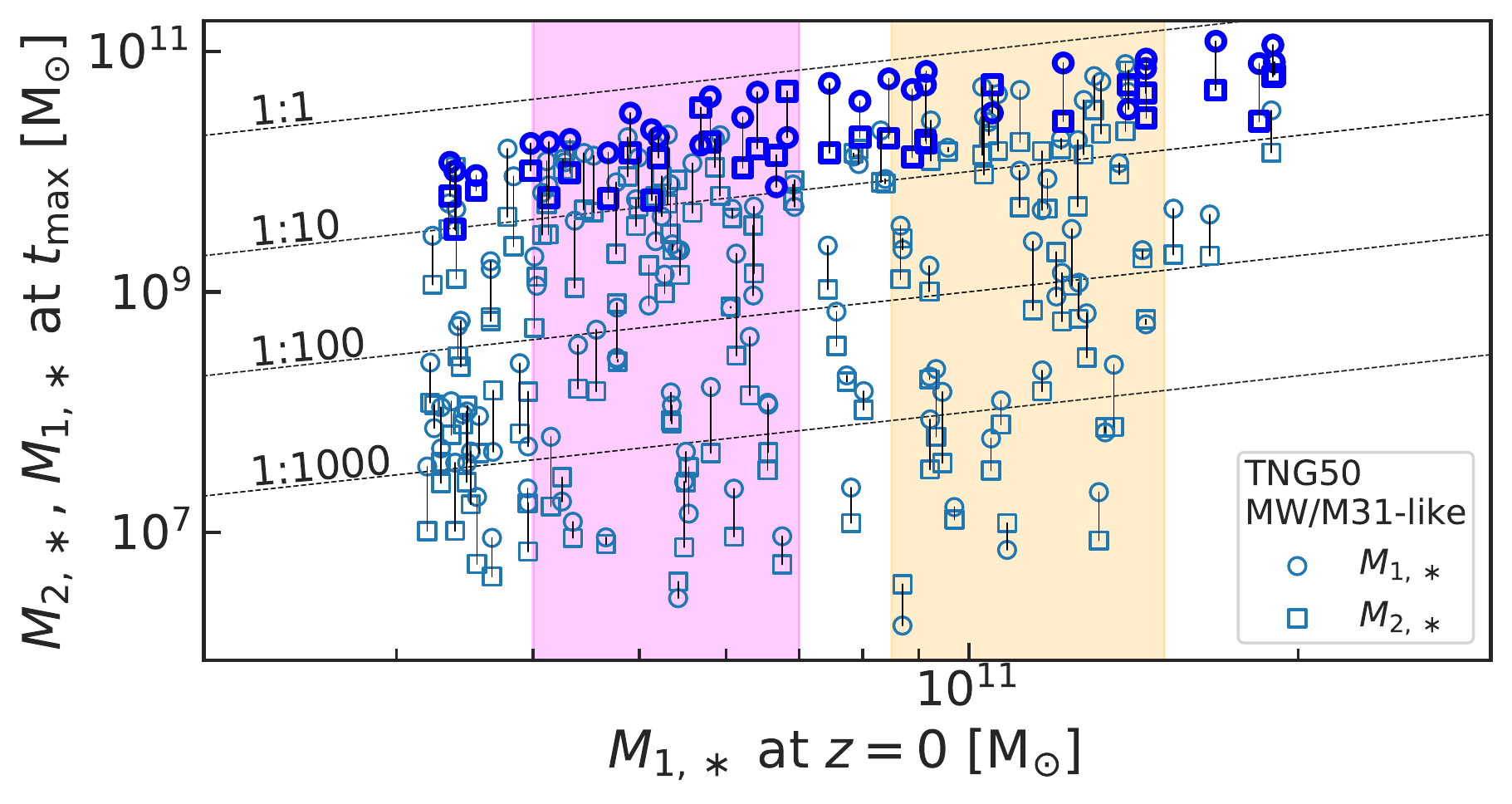}
    \caption{Characteristics of the last major mergers of MW/M31-like galaxies in TNG50 at $z=0$: major mergers. In all panels, MW/M31-like galaxies that experienced their final major merger in the last 5 Gyr are depicted in blue; MW/M31 analogues with their last major merger at any cosmic time are indicated in light blue. 
    \textit{Top left}: Time of the last major merger (in lookback Gyr) for each of the 198 galaxies, in bins of $1\,$Gyr: 30 MW/M31-like galaxies did not undergo any major mergers (orange bin); 95 galaxies (approximately 50 per cent) underwent their last major merger more than 9 Gyr ago ($z\gtrsim1.3)$; 31 galaxies have had a late major merger, i.e. as recent as in the last 5 Gyr. The occurrence of the latter is distributed in an approximately uniform way in the considered period of time.
    \textit{Top right}: Duration of the major mergers, defined as the time elapsed since the secondary reached its maximum in stellar mass ($t_\rmn{max}$) and the moment of coalescence -- bins span 330 million years. 77 per cent of the major mergers lasting longer than $1.5\,$Gyr are recent. 
    \textit{Bottom left}: Stellar-mass ratios between the secondary and the primary, at $t_\rmn{max}$. \textit{Bottom right}: Stellar mass of both progenitors, at $t_\rmn{max}$, versus the stellar mass of the galaxy at $z=0$. For the Galaxy and Andromeda, we show observational estimates with magenta and orange annotations, respectively: the mass-ratio and the time of the known last major merger are taken for the case of GES \citep[][]{Helmi2018, Gallart2019, Naidu2021} and for M32 \citep[][]{DSouza2018}.
    }
    \label{fig:lastMajorMerger}
\end{figure*}

\section{Mergers and disk survival}
\label{sec:disc_survival}
The findings in the previous Section imply that a selection of TNG50 galaxies at $z=0$ have global properties -- stellar mass, stellar diskyness, and large-scale environment -- similar to those of the Galaxy and Andromeda and yet have undergone at least one major merger as recently as over the last 5 Gyr. In particular, 31 of the 198 MW/M31 analogues of TNG50 (16 per cent) are found in this subsample and are hence the focus of the rest of the paper. The choice to focus on galaxies with the time of their last major merger at 5 billion years ago, instead of e.g. 6 or 4 billion years ago, is somewhat arbitrary but is simply meant to qualitatively encapsulate the phenomenology of {\it relatively recent} major mergers. We will drop the binary classification whenever instructive and possible.

\subsection{MW/M31-like galaxies as survivors of recent major mergers}
\label{sec:images}

In Fig.~\ref{fig:images}, we show stellar-light composite images at $z=0$ for the subsample of TNG50 MW/M31 analogues with a major merger within the last 5 billion years, in face-on and edge-on projections. But for a couple of cases -- Subhalo IDs 400973 and 419618, for which the identification as MW/M31 analog is in fact borderline -- they pass the morphological selection criterion based on the shape of the inner stellar mass distribution but have D/T mass ratio of $\sim0.2$ --, these galaxies exhibit clear stellar disk morphologies, although with a great variety in structural properties and extents. Their median D/T mass ratio at $z=0$ (as defined in Section~\ref{sec:galStarProperties}) is 0.40, but ranges between $\sim0.12$ and $\sim0.78$ across the whole sample of Fig.~\ref{fig:images}. For comparison, the D/T ratios of the full MW/M31-like sample range from $\sim$0.10 -- 0.90  with a median of 0.55 and six systems below 0.1, all with counter-rotating structures \citep[see also][]{Joshi2020}.

The geometrically-thin stellar disks of MW/M31-like galaxies with recent major mergers have sech$^2$ heights measured at 4 times the disk scale length spanning between $\sim$50 pc and 2.9 kpc, with median and average of 0.93 and 1.14 kpc (see Section~ \ref{sec:properties} for more details). In numerous cases, the stars are organized in very prominent spiral arms and grand-design spiral systems, which are sometimes distributed asymmetrically. There are also central bars with an extension of $\sim1-4$ kpc (e.g. Subhalo IDs 522983, 523548 and 540920). From the edge-on views, we can appreciate that the stellar haloes are also diverse, even among galaxies with similar stellar disk size: from faint haloes of stars that barely extend beyond the disk, to others that are appreciable in the images for up to a few tens of kpc above and below the disks. A quantification of the mass in the different stellar components (disks, bulges, and haloes) will be given in Section~\ref{sec:properties}.

In Fig.~\ref{fig:lastMajorMerger}, we provide additional statistics and properties of the last major mergers experienced by the MW/M31-like galaxies with a recent major merger (blue), in comparison to those of the whole MW/M31-like sample. Firstly, as appreciable also from Fig.~\ref{fig:mwlike_mergersvstime_tables_plot}, 15 per cent, i.e. 30 galaxies among the TNG50 MW/M31 analogues have never undergone a major merger, i.e. never since $z\sim5$, see Method Section for details. In the top left panel of Fig.~\ref{fig:lastMajorMerger}, these are reported as an orange bar: 6 of those are M31-mass galaxies.

If we look at the times when the last major mergers of MW/M31-like galaxies occurred (time of coalescence, top left panel of Fig.~\ref{fig:lastMajorMerger}), we see that 83 galaxies (42  per cent) experienced their last major merger more than 10 Gyr ago, whereas 95 galaxies (48 per cent) had their last major merger more than 9 Gyr ago -- this compares to $\sim$35 per cent for $M_{*}\sim10^{10}\, \MS$ disk galaxies according to \cite{Font2017}). Importantly, the MW/M31-like galaxies with a recent major merger are not particularly biased in their merger times and span the entire final 5 billion years of cosmic evolution. Magenta and orange shaded vertical bands denote the current estimates of the last major mergers of the Galaxy, namely of GSE \citep{Helmi2018, Myeong2018, Chaplin2020}, and of Andromeda \citep{DSouza2018}, respectively.

In the top right panel, we see that the durations of the last major mergers (see Section~\ref{sec:merger_trees}) span from $\sim$ 0.04 to $\sim$ 3.5 Gyr. The longest mergers are more frequently recent, owing to the shorter dynamical times when the Universe was younger: 77 per cent of the mergers lasting longer than 1.5 Gyr occurred within the last 5 billion years. On average, TNG50 MW/M31-like galaxies experienced major mergers that carried on for 0.7 Gyr (median of 0.5 Gyr), this duration increasing to 1.4 Gyr for those happening since $z\lesssim0.5$ (median of 1.3 Gyr).
Dividing the TNG50 MW/M31-like sample according to stellar mass, and considering only the galaxies that underwent at least one major merger over their history, the last major merger of MW-mass (M31-mass) galaxies occurred on average about 10 (9) billion years ago. The median merger duration is approximately 0.61 and 0.46 Gyr for the two sub-samples. 

The stellar mass ratios of the merging galaxies are quantified in the bottom panels of Fig.~\ref{fig:lastMajorMerger}, evaluated at $t_\rmn{max}$ (the time of the maximum stellar mass of the secondary, see Section~\ref{sec:merger_trees}). All ratios in the bottom left panel are above 0.25, by our definition of a major merger. The mass ratios closer to 1 are more frequent for older mergers (see our definitions of merger mass ratios in Section \ref{sec:merger_trees}). Only one of the 31 recent major mergers has a mass ratio larger than 0.75. Consistently with the hierarchical model of galaxy assembly, the most recent major mergers have in general lower stellar-mass ratios: galaxies at recent times are more massive 
but massive galaxies that could merge with them are rarer.
The absolute values of the galaxy stellar mass of the merging progenitors are shown in the bottom right panel of Fig.~\ref{fig:lastMajorMerger}. MW/M31-like galaxies with recent major mergers (blue symbols) occupy the highest part of the range, at fixed stellar mass, and are compared to the progenitors of all MW/M31-like galaxies with the last major merger happening at any time (light blue symbols): the former subset of galaxies increases their mass by a small factor in the last few Gyrs, so we do not find galaxies with a recent major merger where both progenitors have low stellar mass and the descendant grows then rapidly enough to be included in our MW/M31-like mass cut. Magenta and orange vertical shaded areas denote the current stellar mass constraints of the Galaxy and Andromeda, respectively: according to TNG50, even galaxies with lower, MW-like mass may experience a recent major merger and still be disky at $z=0$.

The figures above show that, according to TNG50, MW/M31-like galaxies with recent major mergers have interacted with relatively massive companions for substantial amounts of times, i.e. on average for $\sim$ 1.4 Gyr with secondaries of $M_\rmn{\ast} \approx 2\times10^{10} \rmn{M}_{\odot}$, resulting in mergers of median stellar mass ratio of 0.41 (i.e. 1:2.5).

\subsection{Gas availability during the mergers}
\label{sec:gas_availability}

But how is it possible that, despite having undergone major mergers as recently as in the last 5 billion years, galaxies can nevertheless exhibit marked disky stellar morphologies at $z=0$?

\begin{figure}
	\includegraphics[width=\columnwidth]{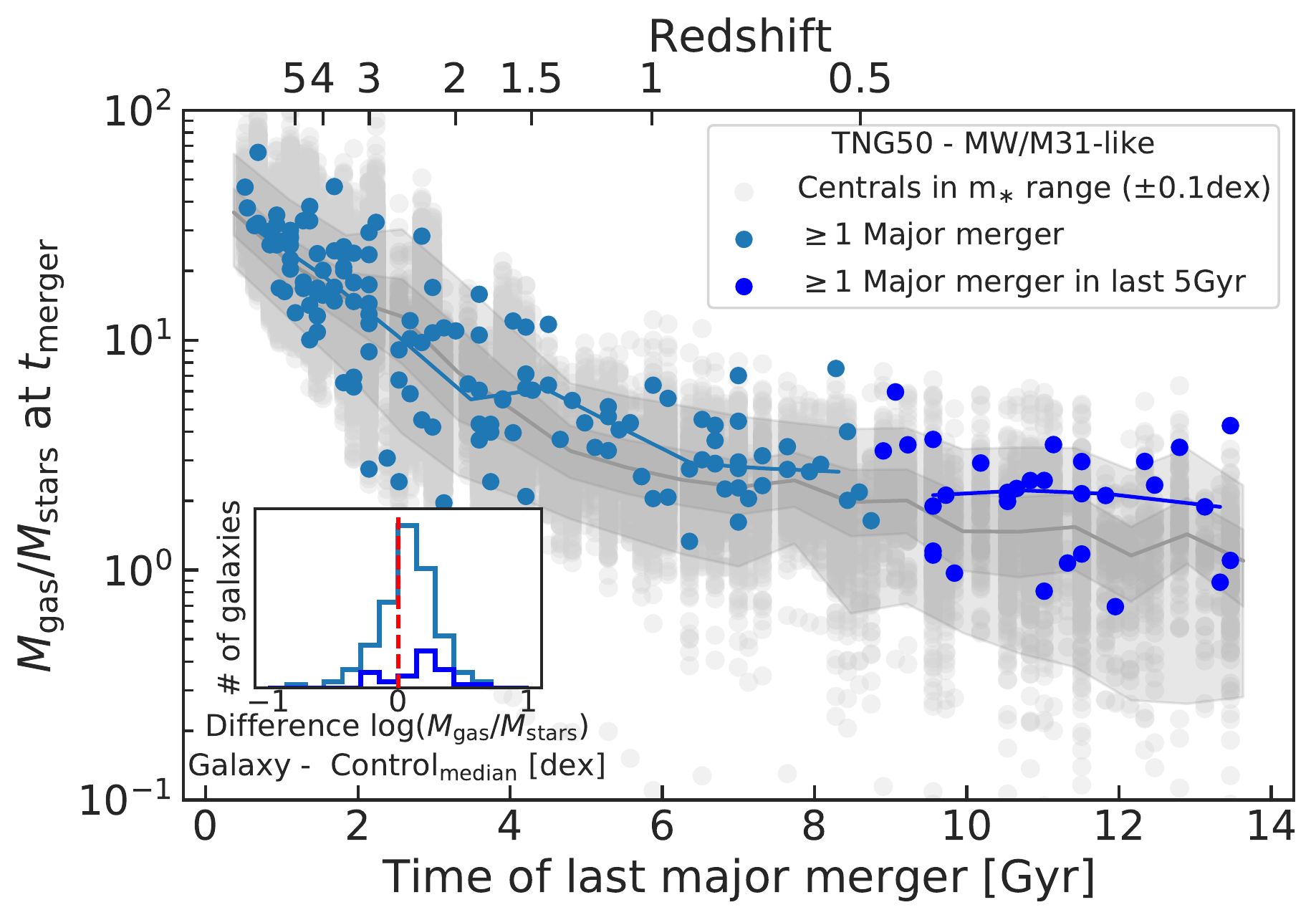}
    \caption{Gas mass fraction ($M_\rmn{gas}$/$M_\rmn{stars}$) of TNG50 MW/M31-like galaxies at the time of their last major merger (time of coalescence). For comparison, in gray, we show the gas fractions of central galaxies in the same stellar mass range ($\pm 0.1$ dex) at the corresponding merger times: the median is the solid curve. MW/M31-like galaxies with recent major mergers (blue symbols) have lower gas-to-star mass fractions, in absolute terms, than those with more ancient mergers -- i.e. the gas mass fraction of galaxies decreases with time. However, they are more gas rich ($0.1-0.2$ dex) than the average central galaxies in TNG50. This gas richness can also be appreciated in the inset histogram, where we show the logarithmic difference of the gas mass fraction, for each of the galaxies, with respect to the median of the control sample in the considered time of the last major merger.}
    \label{fig:gasFracTmerger}
\end{figure}

Previous studies \citep[][]{Hopkins2009, Stewart2009, Hoffman2010} had shown with idealized simulations of mergers that the gas content in the merging progenitors is a determinant factor for the outcome of a merger -- see Introduction: it both conditions how destructive a merger is for the stellar component of the involved galaxies and how actively stars can form in the descendant galaxy. In Fig.~\ref{fig:gasFracTmerger}, we hence analyze the gas availability during merger events for all the  TNG50 MW/M31 analogues undergoing major mergers across cosmic epochs. Firstly, we estimate the gas content of each galaxy by accounting for all the gravitationally-bound gas at the time of the merger: this choice is supported e.g. by \citet[][]{Sparre2021}, who show with simulations that the gas in the circumgalactic medium and even in the outer halo can contribute to the star formation after major merger events between $z=0.3-0.8$ that produce disky, MW-like galaxies. Secondly, we compare the gas mass fractions of each system at coalescence (i.e. the gas mass over the total stellar mass) with that of central galaxies that, at the corresponding major-merger time of TNG50 MW/M31-like galaxies, have the same stellar mass (within a range of $\pm0.1\,\rmn{dex}$) as the galaxy that has resulted from the two progenitors.
%
These galaxies that serve as reference are plotted in gray, with the median gas fractions marked with a gray solid line and the narrower and the broader shaded regions representing, respectively, the 25th-75th and the 5th-95th percentiles. MW/M31 analogues are marked as blue symbols and a median is added to facilitate comparison.

Two facts are noticeable. First, as anticipated (see also e.g. \citealt{Pillepich2019} for TNG50 results) and as it is expected from observations of high-redshift galaxies, the average mass fraction of gas in galaxies decreases with cosmic time -- for the control sample, the gas mass fraction drops by $\sim$1 dex from $z\sim3$ (about 2 Gyr after the Big Bang) to $z\sim0.5$ (about 5 Gyr ago). Generally, the MW/M31 analogues show a gas fraction, on average, comparable to that of the control sample of central galaxies in the similar mass range. Second, and importantly, the difference between samples increases slowly with cosmic time: we think that this is the very manifestation of the fact that galaxies that are selected to be disky at $z=0$ while having undergone a recent major merger constitute somewhat a biased subset. Namely, MW/M31-like galaxies with recent major mergers exhibit somewhat higher gas fractions ($\sim 0.1-0.2$ dex) than the average of other non-merging galaxies with similar mass at the corresponding epoch: 23 of 31 of them lie above the median value of the control sample. Whereas the gas-mass difference is not large, it appears sufficient to trigger star formation in TNG50 galaxies during the merger events, as we show next.

\subsection{Star formation bursts triggered by gas-rich major mergers}
\label{sec:SFRsparks}

The availability of gas enables star formation during the recent major mergers of TNG50 MW/M31-like galaxies. 

\begin{figure}
	\includegraphics[width=\columnwidth]{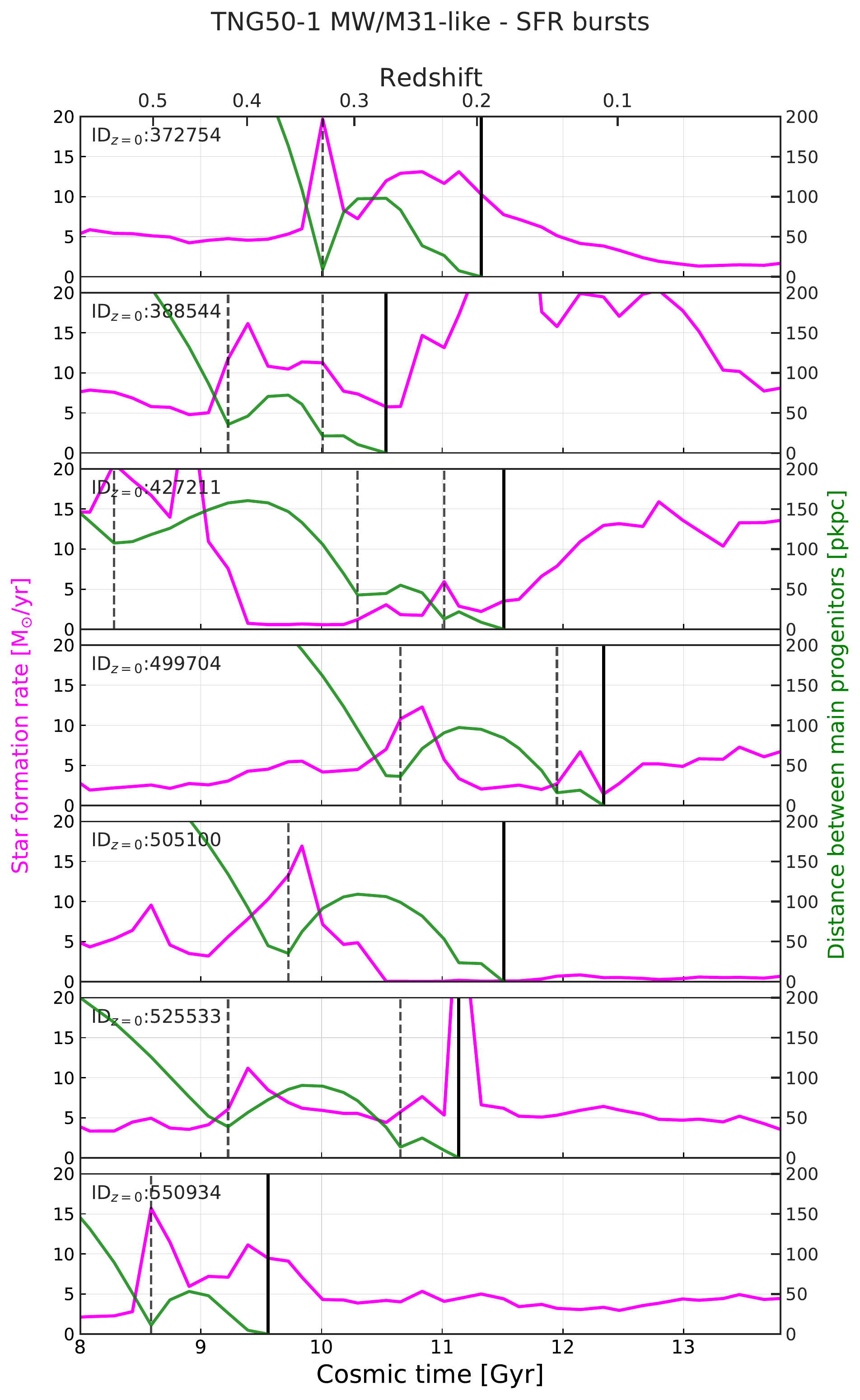}
    \caption{Bursts of star formation in a selection of MW/M31-like galaxies in TNG50 with recent major mergers. SFR in the main progenitor (magenta) and distance between both progenitors (green) are plotted for seven example galaxies, one panel each. The time of the last major merger is marked with a solid vertical line; the last pericentric passages of the secondary progenitor are marked with dotted vertical lines. The galaxy mergers and the close galaxy-galaxy interactions prior to coalescence can trigger substantial bursts of star formation. The SFR after the merger varies depending on the galaxy.}
    \label{fig:SFRbursts}
\end{figure}

We have examined the SFR evolution of the TNG50 MW/M31 analogues with recent major mergers and find that their SFRs can increase substantially in correspondence to their major merger events. Such SF bursts may occur at the time of coalescence or shortly after, and sustained SF may be in place in the descendants through $z=0$. In fact, SF bursts may happen even before the merging galaxies coalesce, specifically at the close pericentric passages of the secondary progenitor in its approaching orbit. About 24 of the 31 MW/M31 analogues with recent major mergers show appreciable bursts of SF triggered by the merger, seven of which are the depicted in Fig.~\ref{fig:SFRbursts}.

There we show the evolution of the instantenous SFR along the main progenitor branch of selected galaxies (magenta)
as a function of cosmic time (starting at 8 Gyr after the Big Bang), compared with their distance (in physical kpc) to the secondary progenitor involved in their last major merger: green curves. For each galaxy, i.e. row, the time of merger is represented with a vertical black solid line, whereas the last pericentric passages are distinguishable as the minima in the distance curves: dotted black vertical lines. It can be clearly seen that, not only coalescence, but also close galaxy-galaxy interactions prior to the merger time can trigger important events of star formation in the main galaxy and consequently can plausibly alter its structure, morphology, and stellar mass content.

It was previously shown that highly ``bursty'' SF was suppressed in galaxy major mergers of the original Illustris simulation due to its insufficient numerical resolution: in particular, this was demonstrated via zoom-in simulations of selected merger events with 40 times better mass resolution \citep{Sparre2016}. Fig.~\ref{fig:SFRbursts} shows that the numerical resolution of TNG50 is sufficient to capture the compression of gas possibly due to the galaxy-galaxy interactions, the funnelling of gas towards the galaxy centers, higher gas density and hence to reproduce bursts of star formation triggered by mergers and galaxy interactions.

\subsection{The cases of disks destroyed during the mergers and reformed vs. those surviving during the merger}
\label{sec:merger_cases}

By inspecting the time evolution of the SFR of the main progenitors of each MW/M31 analogue with recent major mergers and the time evolution and distributions of the orbital properties of their stars (as in Figs.~\ref{fig:circularitiesLMMdestroysDisk} and \ref{fig:circularitiesLMMDiskSurvives}), we find that two main scenarios or pathways are followed by the 31 TNG50 galaxies to be disky at $z=0$ after undergoing a recent major merger. These are as follows:
\begin{enumerate}
  \item in 18 cases (58 per cent), the galaxy's stellar disk, being in place prior to the time of the merger or prior to the pericentric passages, is destroyed by the merger; nevertheless, the galaxy by $z=0$ reforms a stellar disk;
  \item in 11 cases (35 per cent), the stellar disk is not completely destroyed by the merger and therefore the galaxy retains its disky morphology down to $z=0$;
\end{enumerate}

Two remaining galaxies of the sample are difficult to classify. In one case (TNG50 Subhalo ID 419618 at $z=0$), the D/T mass ratio is actually quite low both prior to and after the merger as well as at $z=0$: $\sim0.1-0.2$ -- we noticed this galaxy in Fig.~\ref{fig:images} and possibly this should not have been included in the TNG50 MW/M31-like sample in the first place. In the case of TNG50 Subhalo ID 400973 at $z=0$, the D/T mass ratio around the time of the merger is biased low by the fact that (possibly another) merger event has triggered the production of new stars in counter-rotating orbits, making the study of its evolution rather specific. In both cases, the galaxies have experienced a major merger as recently as in the last one or two snaphots, i.e. in the last $150-300$ million years, so much so that stellar shells are somewhat visible in their stellar light maps. 

\begin{figure}
	\includegraphics[width=1.1\columnwidth]{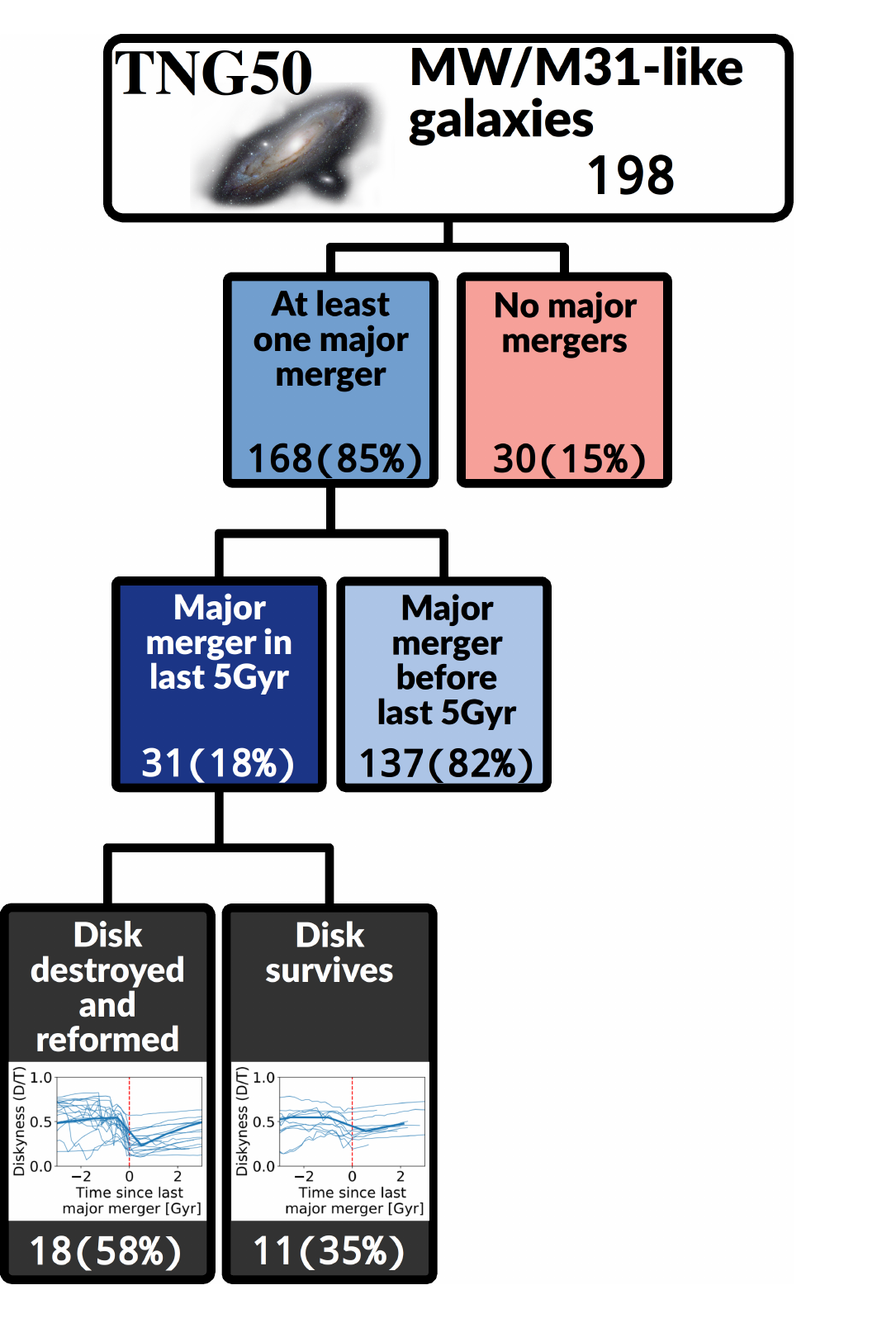}
    \caption{Summary of the evolutionary pathways uncovered via the TNG50 simulation for a sample of galaxies selected at $z=0$ as MW/M31 analogues. MW/M31-like galaxies may have had at least one major merger throughout their history or not at all: 85 and 15 per cent, respectively -- second panels from the top). Among the former, third layer from the top, 31 galaxies, i.e. 18 per cent of the TNG50 MW/M31-like sample with at least one major merger ever, actually underwent one or more major mergers as recently as in the last 5 billion years. For most of them (58 per cent), the previously-existing stellar disk was destroyed but reformed; for another substantial fraction (35 per cent), the previously-existing stellar disk survived through the merger event. Inset diagrams show the evolution of D/T with time, centered at the time of the merger. Thin (thick) curves represent individual galaxies (medians). For the destroyed disk the drop of diskyness is much more noticeable. Two of 198 galaxies cannot be placed in either scenario -- see text for details.  
    In each box, we report the number of galaxies in absolute and relative values, being the latter calculated within the downstream sub-samples.
    }
    \label{fig:diagramLMM_NumGalaxies}
\end{figure}

We have also explicitly checked whether cases exist whereby the recent major merger occurs at a time when the progenitor of the selected MW/M31-like galaxies does not exhibit yet a clear disky stellar morphology (or not anymore due to a previous merger). But for the two galaxies described above, among the 31 TNG50 MW/M31 analogues, this does not occur. 

The frequency of major merger events in MW/M31 analogues and the different evolutionary pathways that the TNG50 simulation has allowed us to uncover are summarized in Fig. \ref{fig:diagramLMM_NumGalaxies}. In the various boxes from top to bottom we show the number of TNG50 MW/M31-like galaxies undergoing certain conditions and paths. 

Somewhat similar scenarios had been found in previous studies, albeit without a specific focus, and hence selection, on the MW/M31 mass scale. For example, \citet{Jackson2020} used the Horizon-AGN cosmological simulation and reported about two main pathways for the existence of massive galaxies with disky stellar morphologies at $z=0$  ($M_{*} \geq 10^{11.4}\,\MS$, above our upper end): surviving disks (30 per cent), owing to anomalously quiet merger histories, and spheroidal galaxies that form a young stellar disk (70 per cent) as a consequence of star formation triggered by a recent gas-rich mergers. According to \citet{Font2017} and the GIMIC simulation, on the other hand, stellar disks reform for one third of the $M_{*}\sim10^{10}\,\MS$ galaxies undergoing a major mergers in the last 7 Gyr. Also, \citet[][]{Sparre2017} showed that four $M_{*}\sim10^{10.5-11}\,\MS$ disk galaxies that were disky before their last major merger (at $z\sim0.5$) were able to reform the disk after the merger destroyed it.

\begin{figure*}
	\includegraphics[width=1.8\columnwidth,height=0.85\textheight]{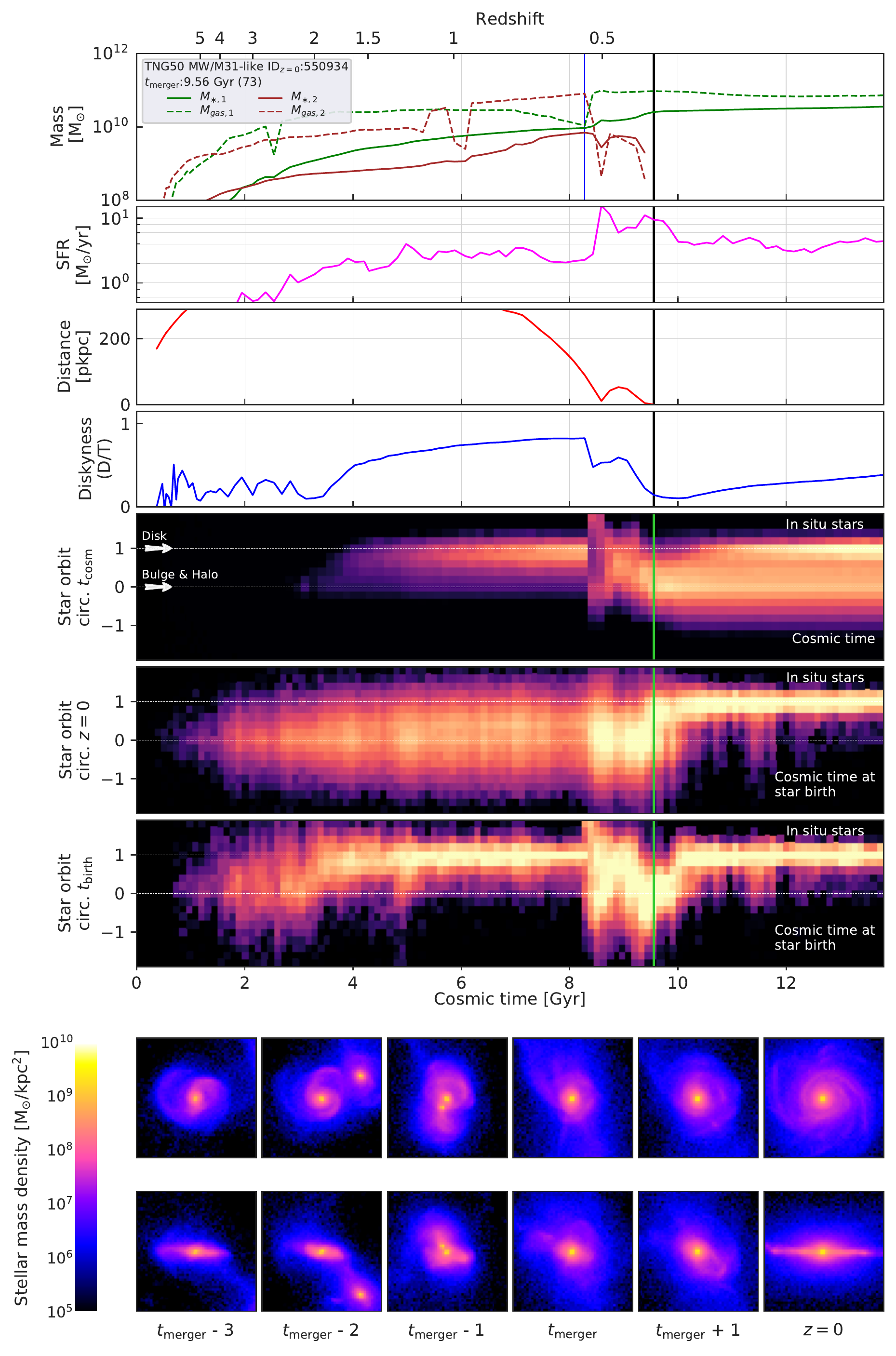}
    \caption{Evolutionary tracks of an example TNG50 MW/M31-like galaxy whose last major merger destroys the stellar disk, but a new one reforms.
    In all panels, but the images at the bottom, values are shown as a function of cosmic time. The vertical thick solid line represents the time of the last major merger; the thin vertical line in the top panel denotes the time when the stellar mass ratio is evaluated.
    \textit{Top panel:} Time evolution of the gravitationaly-bound stellar and gas mass (solid and dashed line, respectively) for the primary and secondary progenitors.
    \textit{Second to fourth panels:} SFR of the main progenitor (fuchsia), distance between progenitors (red), and diskyness i.e. D/T mass ratio (blue -- see definition in Section~\ref{sec:galStarProperties}).
    \textit{Fifth to seventh panels:} Distributions of the orbit circularities of stellar particles. From top to bottom: circularities of the in-situ stars along the main-progenitor branch of the galaxy; in-situ circularities evaluated at $z=0$ and shown as a function of the birth time of the stars; in-situ circularity evaluated at the time of birth and shown as as a function of birth time. 
    \textit{Bottom panels:} Stellar density images of the main progenitors, centered at the main galaxy, face-on and an edge-on views, across snapshots separated by about 150 Myrs.
    }
    \label{fig:circularitiesLMMdestroysDisk}
\end{figure*}
\begin{figure*}
	\includegraphics[width=1.8\columnwidth,height=0.85\textheight]{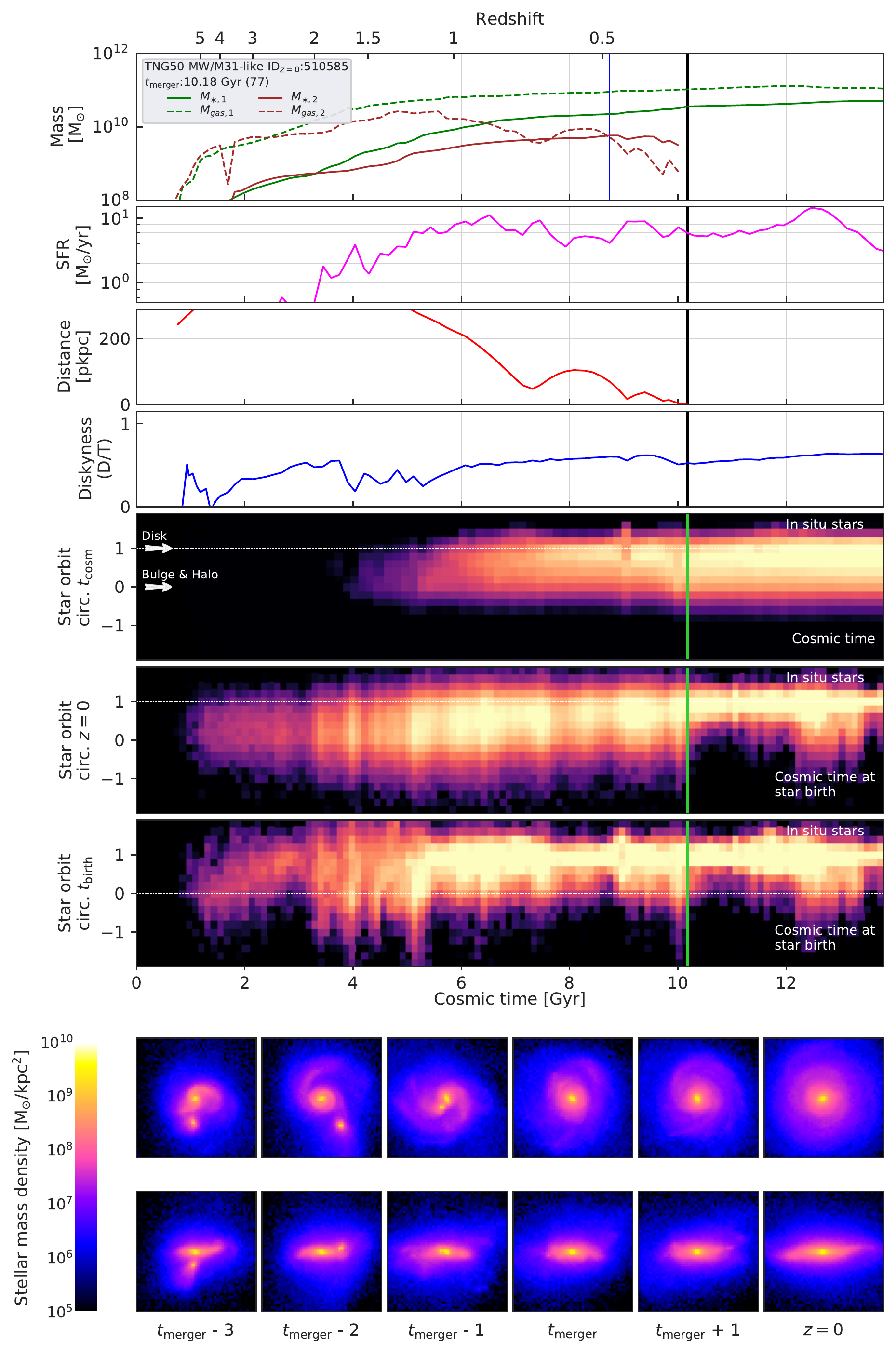}
    \caption{Evolutionary tracks of an example TNG50 MW/M31-like galaxy whose last major merger does not destroy the stellar disk. Panels as in Fig.~\ref{fig:circularitiesLMMdestroysDisk}.
    }
    \label{fig:circularitiesLMMDiskSurvives}
\end{figure*}

Figs.~\ref{fig:circularitiesLMMdestroysDisk} and \ref{fig:circularitiesLMMDiskSurvives} exemplify the time evolution of one prototypical example galaxy for each of the two scenarios identified for TNG50 MW/M31-like galaxies.
Plotted as a function of cosmic time, from top to bottom, are (more details in the caption): stellar and gas mass for the primary and secondary progenitors, SFR of the main progenitor, distance between progenitors, diskyness (D/T ratio) and the distributions of the orbit circularities of the stars formed in situ.
Finally, images of the stellar mass density of the main progenitor are shown from face-on and edge-on projections, at the three snapshots immediately prior to the last major merger, at the snapshot after the merger and at $z=0$ -- which helped in the selection of the galaxies in the MW/M31 sample. In each figure, the vertical solid, thick line represents the time of the last major merger, i.e. the time of coalescence. The blue thin vertical lines in the top panels denote the time when the secondary reaches its maximum stellar mass and the stellar mass ratio is evaluated.

\begin{figure*}
	\includegraphics[width=\columnwidth]{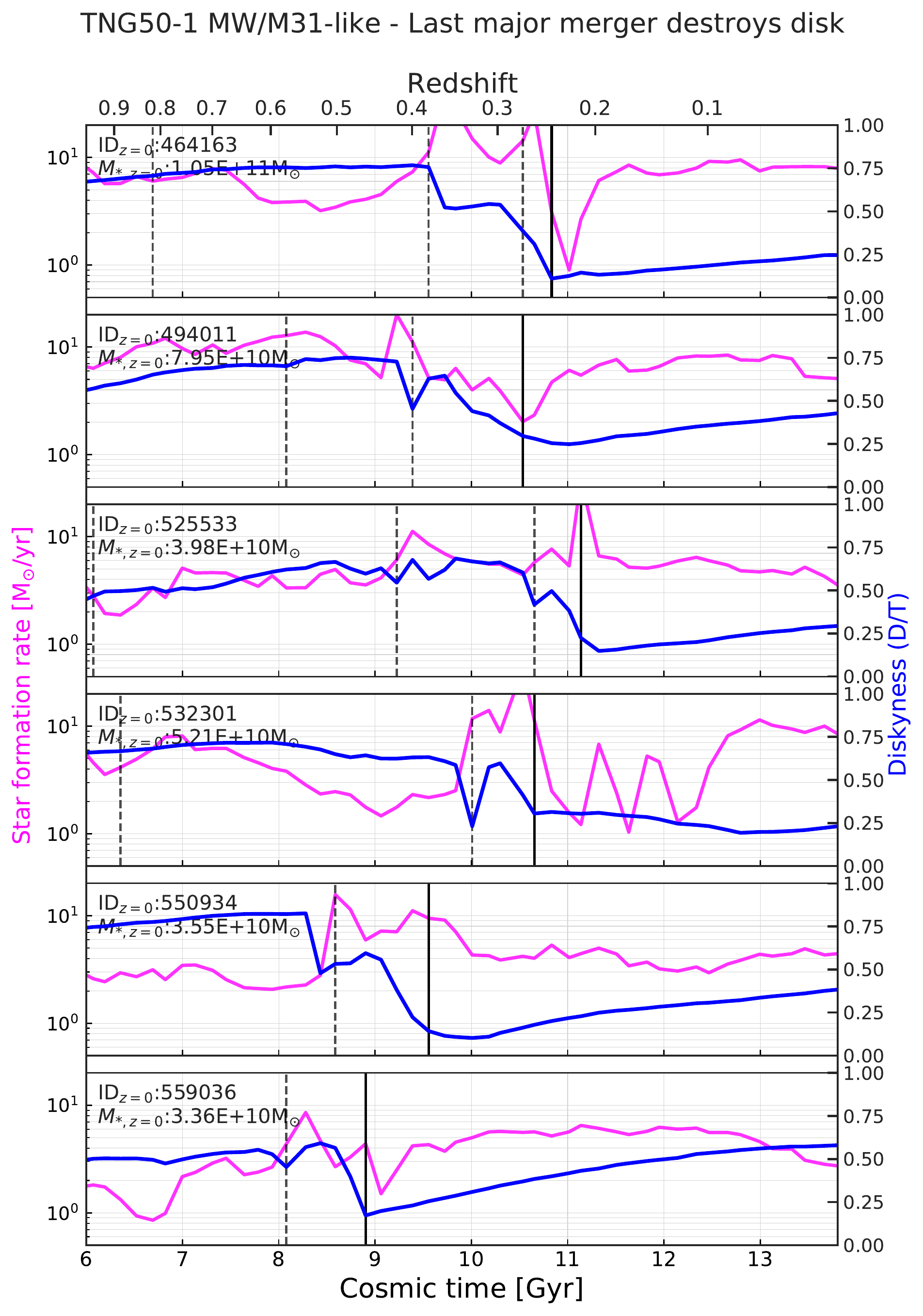}
    \includegraphics[width=\columnwidth]{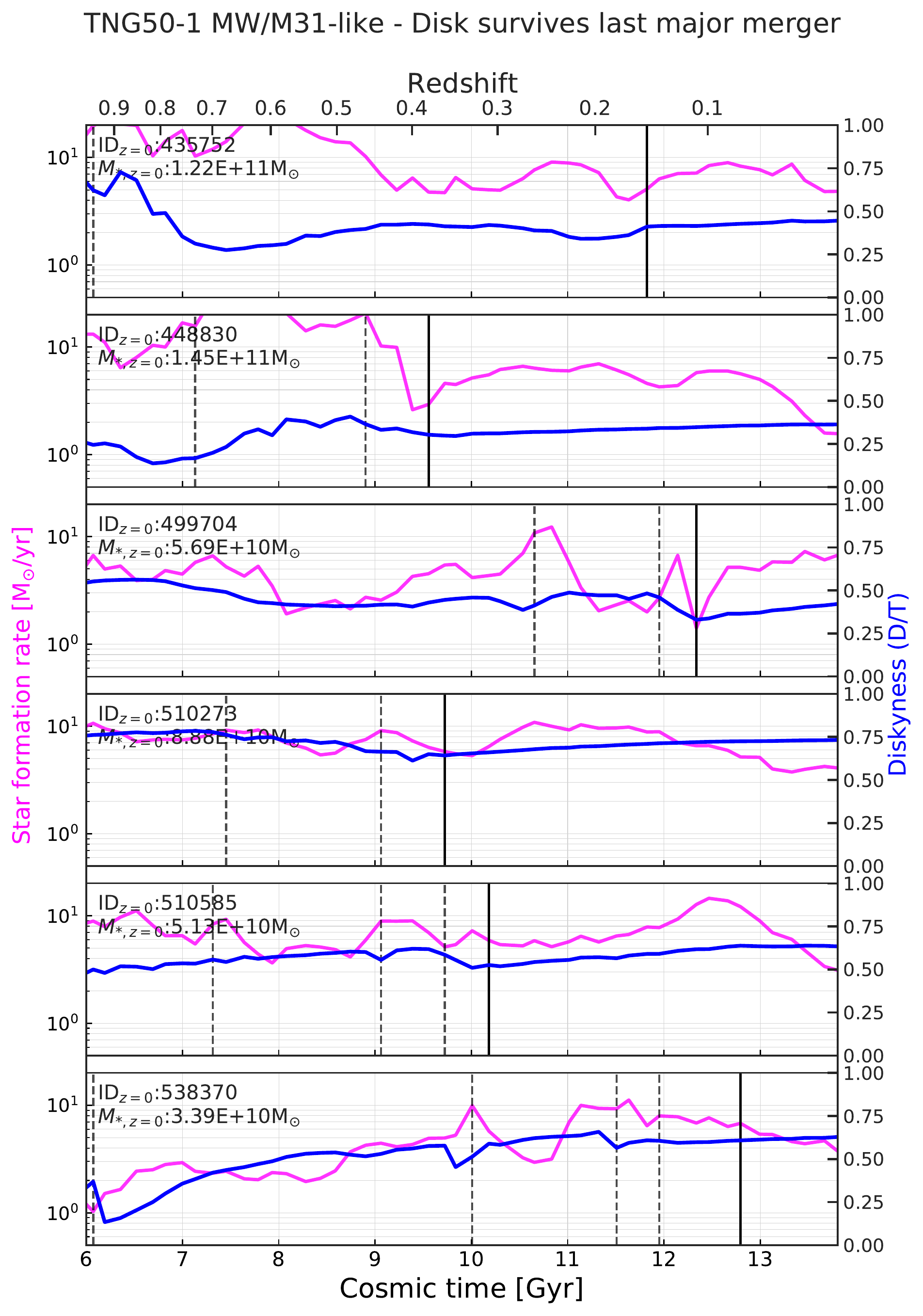}
    \caption{Connection between ongoing star formation and galaxy stellar morphology, i.e. diskyness, for TNG50 MW/M31-like galaxies with recent major mergers. We show the evolution along the main-progenitor branch of individual galaxies of their SFR (fuchsia) and of their D/T ratio (blue), as a function of cosmic time. We show random examples of galaxies whose stellar disk had been destroyed by their last major merger and regenerated (left column: 6 examples of 18 cases) and of galaxies whose stellar disk has not been destroyed (right column: 6 examples of 11 cases).
    The time of the last major merger is marked with a black solid vertical line; the times of the last pericentric passages are marked with black vertical dashed lines. The destruction of the stellar disks in the left column is identifiable in those cases where a galaxy undergoes a sudden drop in diskyness, and usually this occurs in the period between the last pericentric passage and the time of the merger. Compared to the destroyed disks, the galaxies on the right column exhibit milder and gentler drops in diskyness. In all cases, sustained star formation is in place during and after the merger, often with bursts at pericentric passages (see also Fig.~\ref{fig:SFRbursts}) and is temporally coincident with a steady D/T increase after the merger and towards $z=0$, particularly so in the galaxies where the merger destroyed the disk. 
    }
    \label{fig:LMMsfrAndDiskyness}
\end{figure*}

In the first pathway, i.e. Fig. ~\ref{fig:circularitiesLMMdestroysDisk}, a disk of young stars grows in the period of a few billion years between the last major merger and $z=0$. In these cases, the amount of gas available to form new stars is a key factor. From the Figures, it is also clear that, as already alluded to in Fig.~\ref{fig:SFRbursts}, not only does the major merger affect and alter the galaxies, but also the close pericentric passages of the secondary progenitor can have a large impact on the final outcome of the merger and on star formation. In fact, but for periods around coalescence (Fig.~\ref{fig:circularitiesLMMdestroysDisk}) and possibly at pericentric passages, new stars are born in circular orbits, i.e. with circularity at birth typically close to unity -- this is the case at all times and for all galaxies \citep{Pillepich2019}. On the other hand, the current circularity of the in-situ stars that are in $z=0$ MW/M31-like galaxies may be very different, i.e. hotter, than that at birth, particularly for stars that formed at early times: top vs. third circularity panels.

For the galaxy of  Fig.~\ref{fig:circularitiesLMMdestroysDisk}, which is an example of a last major merger that destroys the stellar disk of the main progenitor, the merger produces a drop in the stellar circularities of the galaxy's stars (top and bottom circularity panels). However, stars formed after the merger are born mostly in circular orbits, so that a new stellar disk is present at $z=0$. This sequence is also visible in the stellar density images. Fig.~\ref{fig:circularitiesLMMDiskSurvives} represents a case of a disky main progenitor whose stellar morphology is not affected by the major merger: compared to the previous case, the secondary progenitor approaches the main progenitor more progressively, with multiple pericentric passages prior to the final merger. 

\begin{figure*}
	\includegraphics[width=2.21\columnwidth]{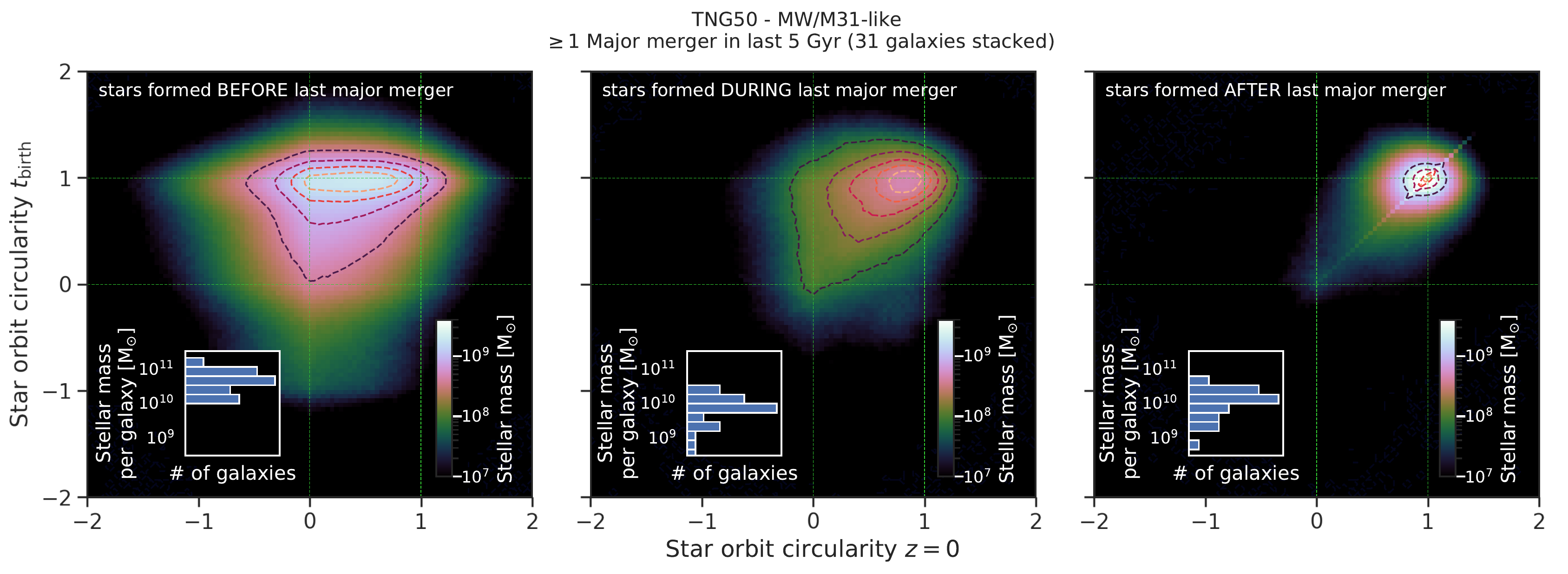}
	\includegraphics[width=1.035\columnwidth]{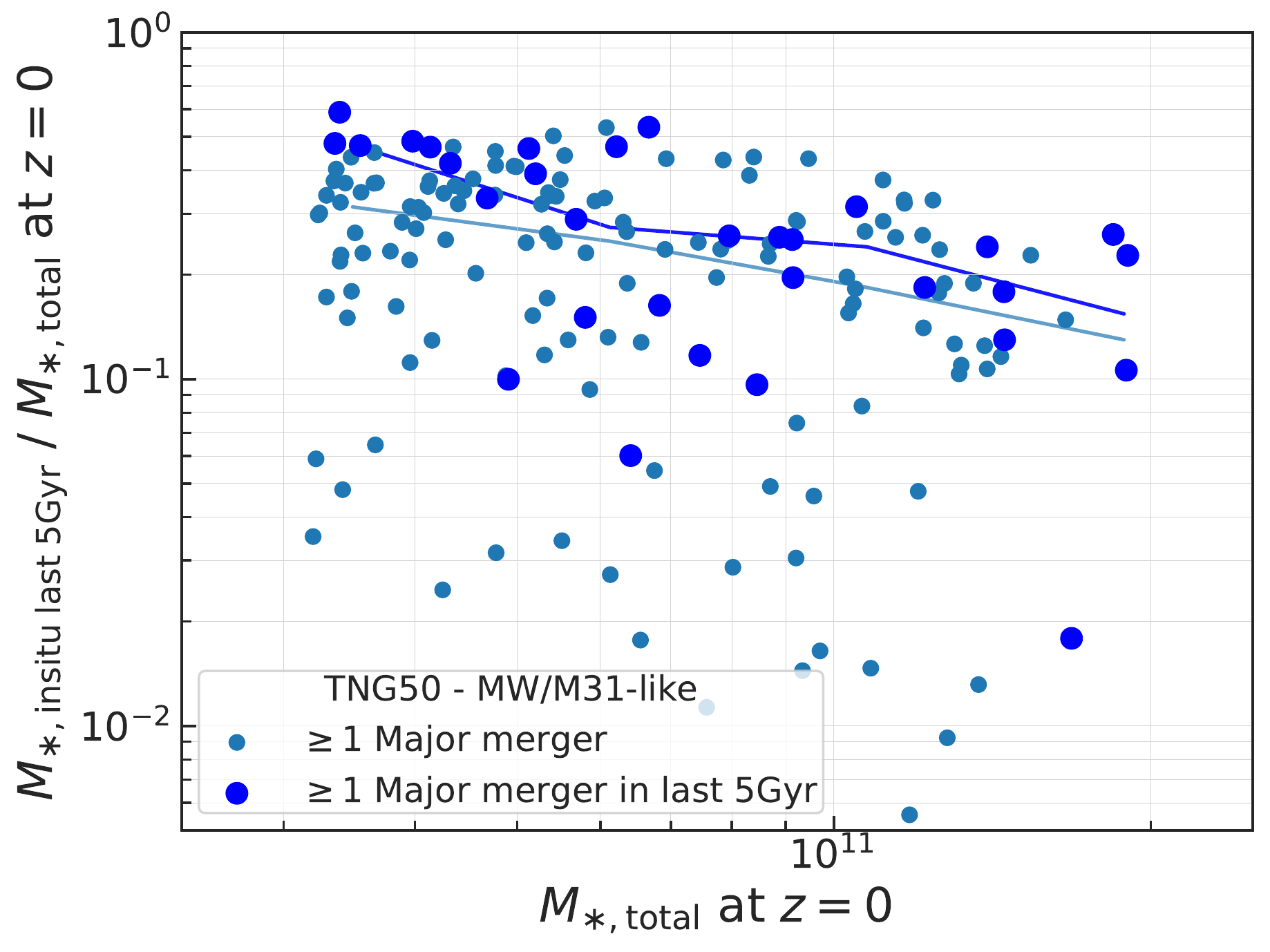}
	\includegraphics[width=1.035\columnwidth]{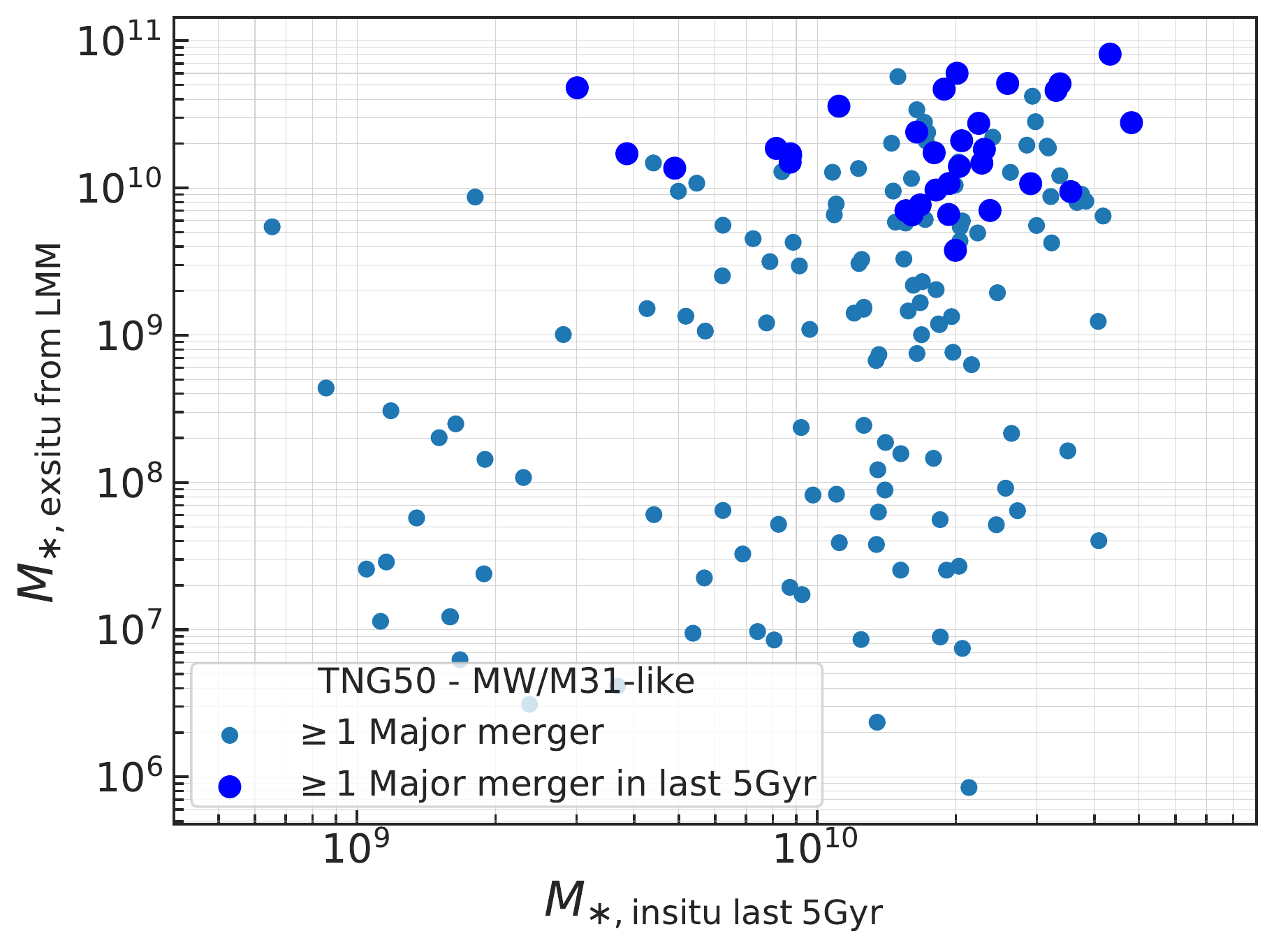}
    \caption{
    Connection between star formation, orbital circularities of the stars and the amount of in-situ star formation triggered by the merger and of the ex-situ stars brought in by the mergers.
    \textit{Top}: circularities of the stellar orbits at the time of formation vs. at $z=0$, stacked for all 31 TNG50 MW/M31-like galaxies with a recent major merger, for stars formed before (before $t_{\rm{max}}$),  during (between $t_{\rm{max}}$ and $t_{\rm{merger}}$), and after the last major merger, from left to right. In all panels, stars in perfectly circular orbits have a circularity $=1$ (or $-1$ if they are counter-rotating); stars in radial orbits have a circularity of $\sim0$. Histograms show the total stellar mass per galaxy born in the respective periods of time.
    \textit{Bottom left}: Fraction of in-situ stellar mass formed in the last 5 Gyr as a function of total stellar mass at $z=0$.
    \textit{Bottom right}: Ex-situ stellar mass brought during the last major merger vs. in situ stellar mass in the last 5 Gyr. In both cases, TNG50 MW/M31-like galaxies with a recent major merger (blue) are contrasted with those with at least one major merger across their history but irrespective of when (light blue). Most of the stars were born in circular orbits, this is generally the case both before and after the times of the last major merger and, to a lesser extent, during the last major merger; however, the orbits of the stars born after the last recent major merger remain mostly unaltered, i.e. in circular orbits, all the way to $z=0$. Furthermore, recent major mergers also seem to trigger more in-situ star formation in the resulting galaxy than in galaxies with more ancient mergers, in addition to bringing large amounts of ex-situ stellar mass.}
    \label{fig:circularitiesPrePostLMM}
\end{figure*}

\subsection{Connection between in-situ star formation and diskyness during and after the mergers, and on the accreted mass }
\label{sec:SF_and_diskyness}

As pointed out above, within the TNG50 model, the orbits of stars at the time of formation are generally circular -- at least at $z\lesssim2-3$, see 7th panels from the top of Figs.~\ref{fig:circularitiesLMMdestroysDisk} and \ref{fig:circularitiesLMMDiskSurvives}. Namely, if star formation occurs, stars naturally form in circular orbits, because they originate from gas that is in rotationally-supported and disky configurations \citep[see also findings by][]{Pillepich2019}. Here we further expand on the connection between star formation and galaxy stellar morphology.

In Fig.~\ref{fig:LMMsfrAndDiskyness}, we examine several examples of TNG50 MW/M31-like galaxies with recent major mergers and for each we show the evolution in time of the instantaneous star-formation rate (fuchsia) and of diskyness D/T (blue) along the main progenitor branch. In these plots the time of the merger is marked with a vertical solid black line and the moments of the last pericentric passages with vertical dashed black lines. 
We show 6 random examples among the cases whose last major merger destroyed the disk but the latter reformed (left panels) and 6 random examples among those galaxies whose last major merger did not destroy the disk. 

First, in the left panels, a noticeable feature is the decrease of the diskyness of the galaxy around the time of the merger or some million years before the merger: in the latter cases, this is due to the effects of the proximity of the secondary progenitor at the pericentric passages. The drop of diskyness is roughly at least one third of the D/T value before any interaction with the secondary. This is not the case for the galaxies in the right panel, where the changes in circularity fraction, D/T, are less pronounced. 

Second, the evolutionary tracks of Fig.~\ref{fig:LMMsfrAndDiskyness} confirm that, after the merger, a disk of young stars can form again: this can be appreciated in the fact that the D/T ratio increases progressively without new drops (the galaxies can undergo additional minor or mini mergers, but no major ones). A progressive increase in the diskyness is here correlated with sustained phases of star formation. 


The connection between ongoing and recent star formation and stellar morphology is quantified and extended to the entire galaxy sample in Fig.~\ref{fig:circularitiesPrePostLMM}. In the top panels, we compare the circularities of the stellar orbits at birth and at $z=0$, for the stars formed before, during, and after the last major merger: namely, before $t_{\rm{max}}$, between $t_{\rm{max}}$ and $t_{\rm{merger}}$, and after $t_{\rm{merger}}$ , i.e. after coalescence, respectively. Results are shown by stacking the orbital properties of all the stars in all the TNG50 MW/M31 analogues with a recent major merger -- therefore the panel on the top left (right) depicts the properties of stars that are mostly older (younger) than 5 billion years -- but see the distribution of the merger times in Fig.~\ref{fig:lastMajorMerger}. Inset histograms show the total stellar mass per galaxy that is formed in-situ in the corresponding periods.

The stars formed before the last major merger (left panel) mostly formed in perfectly circular orbits (circularity $\sim1$). On the other hand, the circularity of their orbits changed as time passed, all the way down to $z=0$, where they span the whole range of values, with circularity $\sim 0-1$, i.e. with also non rotationally-supported orbits. This effect is called orbital heating and is manifested in radial and vertical directions. In fact, at $z=0$, stars can also exhibit negative circularity values, which corresponds to counter-rotating orbits. 
The heating quantified in the top left panel of Fig.~\ref{fig:circularitiesPrePostLMM} is thought to be due to secular evolution and to more violent orbital changes induced by mergers. In this plot we cannot identify the exact drivers of the orbital alterations nor to assess the timescales when these old stars were heated up to their $z=0$ levels. Yet, we notice that a tail of the stars in the left panel of Fig.~\ref{fig:circularitiesPrePostLMM}, and mostly formed before the last major merger, exhibit circularity values at birth also approaching 0, i.e. random motions: these are at least partially stars that are older than $8-10$ billion years -- and this is qualitatively consistent with the picture suggested by observations and also reproduced by simulations whereby galaxies were dynamically hotter at earlier epochs \citep[$z\gtrsim1.5$,][and references therein]{Pillepich2019}. 

Also stars formed {\it during} the major merger events and at pericentric passages (top, middle panel of Fig.~\ref{fig:circularitiesPrePostLMM}) show broad ranges of circularities at birth, with their circularity distribution at $z=0$ being even more spread out.

Focusing on the two top right panels of Fig.~\ref{fig:circularitiesPrePostLMM}, it is manifest (see also inset histograms) that the stars formed after and during the last major merger are less numerous than the stars formed before it -- even just because they could form over shorter periods of time. Yet they are sufficient to secure a disky stellar morphology to the $z=0$ descendants. Stars formed after the merger (top right) were also born in nearly circular orbits, but, in this case, their circularities barely change during the period between the last major merger and $z=0$: whereas the latter is comparatively short for disk heating to have a large effect, it should be noticed that for a fraction of the stars in this panel the period elapsed after the last major merger can be as long as $\sim  5$ Gyr. Yet, the phenomenology of the top right panel of Fig.~\ref{fig:circularitiesPrePostLMM} quantifies the idea that those MW/M31-like galaxies that experienced a recent major merger are disky at $z=0$ because of the sustained star formation and because the newly-formed stars are born in circular orbits and have not yet undergone heating.

These results qualitatively agree with the general picture described by \citet{Peschken2020} for $z=0$ disk galaxies in the Illustris simulation whose last major merger occurred at $z\lesssim1.5$: according to that analysis, the stars born before the last major merger form the $z=0$ spheroidal components and the stars born after the merger constitute a new formed disk.

But how much stellar mass is produced during and after the last recent major mergers? In addition to the insets in the top of Fig.~\ref{fig:circularitiesPrePostLMM}, in the bottom left panels we give the fraction of in-situ stellar mass formed in the last 5 Gyr over the total stellar mass (bottom left) and the amount of ex-situ stellar mass brought by the last major merger vs. the in-situ stellar mass in the last 5 Gyr (bottom right). Galaxies with recent major mergers (blue circles) have formed typically a larger fraction of in-situ stellar mass in the last few billion years than their MW/M31 analogues with more ancient last major mergers -- this ranges in the median, depending on final mass, between $15-30$ per cent of the total $z=0$ mass and the recently formed in-situ stellar masses are, on average, $0.1-0.2$ dex larger in the recently-merged population. This indicates that a larger amount of in-situ star formation has indeed occurred {\it because} of the recent major merger, and would have not occurred at the same levels in the absence of a recent major merger. At the same time, recent major mergers bring large amounts of ex-situ stars (bottom right panel), $\gtrsim$ a few $10^9-10^{11}\,\MS$. This is somewhat necessary, because recent major mergers are more massive in terms of absolute stellar-mass than ancient ones. 

To summarize, recent major mergers bring both larger amounts of ex-situ stellar mass as well as trigger relatively more in-situ star formation in the last few billion years than in the absence of a major merger. This nicely connects to the average stellar mass growth with time of WM/M31-like galaxies -- see Fig.~\ref{fig:assemblyHistory}, left, and compare black thick vs. blue thick curves (i.e. all TNG50 MW/M31-like galaxies vs. those with a major merger in the last 5 billion years): the latter exhibit a suppressed stellar mass growth at $z\gtrsim0.5$ than the typical MW/M31-like galaxy, and de facto `managed' to enter in the stellar mass selection thanks to the mass boost (both in-situ and ex-situ) provided by the recent major merger.

\begin{figure}
	\includegraphics[width=\columnwidth]{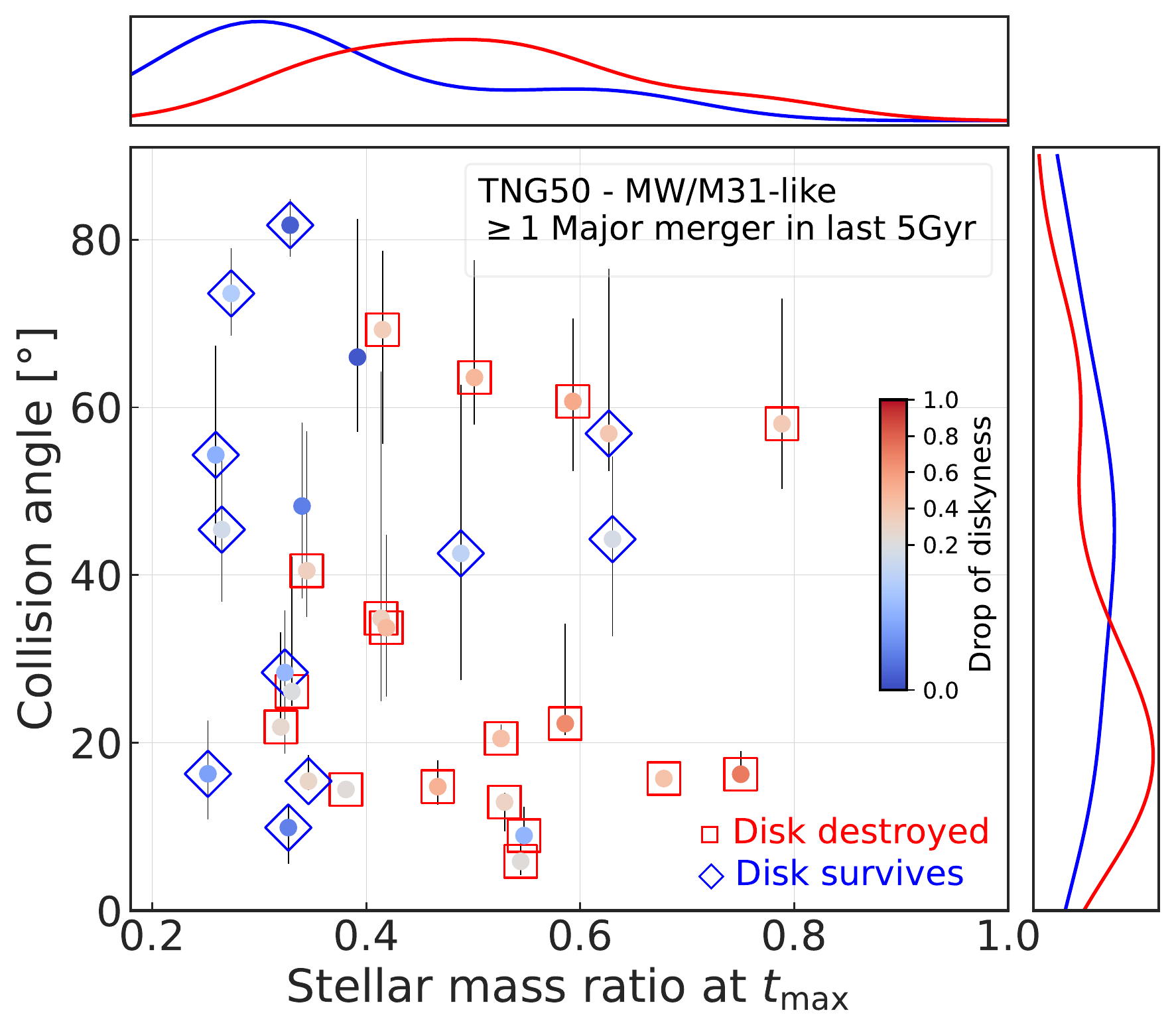}
    \caption{
    Collision angle of the merger orbit vs. mass ratio at $t_{\rm{max}}$, color coded by the absolute drop in diskyness, D/T, during the merger event, for TNG50 MW/M31-like galaxies that experienced a major merger in the last 5 billion years. The points denote the median of the angles across the 5 to 15 snapshots previous to the merger, whereas the errorbars are the 25th-75th percentiles of such angle distribution). The symbols denote whether the last major merger destroyed (squares) or not (diamonds) the previously-existing stellar disk, based on the inspection of the time evolution of the D/T ratios along the main-progenitor branch of each galaxy (see Figs.~\ref{fig:circularitiesLMMdestroysDisk} and  \ref{fig:circularitiesLMMDiskSurvives}). The density plots are calculated with a gaussian kernel. Mergers that do not destroy the pre-existing stellar disks tend to populate the parameter space of smaller stellar-mass ratios and larger collisional angles, i.e. spiralling or large impact-parameter orbits (although the stellar mass ratio has a larger impact).
    }
    \label{fig:orbits}
\end{figure}

\subsection{Orbits of the merging  galaxies}
\label{sec:gal_orbits}

To understand what determines the disruption or survival of a stellar disk during the merger, we inspect the orbital properties the galaxies follow in their merging process: these can be very diverse. We characterize the orbits of the mergers according to two angles: 1) the angle between the plane of the orbit and the stellar angular momentum of the main progenitor (``orbit plane angle''), and 2) the (acute) angle between the velocity vector of the secondary progenitor with respect to the main galaxy and the position vector between them (``collision angle'', as presented in \citealt[][]{Zeng2021}). These angles are measured at individual snapshots and averaged throughout the 5 last snapshots prior to the merger time (approximately over 800 Myr, to minimize the effect of fluctuations that are sometimes caused by the \textsc{Subfind} halo-finder algorithm). 
The orbit plane angle allows us to determine whether the secondary galaxy orbit is prograde (angle: 0-90$^{\circ}$) or retrograde (90-180$^{\circ}$) with respect to the rotation of the main galaxy. The collision angle determines whether the merging orbit resembles a smooth spiraling (high values) or a head-on collision (low values). 

We find (although do not show) that, throughout all the recent major merger events that involve the progenitors of the TNG50 MW/M31-like galaxies, the majority of the mergers that unfold in retrograde orbits (13 from 31 mergers) end up destroying the stellar disks. Also, within our 31 cases, a disk that survives a major merger is more likely to be the result of a prograde than a retrograde orbit. However, these results are tentative and warrant further investigation.

Instead, we can show that it is the combination of orientation of the collision and of stellar mass ratio of the merger that is a good predictor of whether stellar disks are destroyed or not during a recent major merger. This is shown in Fig.~\ref{fig:orbits}, where galaxies with recent major merger are color-coded by the drop (in absolute value) of the D/T mass ratio because of the major merger and the latter is characterized by the collision angle and the stellar mass ratio -- all of them are major, but the ratios go from 1:4 to 1:1. Galaxies whose stellar disks are destroyed by the major merger and then reform (red squares) are more frequent towards larger stellar mass ratios and lower collision angles, i.e. towards mergers with small impact parameters. On the other hand, the lower the stellar mass ratio, the higher the probability that the disk survives the merger event (see blue diamonds, in general with smaller drops in diskyness)\footnote{The reader may notice a galaxy classified as ``disk destroyed'' but with a very low drop in diskyness. This is Subhalo ID 372754 (top left stamp of Fig.~\ref{fig:images}): in this case, the evolution of the total D/T does not fully capture what can instead be evinced by inspecting the full distribution of the stellar orbits at the time of birth vs. $z=0$, as in the corresponding 5th and 7th panels of Figs.~\ref{fig:circularitiesLMMdestroysDisk} and \ref{fig:circularitiesLMMDiskSurvives} for this galaxy.}. To separately quantify the effects of the collision angle and of the stellar mass ratio, we perform Anderson-Darling (AD) and Kolmogorov-Smirnov (KS) tests with the null-hypothesis that the two galaxy samples (destroyed and surviving disks) follow the same distribution. For the angles, we cannot reject the null-hypothesis, namely we cannot exclude that the two distributions are in fact indistinguishable (14 (KS) and 18 (AD) per cent significance levels). On the other hand, for the stellar mass ratios, we can reject the null-hypothesis (as their equality is given with very low significance levels, i.e. 1 (KS) and 0.7 (AD) per cent). We conclude that the stellar mass ratios have more significant effects than the collision angle in the survival of stellar disks.

\section{The properties of MW/M31 analogues with recent major mergers}
\label{sec:properties}
Are the $z=0$ properties of recently-merged MW/M31-like galaxies different from the other galaxies in the sample, or from the rest of the galaxies in TNG50 in the same mass range? In the following we inspect a selection of global and stellar structural properties of the TNG50 MW/M31-like galaxies and contrast those with recent major mergers to the whole sample. We also juxtapose the measurements from TNG50 galaxies to analogous constraints from observations of the Galaxy and Andromeda: these are indicated in the following plots as magenta and orange areas and bands, respectively, and encompass the estimates available from the literature (see Table A1 in Pillepich et al in preparation), including systematic differences and measurement errorbars.

\begin{figure}
	\includegraphics[width=\columnwidth]{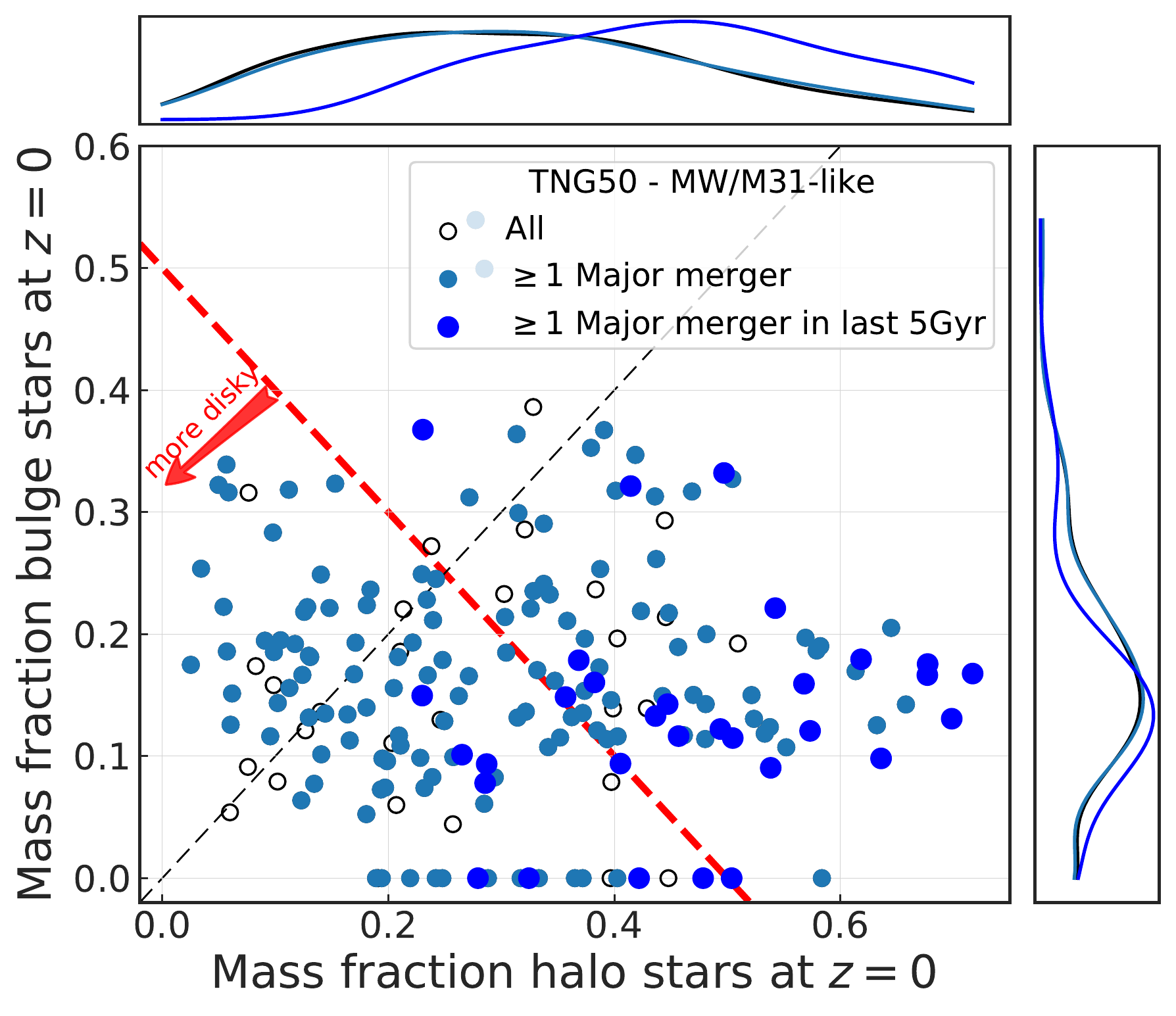}
    \caption{Mass fraction in kinematically-defined bulges vs. mass fraction in kinematically-defined stellar haloes, for TNG50 MW/M31-like galaxies at $z=0$. We compare three sub-samples: all MW/M31 analogues (white circles), analogues that experienced at least one major merger throughout their history (light blue), and analogues with recent major mergers (deep blue). Galaxies below the dashed red line have a total stellar mass fraction in both spheroidal components smaller than 0.5 and are therefore diskier. The black dashed line represents an equal fraction of stars in the bulge and the halo. The density plots are calculated with a gaussian kernel. For most TNG50 MW/M31 analogues the stellar halo is more massive than the bulge. Analogues that underwent recent major mergers have average-mass bulges but more massive stellar haloes, compared to the rest of the MW/M31-like galaxies.
    }
    \label{fig:BulgeHaloZ0}
\end{figure}

\subsection{Average bulges but more massive and shallower stellar haloes}
\label{sec:bulges_haloes}

The origin of stellar bulges in disky galaxies has been debated for at least two decades, with recent observational and theoretical results pointing towards a) a distinction between photometrically- vs. kinematically-defined bulges \citep[e.g.][and references therein]{Du2020} and b) an origin of galactic bulges that is not necessarily always connected to mergers, with different pathways for so-called classical and pseudo-bulges \citep[see e.g. ][and \textcolor{blue}{Gargiulo et al. submitted}, for TNG50-based results and references therein]{Du2021}.

Here we use the kinematic decomposition proposed by \citet{Du2019}, based on a Gaussian mixture separation of the stellar particles in the kinematic phase-space, to separate the stellar structural components of TNG50 galaxies in kinematically-defined stellar bulges and stellar haloes. In Fig. \ref{fig:BulgeHaloZ0}, we account for all stellar particles that are gravitationally bound and we give the stellar mass fractions in such bulges and stellar haloes, by denoting TNG50 MW/M31 analogues that underwent recent major mergers with blue circles and others with more ancient or no major mergers with light blue and white circles. Galaxies that fall below the red dashed line have a total star fraction in spheroidal components lower than 0.5 and are therefore diskier (106 over 198 MW/M31 analogues, i.e. 54 per cent).

For the entire sample of TNG50 MW/M31-like galaxies the bulge fraction is always (except for one galaxy) below 0.5, and below 0.25 in most cases ($\sim$85 per cent). The mass fractions in stellar haloes cover a wide range, with values also greater than 0.5 ($\sim$ 12 per cent) and reaching values of $\sim$0.7.
When we consider both spheroidal components, there is a non-negligible fraction of TNG50 MW/M31-like galaxies (44/198, $\sim$22 per cent) where the bulge is more massive than the stellar halo. But, TNG50 returns also bulgeless galaxies (25 galaxies, $\sim$13 per cent): among these, there are galaxies with major mergers (including recent ones) and without. These latter numbers seem to reasonably agree with previous observational studies of edge-on galaxies \citep[16 per cent found by][albeit differently selected]{Kautsch2006}. MW-mass galaxies have lower stellar halo mass fractions than M31-mass analogues (median values of 0.27 vs. 0.37), whereas their bulge mass fractions are similar (median of $\sim$0.15).

For the galaxies with recent majors mergers, the average bulge fraction is comparable to the average values for the rest of the MW/M31 analogues -- in disagreement with e.g. the predictions of \citet{Puech2012} of more bulge dominant galaxies with $z\sim0.6$ major mergers. A possible interpretative scenario was offered by \cite{Hopkins2010}, who pointed to gas richness as a crucial element in limiting bulge growth. However, their stellar halo mass fraction is on average larger, with median halo mass fractions of 0.46 for MW/M31 analogues with recent mergers vs. 0.31 for the complete sample. Again we perform AD and KS to test the null-hypothesis that the two samples (all MW/M31-like galaxies and galaxies having a recent major merger) follow the same distribution, separately for halo and bulge mass fractions. For the halo fraction, the galaxies with recent major mergers have a statistically-distinct stellar halo mass fraction distribution from that of all MW/M31-like galaxies (as the null-hypothesis of equality is given with 0.04 (KS) and 0.1 (AD) per cent significance levels). For the bulge ratios, with 6.9 (KS) and 14.6 (AD) per cent significance levels, we cannot reject the possibility that the two distributions are in fact identical.

\begin{figure}
    \includegraphics[width=\columnwidth]{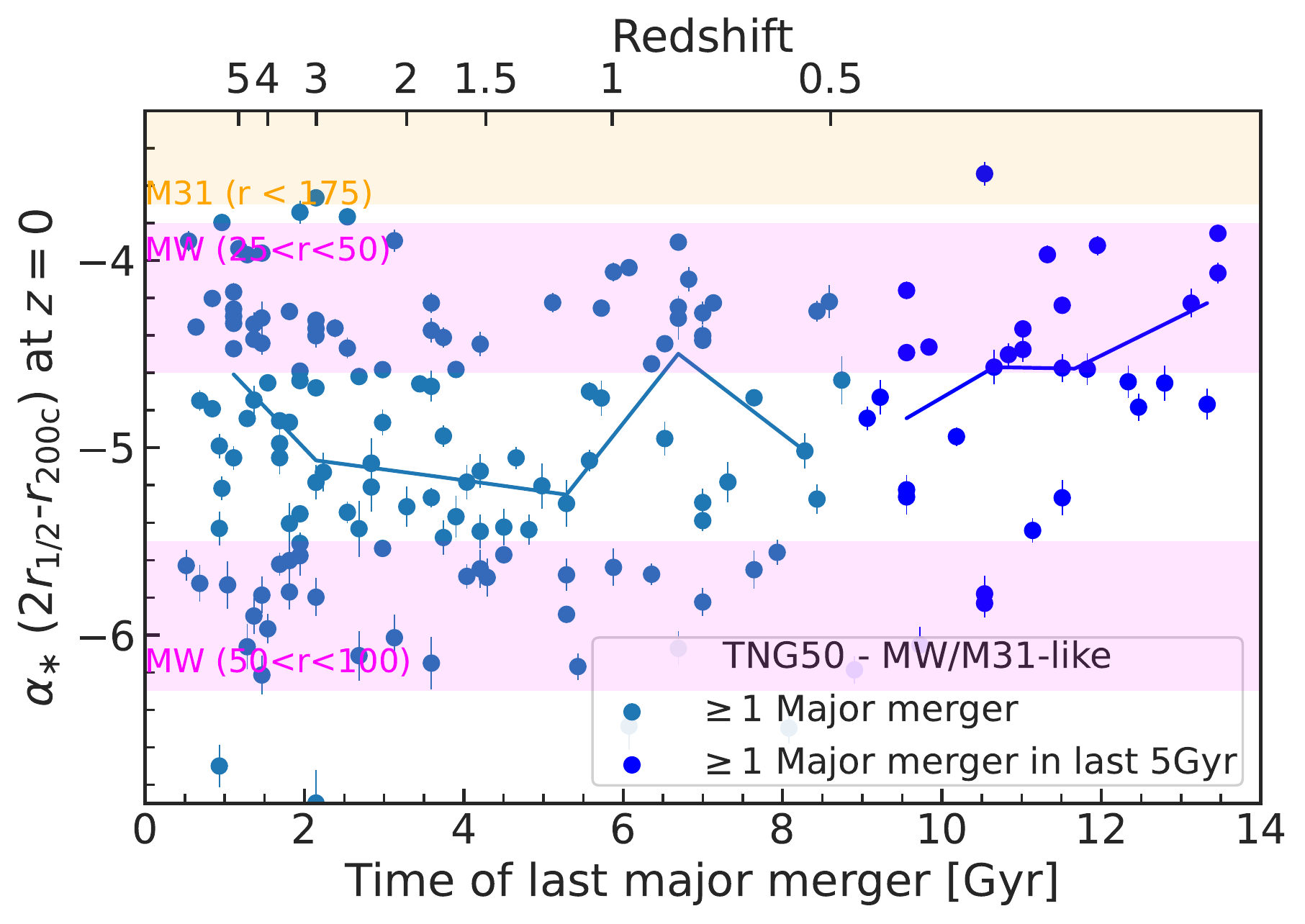}
    \includegraphics[width=\columnwidth]{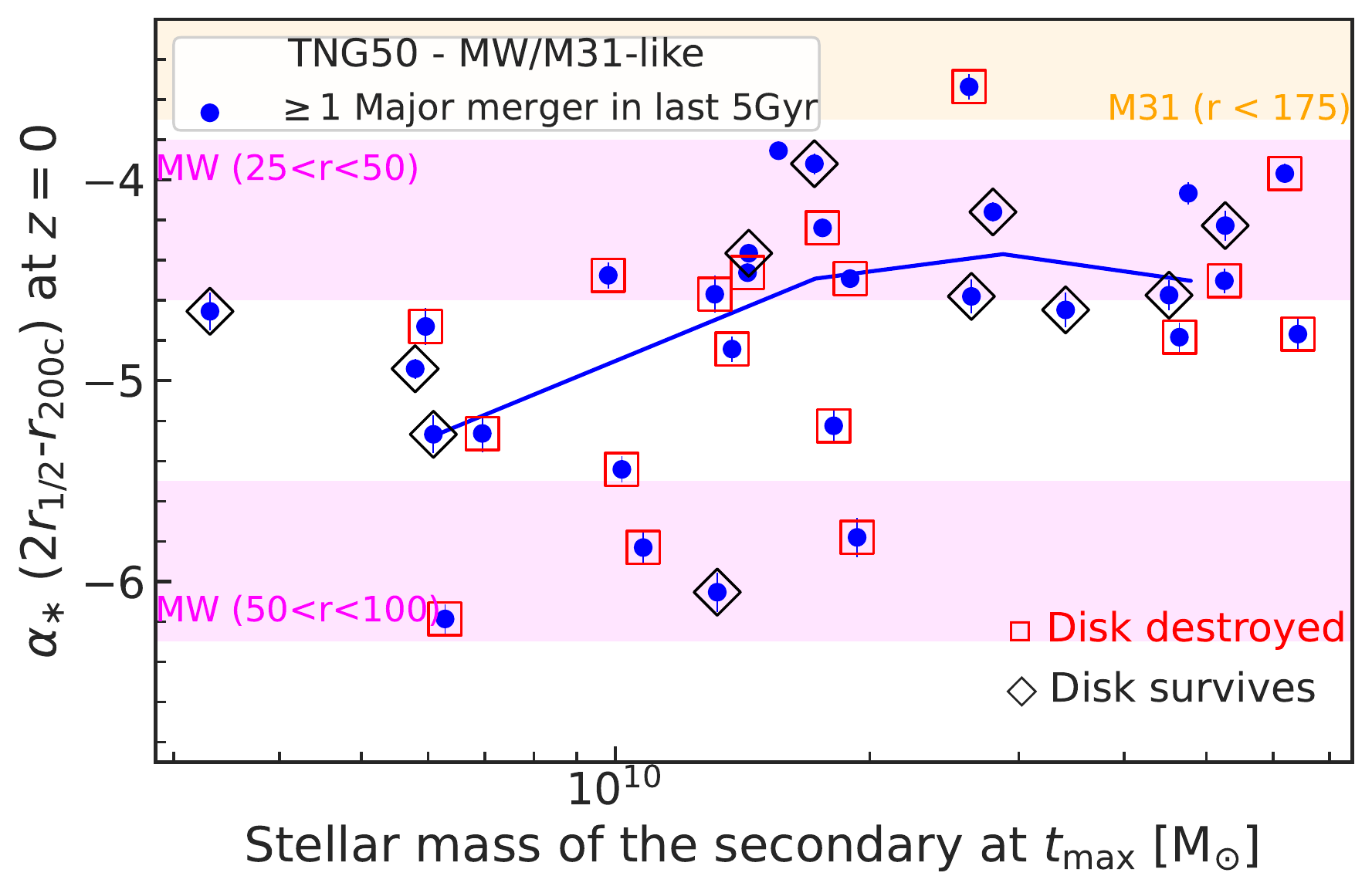}
    \caption{3D slopes of the stellar mass density profiles at large galactocentric distances, i.e. in the stellar haloes, in TNG50 MW/M31-like galaxies. In the top, we show the stellar halo 3D slopes as a function of the time of the merger, in the bottom as a function of the stellar mass of the secondary progenitor. In both panels, MW/M31-like galaxies with recent major mergers are denoted by blue circles and compared in the top to the rest of the sample (light blue). Error bars are one standard deviation errors of the parameter provided by the non-linear least squares fitting routine (python curve\_fit). Magenta and orange shaded area denote current observational constraints in the slopes of the stellar haloes of the MW and M31. In the bottom, we additionally indicate with red squares and black diamonds those recently-merged galaxies whose stellar disk had been destroyed or not by the merger. The stellar haloes of galaxies with recent major mergers have shallower profiles, and among these, the ones whose disk survived the merger but that merged with more massive secondaries are, in general, shallower. We show observational estimates for the MW: for a radius between 25 and 50 kpc, \citet[][]{Bell2008, Watkins2009, Sesar2011, Deason2011}, and for a radius between 50 and 100 kpc \citet[][]{Deason2014}; for M31: \citet[][]{Gilbert2012,Ibata2014}.}
    \label{fig:alphaStellarHalo}
\end{figure}

The stellar haloes of recently-merged MW/M31-like galaxies are not only more massive, but their stars are also less centrally concentrated, albeit to a somewhat weaker degree. Fig.~\ref{fig:alphaStellarHalo} gives the 3D slopes of the spherically-symmetric stellar profilesfitted with a power law ($\rho(r) = \rho_0 r^{\alpha}$) between twice the stellar half-mass radius and $r_{\rm{200c}}$ of TNG50 MW/M31-like galaxies as a function of the time of their last major merger (top) and of the stellar mass of the secondary galaxy involved in the merger (bottom). Error bars cover one standard deviation errors of the parameter provided by the non-linear least squares fitting routine (python curve\_fit). Galaxies with recent major mergers have slightly shallower slopes, with median ([25th-75th] percentiles) of about $-4.6\,(-[4.3,5.1])$ vs. $-4.8\, (-[4.4,5.4])$ for the blue and light blue sub samples, respectively.  However, this difference is only mildly significant (the significance levels are of 17 per cent and 22 per cent for the KS and AS tests of consistent distributions). This is a distinction that is qualitatively in line but weaker than the findings based on Illustris galaxies by \cite{Pillepich2014}. Similarly as found there, TNG50 also predicts the 3D stellar halo slope to depend on galaxy stellar mass: among the MW/M31 analogues, MW-mass galaxies have steeper stellar haloes ($-5.0\pm0.7$) than M31-mass ones ($-4.5\pm0.7$), a fact that is qualitatively consistent with the estimates of the stellar halo profiles of the Galaxy and Andromeda -- see magenta and shaded areas. Furthermore, the more massive the merging companion (within the major merger mass ratios), the shallower the stellar haloes of the descendant galaxy. 

\subsection{Larger fractions and amounts of ex-situ stellar mass}
\label{sec:exsitu}

As in the hierarchical growth of structure scenario stellar haloes of MW-like galaxies are mostly formed by accretion \citep{Zolotov2011, Font2011, Pillepich2015}, it is not surprising then that MW/M31-like galaxies with recent major mergers also exhibit larger amounts and fractions of ex-situ, i.e. accreted, stars at $z=0$. 

\begin{figure}
	\includegraphics[width=\columnwidth]{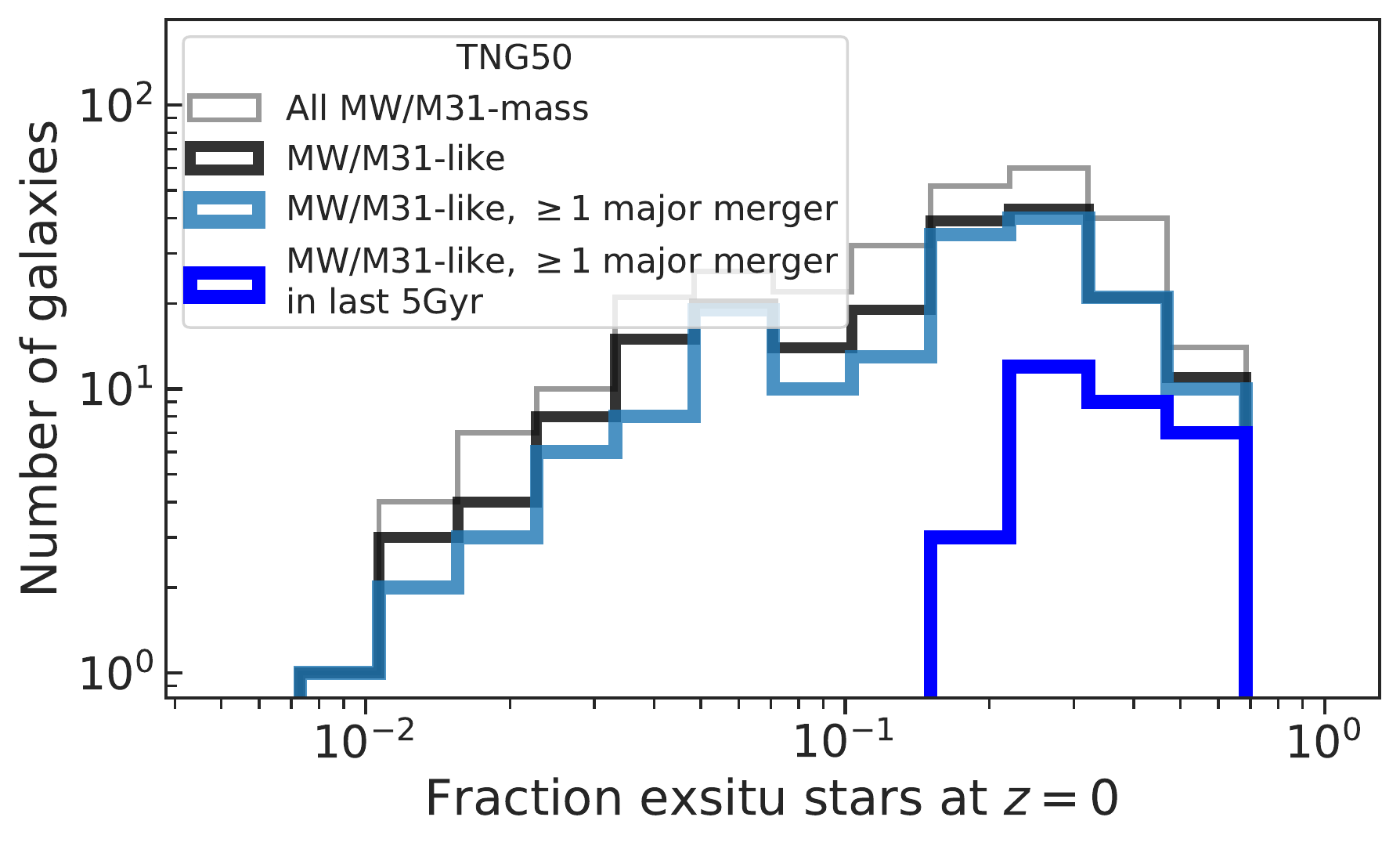}
	\includegraphics[width=\columnwidth]{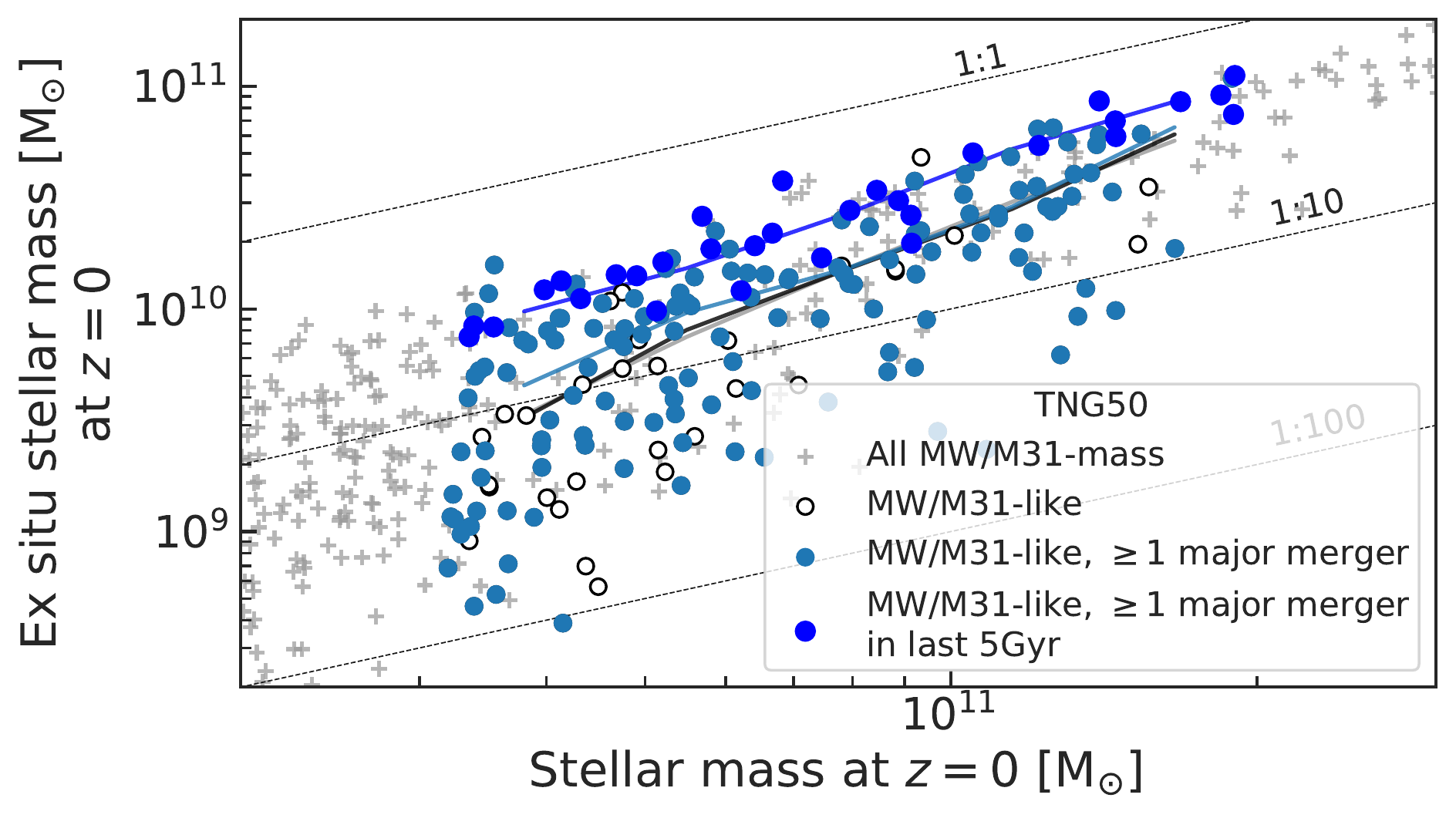}
    \caption{
    Ex-situ, i.e. accreted, stellar mass and stellar mass fractions of TNG50 MW/M31-like galaxies at $z=0$. In the top, we show the distributions of the ex-situ fractions, in the bottom the accreted stellar mass vs. galaxy stellar mass, for different sub-samples. Solid curves in the bottom panel denote medians at fixed stellar mass. Galaxies with recent major mergers have on average larger values of ex-situ stellar mass (absolute and relative). At the opposite end, TNG50 MW/M31-like galaxies that never experienced a major merger (white circles) exhibit lower-than average ex-situ stellar masses, given their galaxy mass.}
    \label{fig:exsituZ0}
\end{figure}

This is quantified in Fig.~\ref{fig:exsituZ0}, in terms of the distributions of the ex-situ stellar mass fraction across different galaxy samples (top) and the total ex-situ stellar mass as a function of galaxy stellar mass at $z=0$ (bottom). In both panels, we compare all TNG50 galaxies in the depicted mass range (gray annotations) to TNG50 MW/M31-like galaxies (black and light blue lines in the top and white and light blue circles in the bottom) and MW/M31-like galaxies that underwent recent major mergers (blue lines and circles). 

The ex-situ fractions range widely: from a few percents to $\sim65$ per cent, with a median value of $\sim0.19$ for TNG50 MW/M31-like galaxies. This is in the ball park of the findings by \cite{RodGom2016} and \cite{Pillepich2018b}, with Illustris and TNG100/TNG300 MW-mass galaxies and haloes, respectively. For the sub-sample with a recent major merger, the ex-situ fraction is biased towards the higher end of the distribution, with a median ex-situ mass fraction throughout the galaxy bodies of $\sim33$ per cent. Moreover, at a fixed stellar mass, galaxies with a recent major merger have more ex-situ stellar mass than the average, by $0.3-0.5$ dex. For the TNG50 MW/M31 analogues with recent major mergers, the total ex-situ mass is always larger than $10^{9.8-10}\,\MS$. Again, the recent major mergers not only trigger star formation (see Figures~\ref{fig:LMMsfrAndDiskyness} and \ref{fig:circularitiesPrePostLMM}) but also bring in higher-than average amounts of accreted stars. 


\begin{figure*}
\includegraphics[width=\columnwidth]{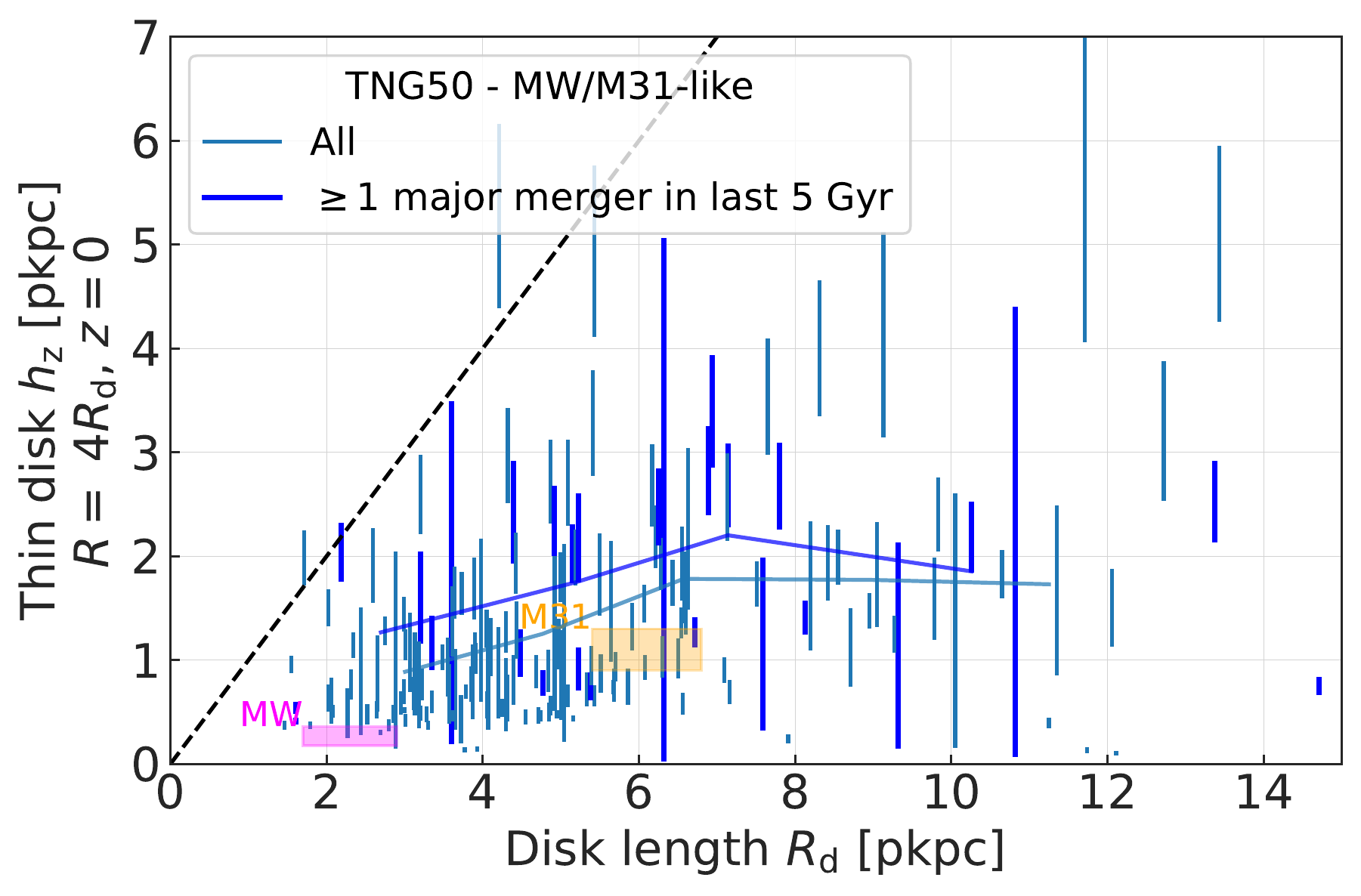}
\includegraphics[width=\columnwidth]{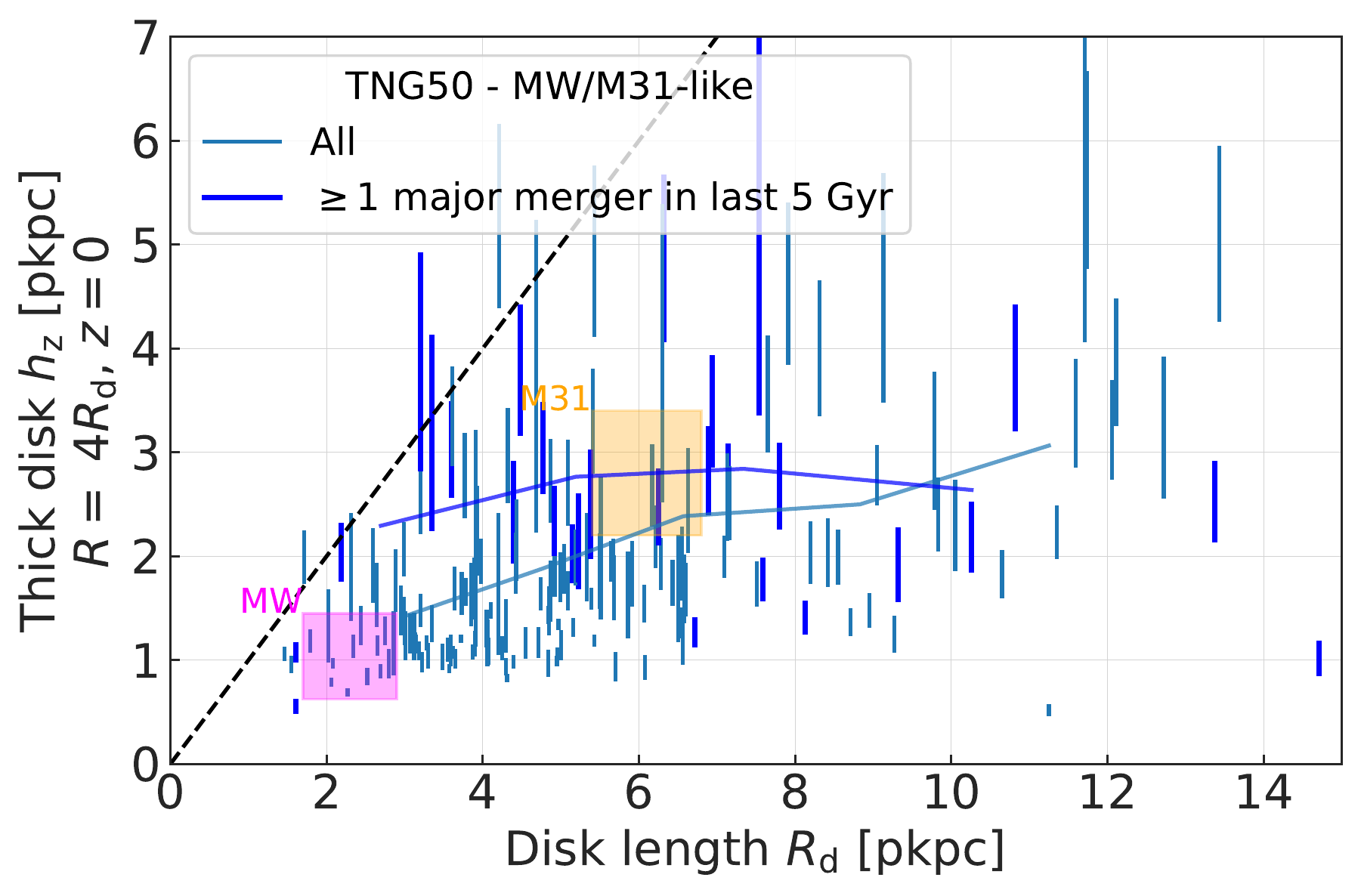}
\includegraphics[width=\columnwidth]{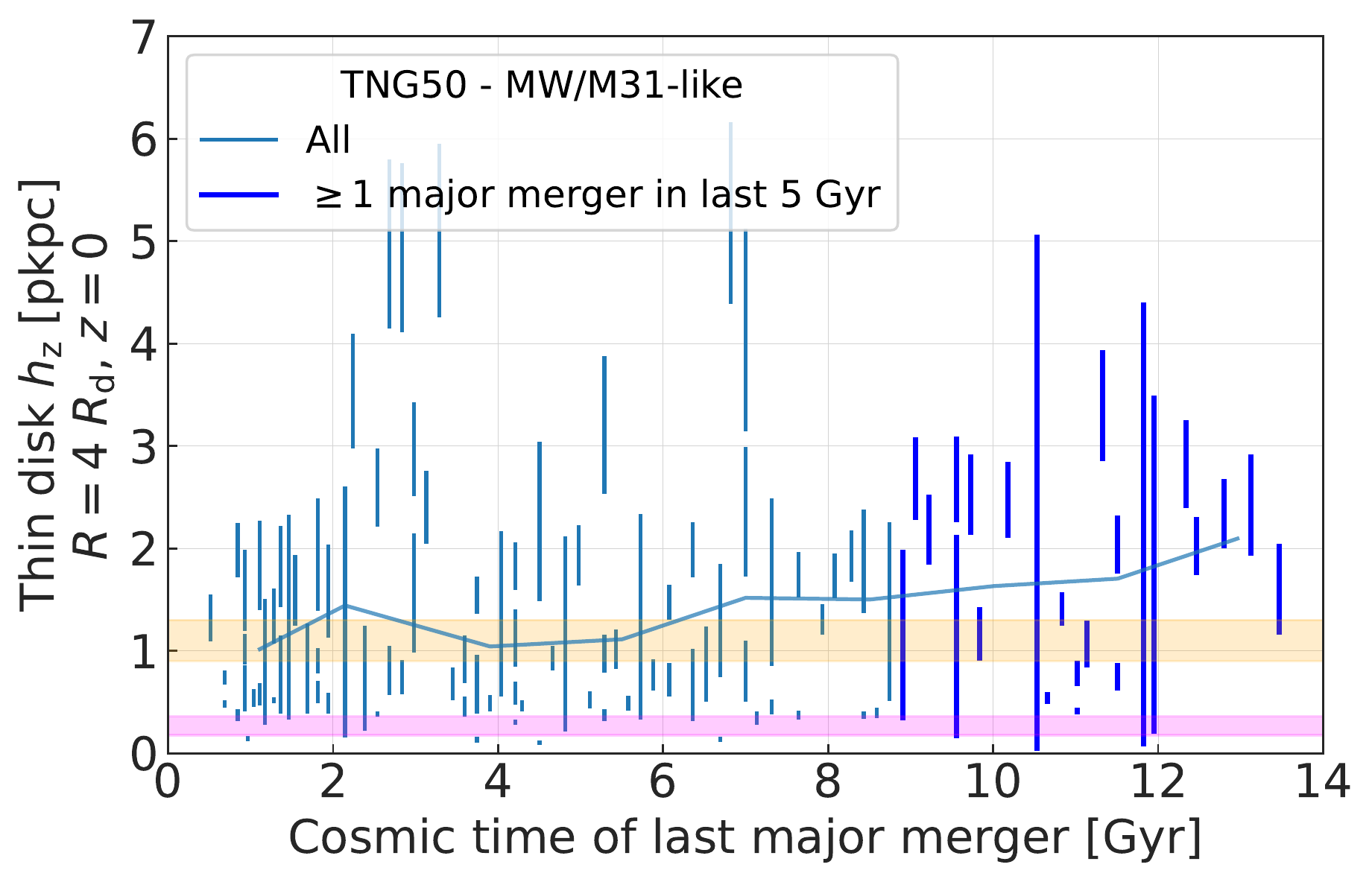}
\includegraphics[width=\columnwidth]{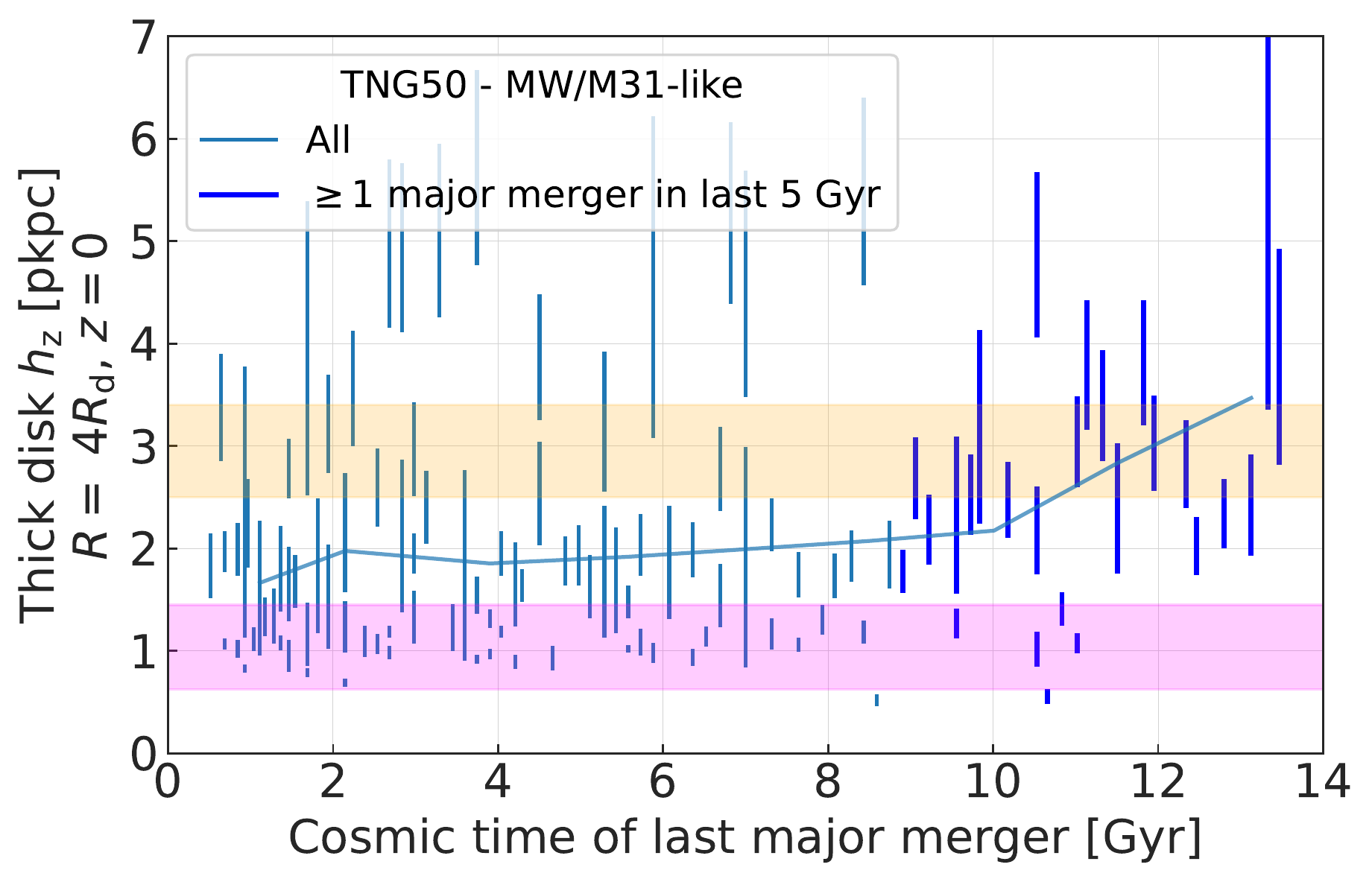}
    \caption{Stellar disk heights of TNG50 MW/M31-like galaxies at $z=0$. We show thin (left) and thick (right) vertical heights vs. disk length (top row) and time of the last major merger (bottom row). In all panels, vertical colored bars span, for each galaxy, the best-fit values for the scale height obtained from two complementary functional forms: double sech and double sech$^2$: light blue vertical bars represent TNG50 MW/M31-like galaxies, blue vertical bars denote those that underwent a recent major merger.
    Solid thick curves are medians  in bins of the quantity on the x-axes of the mean between the best-fit values from the two functional forms. At fixed disk length, galaxies with a recent major merger have somewhat thicker thin and thick stellar disks.
    Observational estimates for the Galaxy and Andromeda are given as magenta and orange boxes, which encompass, and hence marginalize over, diverse measurements including their errorbars: $1.7-2.9$ kpc, $175-360$ pc, and $625-1450$ pc for the disk length, geometrically thin- and thick-disk heights of the MW \citep[][]{Gould1996, Ohja2001, Siegel2002, Juric2008, RixBovy2013, Bland-Hawthorn2016} and $4.8-6.8$ kpc, $900-1300$ pc, and $2200-3400$ pc for the disk length, geometrically thin- and thick-disk heights of M31 \citep[][]{Worthey2005, Barmby2006, Hammer2007, Collins2011}.
    }
    \label{fig:disksSizes}
\end{figure*}

\begin{figure}
\includegraphics[width=0.95\columnwidth]{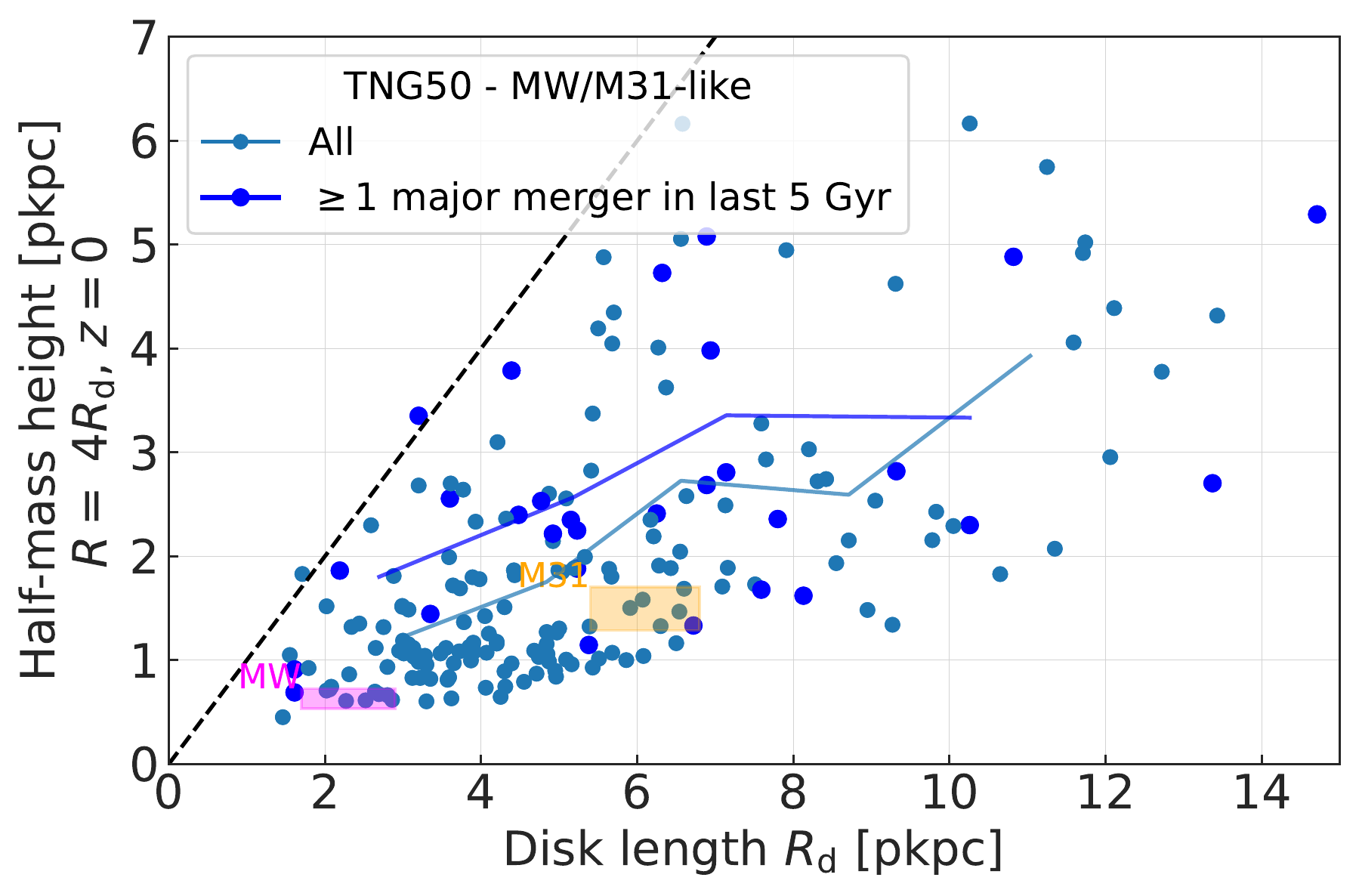}
\includegraphics[width=\columnwidth]{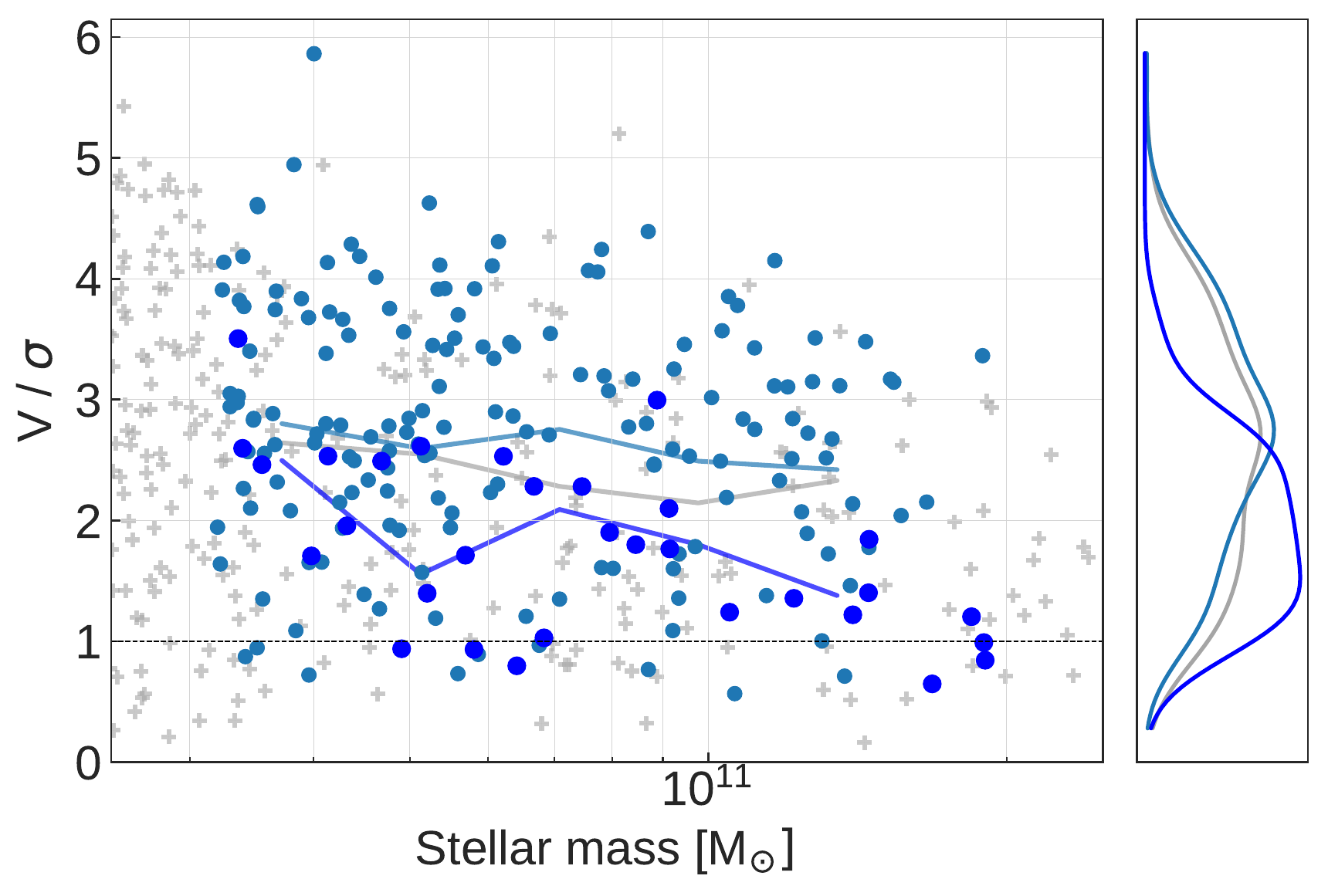}
    \caption{Stellar non-parametric disk heights and stellar vertical kinematics of TNG50 MW/M31-like galaxies at $z=0$. In the top panel, we measure disk heights in a non-parametric way, namely by measuring the stellar half-mass disk heights at $3.5-4.5$ times the disk length, for comparison to the heights in Fig.~\ref{fig:disksSizes}. Observational constraints on the Galaxy and Andromeda (magenta and yellow areas) are derived from the average relationships in TNG50 between half-mass heights and thin and thick disk heights. In the bottom panel, we show the ratio between the maximum rotational velocity and the vertical velocity dispersion of the stars. Light blue circles represent TNG50 MW/M31-like galaxies, blue circles denote those that underwent a recent major merger, and gray crosses represent the rest of the galaxies in TNG50 in the depicted stellar mass range. Solid curves are medians in bins of the quantity on the x-axes. At fixed stellar mass, galaxies with a recent major merger exhibit thicker stellar disks and lower V$/ \sigma$ ratios, i.e. hotter orbits.
    }
    \label{fig:vsigma}
\end{figure}

\subsection{Somewhat thicker and hotter stellar disks}
\label{sec:thickDisks}

Whether $z=0$ galaxies with thick stellar disks have been born with large stellar velocity dispersions and less flat stellar mass distributions and/or whether their stellar disks puffed up across cosmic epochs, because of secular or externally-triggered processes, is an active field of research \citep[see e.g.][for recent works with state-of-the-art cosmological galaxy simulations]{Pillepich2019, Park2019}. Here we do not address this general issue but show that a recent major merger may indeed affect the vertical structure of the stellar disk of a specific selection of galaxies, those with $z=0$ global properties similar to the Galaxy and Andromeda. 

In Fig.~\ref{fig:disksSizes}, we plot the stellar disk heights (top two rows) of all TNG50 MW/M31-like galaxies (light blue circles) and of TNG50 MW/M31-like galaxies that underwent a recent major merger (blue circles). These are shown as a function of stellar disk length (top panels) and of the time of the last major merger (middle panels) for two reasons: a) to control for the influence of galaxy stellar mass and hence disk size; and b) to judge the effects in terms of how long ago a major merger may have perturbed a possibly pre-existing stellar disk. For the disk lengths, for each galaxy projected face on, we fit an exponential profile to the radial distribution of the disk stellar mass surface density in a certain aperture (see Appendix \ref{app:fits} for details). For the disk heights, for each galaxy projected edge-on, we select only stars in circular orbits in the cylindrical shell between 3.5 and 4.5 times the disk length and fit a double parametric formula to the vertical stellar mass density profile, to allow for both geometrical thin and thick disks. More details about the fitting technique are given in the Appendix \ref{app:fits}: for each galaxy, we both fit a double hyperbolic secant and a double squared hyperbolic secant profile and represent in the figure the range spanned by both best-fit values. The averages in bins of disk length or time of last major merger (solid thick curves, top and bottom, respectively) run over the mean of the best-fit values from the two functional fitting functions.

According to TNG50, more extended galaxies exhibit thicker stellar disks, particularly for the geometrically thick component. 
Yet, the galaxy-to-galaxy variation is very large, and in TNG50 we recover example galaxies whose stellar disk structural properties are very similar to both the Galaxy's (magenta shaded areas) and Andromeda's (orange). 

From the top panels of Fig.~\ref{fig:disksSizes}, we uncover that galaxies with a recent major merger have, on average, somewhat larger stellar disk heights, at fixed disk length, but this effect is more pronounced for the thick (right) rather than the thin (left) component and for smaller galaxies. From the bottom panels, we also see an increase in disk thickness that mildly correlates with the time of the last major merger, with thicker disks the more recent the last major merger. Although these relationships have significant scatter, they are in place despite the fact that more recent major mergers impact galactic stellar structures that are expected to be thinner and colder (because they exist at lower redshift) than those impacted by merger at higher redshifts, at least at fixed stellar mass \citep{Pillepich2019}. According to AD and KS tests, we find that, for the geometrically-thin component, we cannot reject the possibility that the distributions of the heights of all MW/M31-like galaxies and of those with recent major mergers are indistinguishable (15 (AD) and 20 (KS) per cent significance levels for the null-hypothesis of equality in the case of the double sech$^2$; 3 (AD) and 2 (KS) for the double sech). On the other hand, for the geometrically-thick component, the distributions of disk heights are clearly different (0.7 (AD) and 0.2 (KS) per cent significance levels for the double sech$^2$; 0.3 (AD) and 0.03 (KS) for the double sech).

A similar general picture is in place also when the thickness of the stellar disk is evaluated in a non-parametric way. In Fig.~\ref{fig:vsigma}, top, we show the stellar half-mass heights of TNG50 MW/M31-like galaxies, measured in the annuli between 3.5 and 4.5 times the disk length: TNG50 MW/M31-like galaxies with recent major mergers have larger stellar half-mass heights than the whole population -- and the stellar half-mass height correlates better with the thick-disk height of galaxies than with the thin-disk height. Moreover, when turning to quantify the {\it kinematics} of the stellar disks (Fig.~\ref{fig:vsigma}, bottom), we find that the ratio between the maximum rotational velocity and the vertical velocity dispersion of the stars \citep[as in][]{Pillepich2019} is lower for MW/M31 analogues with recent major mergers, namely, they are kinematically hotter than the rest of the sample, which concurs at least qualitatively with the structural analysis.

We conclude that MW/M31-like galaxies that have undergone recent major mergers exhibit somewhat thicker and clearly hotter stellar disks, but with a significant scatter. TNG50 does return example galaxies that experienced a recent major merger and have thin and thick disk heights similar to those of Andromeda (orange areas).
Interestingly, however, for galaxies with disk length smaller than $3-4$ kpc and hence smaller than Andromeda (see magenta vs. orange observational constraints), TNG50 seems to imply that it is improbable, but not impossible, for a galaxy to have emerged from a major merger over the last few billion year and exhibit at $z=0$ thin-disk heights as low as that of the Galaxy. This is qualitatively in line with the ideas of \citet{Toth1992, Wyse2001, Hammer2007}, who however excluded the possibility. In fact, such statements strongly depend on how and where the vertical structure of the stellar disks is assessed. For example, in terms of thick-disk or stellar half-mass heights, a few TNG50 galaxies appear as compact and similarly thick as the Galaxy. Moreover, there are two TNG50 galaxies that underwent a recent major merger and whose thin disks has a similar disk length and height as the Galaxy's (Subhalo IDs 528322 and 532301). Finally, when we measure disk heights of TNG50 galaxies at a fixed galactocentric distance of e.g. 8 kpc (irrespective of disk size), we do find a few galaxies that underwent recent major mergers and have a stellar disk as thin as $\lesssim350-400$ pc, i.e. similar to the MWs.

These last numbers and considerations indicate that the limited numerical resolution of TNG50 should not affect our scientific conclusions, in that simulated stellar disks can be thinner than the simulation's softening length (288 pc for DM and stellar particles, $\geq$72 pc for the gas cells -- see Section~\ref{sec:methods}). An extended study of the resolution effects on galaxy sizes and heights in TNG50 is presented in \citet[][]{Pillepich2019}, where it was shown that, in terms of stellar half-mass heights, stellar disk thickness can be considered to be converged in TNG50 to better than 20–40 per cent.

\subsection{Hints of more massive SMBHs, larger gas reservoirs and star formation rates}
\label{sec:props_SMBH}

We conclude this analysis by enunciating (without showing) the differences for a few additional global properties at $z=0$ of TNG50 MW/M31-like galaxies with and without a recent major merger. 

Whereas the average stellar-to-halo-mass relation of TNG50 MW/M31-like galaxies with recent major mergers is comparable to the entire sample of MW/M31 analogues, the former seem to be somewhat biased towards larger SMBH masses, at fixed stellar mass. Two effects can explain this: the SMBHs present in the two progenitors were on average larger, even just because they merged at more recent epochs than the others, or the SMBH of the resulting galaxy has accreted more gas and grown more quickly in the post-merger epoch than the SMBHs in the control-sample galaxies. As the effect is not particularly large, we do not investigate further on this here and we refer the interested reader to \citealt{PillepichInPrep} for a discussion on the masses of TNG50 SMBHs vs. that of Sgr A$^*$ in the Galaxy.

On the other hand, at fixed galaxy stellar mass, the gas mass fraction and SFR (within 2$R_\rmn{1/2,\ast}$ at $z=0$) are higher for the galaxies with recent major mergers. This is overall consistent with the picture described above, whereby the recent major merger brought in gas, which in turn cooled down, becoming eligible for star formation. Therefore, not only the gas fractions were already larger at the time of coalescence (Fig.~\ref{fig:gasFracTmerger}), but we also find that MW/M31-like galaxies with recent major mergers are biased high in terms of gas availability with respect to the rest of MW/M31 analogues still at $z=0$.

\section{Summary and conclusions}
\label{sec:conclusions}
We have used the cosmological magneto-hydrodynamical galaxy simulation TNG50 to quantify the mass assembly and merger histories of 198 MW/M31-like galaxies selected at $z=0$. We have placed a special focus in studying how frequently, and how, MW/M31-like galaxies may undergo a recent major merger (e.g. occurring within the last 5 billion years and with stellar mass ratio > 1:4) and still exhibit a disky  stellar morphology at $z=0$.\\

The main findings that emerge from our analysis are:

\begin{itemize}

    \item The progenitors of TNG50 MW/M31-like galaxies at $z\sim3$ were, on average, $\sim$40 times less massive in stellar mass than today, but with a large past scatter: $\gtrsim$ 2 dex between the 10th-90th percentiles (7 times less massive in total halo mass, with scatter 0.6 dex;  Fig.~\ref{fig:assemblyHistory}).\\

    \item Major mergers are common: 168 of the 198 MW/M31-like galaxies in TNG50 (85 per cent) have undergone at least one major merger throughout their history (or specifically, at $z\lesssim5$;  Fig.~\ref{fig:mwlike_mergersvstime_tables_plot}).\\
    
    \item TNG50 returns galaxies with stellar mass compatible with the Galaxy and Andromeda and with overall disky stellar morphology at $z=0$ (Fig.~\ref{fig:images}) even in the cases when these have undergone a recent major merger: 31 MW/M31-like galaxies in TNG50 (16 per cent) have experienced at least one major merger in the last 5 Gyr (Fig.~\ref{fig:lastMajorMerger}).\\
    
    \item Galaxies with recent major mergers have interacted with relatively massive companions for significant amounts of times, i.e. on average for $\sim$1.4 billion years. The companions, i.e. secondary progenitors, are massive objects, with median stellar mass of $\sim2\times 10^{10}\,\MS$ (i.e. average stellar mass ratio of 1:2.5;  Fig.~\ref{fig:lastMajorMerger}).\\

    \item According to TNG50, there are two main pathways that can lead to a disky MW/M31-mass galaxy at $z=0$ after a recent major merger: i) the pre-existing stellar disk is destroyed during the interactions and merger with the companion, but reforms (Fig.~\ref{fig:circularitiesLMMdestroysDisk}); and ii) somewhat less frequently, the latter does not disrupt the pre-exisisting stellar disk (Fig.~\ref{fig:circularitiesLMMDiskSurvives}). Whether the one or the other occurs depends, for example, on the merger configuration, with mergers with larger stellar mass ratios and smaller impact parameters following more frequently the first scenario (Fig.~\ref{fig:orbits}). \\
    
    \item In both cases, gas was sufficiently available to either trigger star-formation bursts at pericentric passages or at coalescence or both, as well as to sustain prolonged star formation until $z=0$, with the ensuing (re)formation of a disk of young(er) stars (Fig.~\ref{fig:gasFracTmerger}).

\end{itemize}

By comparing the $z=0$ structural and global properties of TNG50 MW/M31 analogues that underwent a recent major merger with those with more ancient last major mergers, we find that the former have, on average:

\begin{itemize}
    \item larger amounts of in-situ stellar mass produced over the last few billion years, namely, as a consequence of the recent merger (Fig.~\ref{fig:circularitiesPrePostLMM});\\
    
    \item similarly massive kinematically-defined bulges (Fig.~\ref{fig:BulgeHaloZ0});\\
    
    \item more massive and somewhat shallower stellar haloes (Fig.~\ref{fig:BulgeHaloZ0}); \\
    
    \item larger amounts and relative fractions of ex-situ, i.e. accreted, stellar mass also at fixed $z=0$ galaxy mass (Fig.~\ref{fig:exsituZ0}); \\
    
    \item thicker and hotter stellar disks, but only for the geometrically thick-disk components and for galaxies with smaller disks (Fig.~\ref{fig:disksSizes}); \\
    
    \item somewhat more massive SMBHs at fixed $z=0$ galaxy mass, and larger gas mass reservoirs and higher star formation rates even at $z=0$. 
\end{itemize}

These results suggest that it may be possible to associate a probability for an observed galaxy to have experienced a recent major merger on the basis of some key structural and global observable properties \citep[][]{Zhu2021, Eisert2022}.

Importantly, the results quantified in this paper align well with the current observational constraints on the properties and on the recent past assembly histories of the Galaxy and Andromeda. As anticipated in the Introduction, whereas the last major merger of the Milky Way may have occurred as early as $8-11$ billion years ago, Andromeda's last major merger may have happened only a few billion years ago. As indicated throughout the paper by reporting known observational properties from the literature, the Milky Way exhibits a steeper stellar halo profile, a thinner and colder stellar disk, and a higher star formation rate -- given its mass -- than Andromeda: all this is consistent with the average phenomenology of TNG50 MW/M31-like galaxies with ancient and recent last major mergers. 

With this paper we clearly signal that, with current state-of-the-art cosmological galaxy simulations that encompass wide ranges of merger histories within the $\Lambda$CDM hierarchical growth of structure, there is no need to simulate galaxies with quiet recent merger histories to obtain galactic stellar disks at $z=0$. This approach had been frequently adopted over the last decade with zoom-in cosmological simulations of {\it MW- or M31-mass} galaxies, whose parent DM haloes had to be chosen from lower-resolution, DM-only volumes for re-simulation. Whereas this choice has almost always been dictated by the difficulties of the overcooling problem and by practical strategies (quieter merger histories imply faster computing times), it has the drawback that, in practice, no past and recent zoom-in simulations could really address the question of whether the absence of a major merger since $z\lesssim1$ is a necessary condition for a galaxy to have a stellar disk as thin as the Galaxy's (i.e. $175-360$ pc and $625-1450$ pc for geometrically thin and thick components, respectively). Additionally, high-resolution galaxy simulations that are suitable for understanding the formation and evolution of Andromeda have been, by imposition or necessity, exceedingly rare, at least until recently and until TNG50. In fact, TNG50 produces a few example galaxies whose more detailed stellar disk structures are also compatible with the Galaxy's and Andromeda's. Even more interesting, TNG50 produces example galaxies that did experience a recent major merger and have thin and thick disk heights similar to those of Andromeda ($0.9-1.3$ and $2.2-3.4$ kpc, respectively).
Moreover, according to TNG50, it does seem improbable, but not impossible, for a galaxy as small (in disk size) as the Galaxy to have emerged from a major merger over the last 5 billion years with thin-disk height as small as the Galaxy's. However, these statements do depend on where within the disk and how the vertical structure is assessed.

We conclude this discussion by highlighting that important structural changes and star formation episodes can be triggered in the progenitors of MW/M31-like galaxies also prior to coalescence with a major companion, i.e. during pericentric passages (see Figs.~\ref{fig:SFRbursts} and \ref{fig:LMMsfrAndDiskyness}). It is hence to be expected, and to be searched for, that the Milky Way's disk and stellar halo may exhibit perturbations (such as waves and vertical spirals) that are linked to the passage of the Large Magellanic Cloud, as recently pointed out with observational data by e.g.  \citet{Vasiliev2021} and \citet{Conroy2021}. Our findings also offer a qualitative glimpse into what our Galaxy may experience in just a a couple of billion years in the future \citep{Cautun2019}, when the Large Magellanic Cloud will eventually merge with it.

\section*{Acknowledgements}
DS, AP, and MD acknowledge support by the Deutsche Forschungsgemeinschaft (DFG, German Research Foundation) -- Project-ID 138713538 -- SFB 881 (``The Milky Way System'', subprojects A01 and A06). The TNG50 simulation was realized with compute time granted by the Gauss Centre for Super- computing (GCS), under the GCS Large-Scale Project GCS-DWAR (2016; PIs Nelson/Pillepich) on the GCS share of the supercomputer Hazel Hen at the High Performance Computing Center Stuttgart (HLRS). DN acknowledges funding from the Deutsche Forschungsgemeinschaft (DFG) through an Emmy Noether Research Group (grant number NE 2441/1-1).

\section*{Data Availability}

Data directly related to this publication and its figures are available on request from the corresponding author. Two catalogs with merger statistics and merger properties for the TNG50 MW/M31-like galaxies can be found at \url{https://www.tng-project.org/results}. The IllustrisTNG simulations, including TNG50, are publicly available and accessible at \url{www.tng-project.org/data} \citep[][]{Nelson2019a}.


\DeclareRobustCommand{\VAN}[3]{#3}
\bibliographystyle{mnras}
\bibliography{MWM31_biblio}

\begin{thebibliography}{}
\makeatletter
\relax
\def\mn@urlcharsother{\let\do\@makeother \do\$\do\&\do\#\do\^\do\_\do\%\do\~}
\def\mn@doi{\begingroup\mn@urlcharsother \@ifnextchar [ {\mn@doi@}
  {\mn@doi@[]}}
\def\mn@doi@[#1]#2{\def\@tempa{#1}\ifx\@tempa\@empty \href
  {http://dx.doi.org/#2} {doi:#2}\else \href {http://dx.doi.org/#2} {#1}\fi
  \endgroup}
\def\mn@eprint#1#2{\mn@eprint@#1:#2::\@nil}
\def\mn@eprint@arXiv#1{\href {http://arxiv.org/abs/#1} {{\tt arXiv:#1}}}
\def\mn@eprint@dblp#1{\href {http://dblp.uni-trier.de/rec/bibtex/#1.xml}
  {dblp:#1}}
\def\mn@eprint@#1:#2:#3:#4\@nil{\def\@tempa {#1}\def\@tempb {#2}\def\@tempc
  {#3}\ifx \@tempc \@empty \let \@tempc \@tempb \let \@tempb \@tempa \fi \ifx
  \@tempb \@empty \def\@tempb {arXiv}\fi \@ifundefined
  {mn@eprint@\@tempb}{\@tempb:\@tempc}{\expandafter \expandafter \csname
  mn@eprint@\@tempb\endcsname \expandafter{\@tempc}}}

\bibitem[\protect\citeauthoryear{{Agertz} \& {Kravtsov}}{{Agertz} \&
  {Kravtsov}}{2015}]{AgertzKravtsov2015}
{Agertz} O.,  {Kravtsov} A.~V.,  2015, \mn@doi [\apj]
  {10.1088/0004-637X/804/1/18}, \href
  {https://ui.adsabs.harvard.edu/abs/2015ApJ...804...18A} {804, 18}

\bibitem[\protect\citeauthoryear{{Agertz}, {Teyssier}  \& {Moore}}{{Agertz}
  et~al.}{2011}]{Agertz2011}
{Agertz} O.,  {Teyssier} R.,   {Moore} B.,  2011, \mn@doi [\mnras]
  {10.1111/j.1365-2966.2010.17530.x}, \href
  {https://ui.adsabs.harvard.edu/abs/2011MNRAS.410.1391A} {410, 1391}

\bibitem[\protect\citeauthoryear{{Aumer}, {White}, {Naab}  \&
  {Scannapieco}}{{Aumer} et~al.}{2013}]{Aumer2013}
{Aumer} M.,  {White} S. D.~M.,  {Naab} T.,   {Scannapieco} C.,  2013, \mn@doi
  [\mnras] {10.1093/mnras/stt1230}, \href
  {https://ui.adsabs.harvard.edu/abs/2013MNRAS.434.3142A} {434, 3142}

\bibitem[\protect\citeauthoryear{{Banerjee} \& {Jog}}{{Banerjee} \&
  {Jog}}{2007}]{Banerjee2007}
{Banerjee} A.,  {Jog} C.~J.,  2007, \mn@doi [\apj] {10.1086/517605}, \href
  {https://ui.adsabs.harvard.edu/abs/2007ApJ...662..335B} {662, 335}

\bibitem[\protect\citeauthoryear{{Barmby} et~al.,}{{Barmby}
  et~al.}{2006}]{Barmby2006}
{Barmby} P.,  et~al., 2006, \mn@doi [\apjl] {10.1086/508626}, \href
  {https://ui.adsabs.harvard.edu/abs/2006ApJ...650L..45B} {650, L45}

\bibitem[\protect\citeauthoryear{{Barnes}}{{Barnes}}{1988}]{Barnes1988}
{Barnes} J.~E.,  1988, \mn@doi [\apj] {10.1086/166593}, \href
  {https://ui.adsabs.harvard.edu/abs/1988ApJ...331..699B} {331, 699}

\bibitem[\protect\citeauthoryear{{Barnes}}{{Barnes}}{1992}]{Barnes1992a}
{Barnes} J.~E.,  1992, \mn@doi [\apj] {10.1086/171522}, \href
  {https://ui.adsabs.harvard.edu/abs/1992ApJ...393..484B} {393, 484}

\bibitem[\protect\citeauthoryear{{Barnes} \& {Hernquist}}{{Barnes} \&
  {Hernquist}}{1996}]{Barnes1996}
{Barnes} J.~E.,  {Hernquist} L.,  1996, \mn@doi [\apj] {10.1086/177957}, \href
  {https://ui.adsabs.harvard.edu/abs/1996ApJ...471..115B} {471, 115}

\bibitem[\protect\citeauthoryear{{Bell} et~al.,}{{Bell}
  et~al.}{2008}]{Bell2008}
{Bell} E.~F.,  et~al., 2008, \mn@doi [\apj] {10.1086/588032}, \href
  {https://ui.adsabs.harvard.edu/abs/2008ApJ...680..295B} {680, 295}

\bibitem[\protect\citeauthoryear{{Belokurov}, {Erkal}, {Evans}, {Koposov}  \&
  {Deason}}{{Belokurov} et~al.}{2018}]{Belokurov2018}
{Belokurov} V.,  {Erkal} D.,  {Evans} N.~W.,  {Koposov} S.~E.,   {Deason}
  A.~J.,  2018, \mn@doi [\mnras] {10.1093/mnras/sty982}, \href
  {https://ui.adsabs.harvard.edu/abs/2018MNRAS.478..611B} {478, 611}

\bibitem[\protect\citeauthoryear{{Bizyaev}, {Kautsch}, {Mosenkov},
  {Reshetnikov}, {Sotnikova}, {Yablokova}  \& {Hillyer}}{{Bizyaev}
  et~al.}{2014}]{2014Bizyaev}
{Bizyaev} D.~V.,  {Kautsch} S.~J.,  {Mosenkov} A.~V.,  {Reshetnikov} V.~P.,
  {Sotnikova} N.~Y.,  {Yablokova} N.~V.,   {Hillyer} R.~W.,  2014, \mn@doi
  [\apj] {10.1088/0004-637X/787/1/24}, \href
  {https://ui.adsabs.harvard.edu/abs/2014ApJ...787...24B} {787, 24}

\bibitem[\protect\citeauthoryear{{Bland-Hawthorn} \&
  {Gerhard}}{{Bland-Hawthorn} \& {Gerhard}}{2016}]{Bland-Hawthorn2016}
{Bland-Hawthorn} J.,  {Gerhard} O.,  2016, \mn@doi [\araa]
  {10.1146/annurev-astro-081915-023441}, \href
  {https://ui.adsabs.harvard.edu/abs/2016ARA&A..54..529B} {54, 529}

\bibitem[\protect\citeauthoryear{{Boardman} et~al.,}{{Boardman}
  et~al.}{2020}]{Boardman2020}
{Boardman} N.,  et~al., 2020, \mn@doi [\mnras] {10.1093/mnras/staa2731}, \href
  {https://ui.adsabs.harvard.edu/abs/2020MNRAS.498.4943B} {498, 4943}

\bibitem[\protect\citeauthoryear{{Bonaca} et~al.,}{{Bonaca}
  et~al.}{2020}]{Bonaca2020}
{Bonaca} A.,  et~al., 2020, \mn@doi [\apjl] {10.3847/2041-8213/ab9caa}, \href
  {https://ui.adsabs.harvard.edu/abs/2020ApJ...897L..18B} {897, L18}

\bibitem[\protect\citeauthoryear{{Buck}, {Obreja}, {Macci{\`o}}, {Minchev},
  {Dutton}  \& {Ostriker}}{{Buck} et~al.}{2020}]{Buck2020}
{Buck} T.,  {Obreja} A.,  {Macci{\`o}} A.~V.,  {Minchev} I.,  {Dutton} A.~A.,
  {Ostriker} J.~P.,  2020, \mn@doi [\mnras] {10.1093/mnras/stz3241}, \href
  {https://ui.adsabs.harvard.edu/abs/2020MNRAS.491.3461B} {491, 3461}

\bibitem[\protect\citeauthoryear{{Cautun}, {Deason}, {Frenk}  \&
  {McAlpine}}{{Cautun} et~al.}{2019}]{Cautun2019}
{Cautun} M.,  {Deason} A.~J.,  {Frenk} C.~S.,   {McAlpine} S.,  2019, \mn@doi
  [\mnras] {10.1093/mnras/sty3084}, \href
  {https://ui.adsabs.harvard.edu/abs/2019MNRAS.483.2185C} {483, 2185}

\bibitem[\protect\citeauthoryear{{Chaplin} et~al.,}{{Chaplin}
  et~al.}{2020}]{Chaplin2020}
{Chaplin} W.~J.,  et~al., 2020, \mn@doi [Nature Astronomy]
  {10.1038/s41550-019-0975-9}, \href
  {https://ui.adsabs.harvard.edu/abs/2020NatAs...4..382C} {4, 382}

\bibitem[\protect\citeauthoryear{{Collins} et~al.,}{{Collins}
  et~al.}{2011}]{Collins2011}
{Collins} M.~L.~M.,  et~al., 2011, \mn@doi [\mnras]
  {10.1111/j.1365-2966.2011.18238.x}, \href
  {https://ui.adsabs.harvard.edu/abs/2011MNRAS.413.1548C} {413, 1548}

\bibitem[\protect\citeauthoryear{{Conroy}, {Naidu}, {Garavito-Camargo},
  {Besla}, {Zaritsky}, {Bonaca}  \& {Johnson}}{{Conroy}
  et~al.}{2021}]{Conroy2021}
{Conroy} C.,  {Naidu} R.~P.,  {Garavito-Camargo} N.,  {Besla} G.,  {Zaritsky}
  D.,  {Bonaca} A.,   {Johnson} B.~D.,  2021, arXiv e-prints, \href
  {https://ui.adsabs.harvard.edu/abs/2021arXiv210409515C} {p. arXiv:2104.09515}

\bibitem[\protect\citeauthoryear{{Crain} et~al.,}{{Crain}
  et~al.}{2015}]{Crain2015}
{Crain} R.~A.,  et~al., 2015, \mn@doi [\mnras] {10.1093/mnras/stv725}, \href
  {https://ui.adsabs.harvard.edu/abs/2015MNRAS.450.1937C} {450, 1937}

\bibitem[\protect\citeauthoryear{{D'Souza} \& {Bell}}{{D'Souza} \&
  {Bell}}{2018}]{DSouza2018}
{D'Souza} R.,  {Bell} E.~F.,  2018, \mn@doi [Nature Astronomy]
  {10.1038/s41550-018-0533-x}, \href
  {https://ui.adsabs.harvard.edu/abs/2018NatAs...2..737D} {2, 737}

\bibitem[\protect\citeauthoryear{{Davis}, {Efstathiou}, {Frenk}  \&
  {White}}{{Davis} et~al.}{1985}]{Davis1985}
{Davis} M.,  {Efstathiou} G.,  {Frenk} C.~S.,   {White} S.~D.~M.,  1985,
  \mn@doi [\apj] {10.1086/163168}, \href
  {https://ui.adsabs.harvard.edu/abs/1985ApJ...292..371D} {292, 371}

\bibitem[\protect\citeauthoryear{{De Silva} et~al.,}{{De Silva}
  et~al.}{2015}]{DeSilva2015}
{De Silva} G.~M.,  et~al., 2015, \mn@doi [\mnras] {10.1093/mnras/stv327}, \href
  {https://ui.adsabs.harvard.edu/abs/2015MNRAS.449.2604D} {449, 2604}

\bibitem[\protect\citeauthoryear{{Deason}, {Belokurov}  \& {Evans}}{{Deason}
  et~al.}{2011}]{Deason2011}
{Deason} A.~J.,  {Belokurov} V.,   {Evans} N.~W.,  2011, \mn@doi [\mnras]
  {10.1111/j.1365-2966.2011.19237.x}, \href
  {https://ui.adsabs.harvard.edu/abs/2011MNRAS.416.2903D} {416, 2903}

\bibitem[\protect\citeauthoryear{{Deason}, {Belokurov}, {Koposov}  \&
  {Rockosi}}{{Deason} et~al.}{2014}]{Deason2014}
{Deason} A.~J.,  {Belokurov} V.,  {Koposov} S.~E.,   {Rockosi} C.~M.,  2014,
  \mn@doi [\apj] {10.1088/0004-637X/787/1/30}, \href
  {https://ui.adsabs.harvard.edu/abs/2014ApJ...787...30D} {787, 30}

\bibitem[\protect\citeauthoryear{{Deng} et~al.,}{{Deng}
  et~al.}{2012}]{Deng2012}
{Deng} L.-C.,  et~al., 2012, \mn@doi [Research in Astronomy and Astrophysics]
  {10.1088/1674-4527/12/7/003}, \href
  {https://ui.adsabs.harvard.edu/abs/2012RAA....12..735D} {12, 735}

\bibitem[\protect\citeauthoryear{{Dierickx} \& {Loeb}}{{Dierickx} \&
  {Loeb}}{2017}]{Dierickx2017}
{Dierickx} M. I.~P.,  {Loeb} A.,  2017, \mn@doi [\apj]
  {10.3847/1538-4357/836/1/92}, \href
  {https://ui.adsabs.harvard.edu/abs/2017ApJ...836...92D} {836, 92}

\bibitem[\protect\citeauthoryear{{\VAN{Dokkum}{van}{van}}~Dokkum
  et~al.,}{{\VAN{Dokkum}{van}{van}}~Dokkum et~al.}{2013}]{vanDokkum2013}
{\VAN{Dokkum}{van}{van}}~Dokkum P.~G.,  et~al., 2013, \mn@doi [\apjl]
  {10.1088/2041-8205/771/2/L35}, \href
  {https://ui.adsabs.harvard.edu/abs/2013ApJ...771L..35V} {771, L35}

\bibitem[\protect\citeauthoryear{{Du}, {Ho}, {Zhao}, {Shi}, {Debattista},
  {Hernquist}  \& {Nelson}}{{Du} et~al.}{2019}]{Du2019}
{Du} M.,  {Ho} L.~C.,  {Zhao} D.,  {Shi} J.,  {Debattista} V.~P.,  {Hernquist}
  L.,   {Nelson} D.,  2019, \mn@doi [\apj] {10.3847/1538-4357/ab43cc}, \href
  {https://ui.adsabs.harvard.edu/abs/2019ApJ...884..129D} {884, 129}

\bibitem[\protect\citeauthoryear{{Du}, {Ho}, {Debattista}, {Pillepich},
  {Nelson}, {Zhao}  \& {Hernquist}}{{Du} et~al.}{2020}]{Du2020}
{Du} M.,  {Ho} L.~C.,  {Debattista} V.~P.,  {Pillepich} A.,  {Nelson} D.,
  {Zhao} D.,   {Hernquist} L.,  2020, \mn@doi [\apj]
  {10.3847/1538-4357/ab8fa8}, \href
  {https://ui.adsabs.harvard.edu/abs/2020ApJ...895..139D} {895, 139}

\bibitem[\protect\citeauthoryear{{Du}, {Ho}, {Debattista}, {Pillepich},
  {Nelson}, {Hernquist}  \& {Weinberger}}{{Du} et~al.}{2021}]{Du2021}
{Du} M.,  {Ho} L.~C.,  {Debattista} V.~P.,  {Pillepich} A.,  {Nelson} D.,
  {Hernquist} L.,   {Weinberger} R.,  2021, arXiv e-prints, \href
  {https://ui.adsabs.harvard.edu/abs/2021arXiv210112373D} {p. arXiv:2101.12373}

\bibitem[\protect\citeauthoryear{{Dubois} et~al.,}{{Dubois}
  et~al.}{2014}]{Dubois2014}
{Dubois} Y.,  et~al., 2014, \mn@doi [\mnras] {10.1093/mnras/stu1227}, \href
  {https://ui.adsabs.harvard.edu/abs/2014MNRAS.444.1453D} {444, 1453}

\bibitem[\protect\citeauthoryear{{Eisert}, {Pillepich}, {Nelson}, {Klessen},
  {Huertas-Company}  \& {Rodriguez-Gomez}}{{Eisert} et~al.}{2022}]{Eisert2022}
{Eisert} L.,  {Pillepich} A.,  {Nelson} D.,  {Klessen} R.~S.,
  {Huertas-Company} M.,   {Rodriguez-Gomez} V.,  2022, arXiv e-prints, \href
  {https://ui.adsabs.harvard.edu/abs/2022arXiv220206967E} {p. arXiv:2202.06967}

\bibitem[\protect\citeauthoryear{{Engler} et~al.,}{{Engler}
  et~al.}{2021}]{Engler2021}
{Engler} C.,  et~al., 2021, \mn@doi [\mnras] {10.1093/mnras/staa3505}, \href
  {https://ui.adsabs.harvard.edu/abs/2021MNRAS.500.3957E} {500, 3957}

\bibitem[\protect\citeauthoryear{{Fakhouri} \& {Ma}}{{Fakhouri} \&
  {Ma}}{2008}]{Fakhouri2008}
{Fakhouri} O.,  {Ma} C.-P.,  2008, \mn@doi [\mnras]
  {10.1111/j.1365-2966.2008.13075.x}, \href
  {https://ui.adsabs.harvard.edu/abs/2008MNRAS.386..577F} {386, 577}

\bibitem[\protect\citeauthoryear{{Fakhouri}, {Ma}  \&
  {Boylan-Kolchin}}{{Fakhouri} et~al.}{2010}]{Fakhouri2010}
{Fakhouri} O.,  {Ma} C.-P.,   {Boylan-Kolchin} M.,  2010, \mn@doi [\mnras]
  {10.1111/j.1365-2966.2010.16859.x}, \href
  {https://ui.adsabs.harvard.edu/abs/2010MNRAS.406.2267F} {406, 2267}

\bibitem[\protect\citeauthoryear{{Flynn}, {Holmberg}, {Portinari}, {Fuchs}  \&
  {Jahrei{\ss}}}{{Flynn} et~al.}{2006}]{Flynn2006}
{Flynn} C.,  {Holmberg} J.,  {Portinari} L.,  {Fuchs} B.,   {Jahrei{\ss}} H.,
  2006, \mn@doi [\mnras] {10.1111/j.1365-2966.2006.10911.x}, \href
  {https://ui.adsabs.harvard.edu/abs/2006MNRAS.372.1149F} {372, 1149}

\bibitem[\protect\citeauthoryear{{Font}, {McCarthy}, {Crain}, {Theuns},
  {Schaye}, {Wiersma}  \& {Dalla Vecchia}}{{Font} et~al.}{2011}]{Font2011}
{Font} A.~S.,  {McCarthy} I.~G.,  {Crain} R.~A.,  {Theuns} T.,  {Schaye} J.,
  {Wiersma} R.~P.~C.,   {Dalla Vecchia} C.,  2011, \mn@doi [\mnras]
  {10.1111/j.1365-2966.2011.19227.x}, \href
  {https://ui.adsabs.harvard.edu/abs/2011MNRAS.416.2802F} {416, 2802}

\bibitem[\protect\citeauthoryear{{Font}, {McCarthy}, {Le Brun}, {Crain}  \&
  {Kelvin}}{{Font} et~al.}{2017}]{Font2017}
{Font} A.~S.,  {McCarthy} I.~G.,  {Le Brun} A. M.~C.,  {Crain} R.~A.,
  {Kelvin} L.~S.,  2017, \mn@doi [\pasa] {10.1017/pasa.2017.50}, \href
  {https://ui.adsabs.harvard.edu/abs/2017PASA...34...50F} {34, e050}

\bibitem[\protect\citeauthoryear{{Font} et~al.,}{{Font}
  et~al.}{2020}]{Font2020}
{Font} A.~S.,  et~al., 2020, \mn@doi [\mnras] {10.1093/mnras/staa2463}, \href
  {https://ui.adsabs.harvard.edu/abs/2020MNRAS.498.1765F} {498, 1765}

\bibitem[\protect\citeauthoryear{{Gaia Collaboration} et~al.,}{{Gaia
  Collaboration} et~al.}{2018}]{GaiaDR22018}
{Gaia Collaboration} et~al., 2018, \mn@doi [\aap]
  {10.1051/0004-6361/201833051}, \href
  {https://ui.adsabs.harvard.edu/abs/2018A&A...616A...1G} {616, A1}

\bibitem[\protect\citeauthoryear{{Gallart}, {Bernard}, {Brook}, {Ruiz-Lara},
  {Cassisi}, {Hill}  \& {Monelli}}{{Gallart} et~al.}{2019}]{Gallart2019}
{Gallart} C.,  {Bernard} E.~J.,  {Brook} C.~B.,  {Ruiz-Lara} T.,  {Cassisi} S.,
   {Hill} V.,   {Monelli} M.,  2019, \mn@doi [Nature Astronomy]
  {10.1038/s41550-019-0829-5}, \href
  {https://ui.adsabs.harvard.edu/abs/2019NatAs...3..932G} {3, 932}

\bibitem[\protect\citeauthoryear{{Garrison-Kimmel} et~al.,}{{Garrison-Kimmel}
  et~al.}{2018}]{Garrison-Kimmel2018}
{Garrison-Kimmel} S.,  et~al., 2018, \mn@doi [\mnras] {10.1093/mnras/sty2513},
  \href {https://ui.adsabs.harvard.edu/abs/2018MNRAS.481.4133G} {481, 4133}

\bibitem[\protect\citeauthoryear{{Geehan}, {Fardal}, {Babul}  \&
  {Guhathakurta}}{{Geehan} et~al.}{2006}]{Geehan2006}
{Geehan} J.~J.,  {Fardal} M.~A.,  {Babul} A.,   {Guhathakurta} P.,  2006,
  \mn@doi [\mnras] {10.1111/j.1365-2966.2005.09863.x}, \href
  {https://ui.adsabs.harvard.edu/abs/2006MNRAS.366..996G} {366, 996}

\bibitem[\protect\citeauthoryear{{Genel}, {Bouch{\'e}}, {Naab}, {Sternberg}  \&
  {Genzel}}{{Genel} et~al.}{2010}]{Genel2010}
{Genel} S.,  {Bouch{\'e}} N.,  {Naab} T.,  {Sternberg} A.,   {Genzel} R.,
  2010, \mn@doi [\apj] {10.1088/0004-637X/719/1/229}, \href
  {https://ui.adsabs.harvard.edu/abs/2010ApJ...719..229G} {719, 229}

\bibitem[\protect\citeauthoryear{{Genel} et~al.,}{{Genel}
  et~al.}{2014}]{Genel2014}
{Genel} S.,  et~al., 2014, \mn@doi [\mnras] {10.1093/mnras/stu1654}, \href
  {https://ui.adsabs.harvard.edu/abs/2014MNRAS.445..175G} {445, 175}

\bibitem[\protect\citeauthoryear{{Gilbert} et~al.,}{{Gilbert}
  et~al.}{2012}]{Gilbert2012}
{Gilbert} K.~M.,  et~al., 2012, \mn@doi [\apj] {10.1088/0004-637X/760/1/76},
  \href {https://ui.adsabs.harvard.edu/abs/2012ApJ...760...76G} {760, 76}

\bibitem[\protect\citeauthoryear{{Gould}, {Bahcall}  \& {Flynn}}{{Gould}
  et~al.}{1996}]{Gould1996}
{Gould} A.,  {Bahcall} J.~N.,   {Flynn} C.,  1996, \mn@doi [\apj]
  {10.1086/177460}, \href
  {https://ui.adsabs.harvard.edu/abs/1996ApJ...465..759G} {465, 759}

\bibitem[\protect\citeauthoryear{{Governato} et~al.,}{{Governato}
  et~al.}{2004}]{Governato2004}
{Governato} F.,  et~al., 2004, \mn@doi [\apj] {10.1086/383516}, \href
  {https://ui.adsabs.harvard.edu/abs/2004ApJ...607..688G} {607, 688}

\bibitem[\protect\citeauthoryear{{Grand} et~al.,}{{Grand}
  et~al.}{2017}]{Grand2017}
{Grand} R. J.~J.,  et~al., 2017, \mn@doi [\mnras] {10.1093/mnras/stx071}, \href
  {https://ui.adsabs.harvard.edu/abs/2017MNRAS.467..179G} {467, 179}

\bibitem[\protect\citeauthoryear{{Grand} et~al.,}{{Grand}
  et~al.}{2020}]{Grand2020}
{Grand} R. J.~J.,  et~al., 2020, \mn@doi [\mnras] {10.1093/mnras/staa2057},
  \href {https://ui.adsabs.harvard.edu/abs/2020MNRAS.497.1603G} {497, 1603}

\bibitem[\protect\citeauthoryear{{Guedes}, {Callegari}, {Madau}  \&
  {Mayer}}{{Guedes} et~al.}{2011}]{Guedes2011}
{Guedes} J.,  {Callegari} S.,  {Madau} P.,   {Mayer} L.,  2011, \mn@doi [\apj]
  {10.1088/0004-637X/742/2/76}, \href
  {https://ui.adsabs.harvard.edu/abs/2011ApJ...742...76G} {742, 76}

\bibitem[\protect\citeauthoryear{{Hammer}, {Puech}, {Chemin}, {Flores}  \&
  {Lehnert}}{{Hammer} et~al.}{2007}]{Hammer2007}
{Hammer} F.,  {Puech} M.,  {Chemin} L.,  {Flores} H.,   {Lehnert} M.~D.,  2007,
  \mn@doi [\apj] {10.1086/516727}, \href
  {https://ui.adsabs.harvard.edu/abs/2007ApJ...662..322H} {662, 322}

\bibitem[\protect\citeauthoryear{{Hau}, {Bower}, {Kilborn}, {Forbes}, {Balogh}
  \& {Oosterloo}}{{Hau} et~al.}{2008}]{Hau2008}
{Hau} G. K.~T.,  {Bower} R.~G.,  {Kilborn} V.,  {Forbes} D.~A.,  {Balogh}
  M.~L.,   {Oosterloo} T.,  2008, \mn@doi [\mnras]
  {10.1111/j.1365-2966.2007.12740.x}, \href
  {https://ui.adsabs.harvard.edu/abs/2008MNRAS.385.1965H} {385, 1965}

\bibitem[\protect\citeauthoryear{{Helmi}, {White}, {de Zeeuw}  \&
  {Zhao}}{{Helmi} et~al.}{1999}]{Helmi1999}
{Helmi} A.,  {White} S. D.~M.,  {de Zeeuw} P.~T.,   {Zhao} H.,  1999, \mn@doi
  [\nat] {10.1038/46980}, \href
  {https://ui.adsabs.harvard.edu/abs/1999Natur.402...53H} {402, 53}

\bibitem[\protect\citeauthoryear{{Helmi}, {Babusiaux}, {Koppelman}, {Massari},
  {Veljanoski}  \& {Brown}}{{Helmi} et~al.}{2018}]{Helmi2018}
{Helmi} A.,  {Babusiaux} C.,  {Koppelman} H.~H.,  {Massari} D.,  {Veljanoski}
  J.,   {Brown} A. G.~A.,  2018, \mn@doi [\nat] {10.1038/s41586-018-0625-x},
  \href {https://ui.adsabs.harvard.edu/abs/2018Natur.563...85H} {563, 85}

\bibitem[\protect\citeauthoryear{{Hernquist}}{{Hernquist}}{1989}]{Hernquist1989}
{Hernquist} L.,  1989, \mn@doi [\nat] {10.1038/340687a0}, \href
  {https://ui.adsabs.harvard.edu/abs/1989Natur.340..687H} {340, 687}

\bibitem[\protect\citeauthoryear{{Hernquist}}{{Hernquist}}{1993}]{Hernquist1993}
{Hernquist} L.,  1993, \mn@doi [\apj] {10.1086/172686}, \href
  {https://ui.adsabs.harvard.edu/abs/1993ApJ...409..548H} {409, 548}

\bibitem[\protect\citeauthoryear{{Hernquist} \& {Barnes}}{{Hernquist} \&
  {Barnes}}{1991}]{Hernquist1991}
{Hernquist} L.,  {Barnes} J.~E.,  1991, \mn@doi [\nat] {10.1038/354210a0},
  \href {https://ui.adsabs.harvard.edu/abs/1991Natur.354..210H} {354, 210}

\bibitem[\protect\citeauthoryear{{Hoffman}, {Cox}, {Dutta}  \&
  {Hernquist}}{{Hoffman} et~al.}{2010}]{Hoffman2010}
{Hoffman} L.,  {Cox} T.~J.,  {Dutta} S.,   {Hernquist} L.,  2010, \mn@doi
  [\apj] {10.1088/0004-637X/723/1/818}, \href
  {https://ui.adsabs.harvard.edu/abs/2010ApJ...723..818H} {723, 818}

\bibitem[\protect\citeauthoryear{{Hopkins}, {Cox}, {Younger}  \&
  {Hernquist}}{{Hopkins} et~al.}{2009}]{Hopkins2009}
{Hopkins} P.~F.,  {Cox} T.~J.,  {Younger} J.~D.,   {Hernquist} L.,  2009,
  \mn@doi [\apj] {10.1088/0004-637X/691/2/1168}, \href
  {https://ui.adsabs.harvard.edu/abs/2009ApJ...691.1168H} {691, 1168}

\bibitem[\protect\citeauthoryear{{Hopkins} et~al.,}{{Hopkins}
  et~al.}{2010}]{Hopkins2010}
{Hopkins} P.~F.,  et~al., 2010, \mn@doi [\apj] {10.1088/0004-637X/715/1/202},
  \href {https://ui.adsabs.harvard.edu/abs/2010ApJ...715..202H} {715, 202}

\bibitem[\protect\citeauthoryear{{Ibata} et~al.,}{{Ibata}
  et~al.}{2014}]{Ibata2014}
{Ibata} R.~A.,  et~al., 2014, \mn@doi [\apj] {10.1088/0004-637X/780/2/128},
  \href {https://ui.adsabs.harvard.edu/abs/2014ApJ...780..128I} {780, 128}

\bibitem[\protect\citeauthoryear{{Ilbert} et~al.,}{{Ilbert}
  et~al.}{2006}]{Ilbert2006}
{Ilbert} O.,  et~al., 2006, \mn@doi [\aap] {10.1051/0004-6361:20053632}, \href
  {https://ui.adsabs.harvard.edu/abs/2006A&A...453..809I} {453, 809}

\bibitem[\protect\citeauthoryear{{Ivezi{\'c}} et~al.,}{{Ivezi{\'c}}
  et~al.}{2000}]{Ivezic2000}
{Ivezi{\'c}} {\v{Z}}.,  et~al., 2000, \mn@doi [\aj] {10.1086/301455}, \href
  {https://ui.adsabs.harvard.edu/abs/2000AJ....120..963I} {120, 963}

\bibitem[\protect\citeauthoryear{{Jackson}, {Martin}, {Kaviraj}, {Laigle},
  {Devriendt}, {Dubois}  \& {Pichon}}{{Jackson} et~al.}{2020}]{Jackson2020}
{Jackson} R.~A.,  {Martin} G.,  {Kaviraj} S.,  {Laigle} C.,  {Devriendt}
  J.~E.~G.,  {Dubois} Y.,   {Pichon} C.,  2020, \mn@doi [\mnras]
  {10.1093/mnras/staa970}, \href
  {https://ui.adsabs.harvard.edu/abs/2020MNRAS.494.5568J} {494, 5568}

\bibitem[\protect\citeauthoryear{{Joshi}, {Pillepich}, {Nelson}, {Marinacci},
  {Springel}, {Rodriguez-Gomez}, {Vogelsberger}  \& {Hernquist}}{{Joshi}
  et~al.}{2020}]{Joshi2020}
{Joshi} G.~D.,  {Pillepich} A.,  {Nelson} D.,  {Marinacci} F.,  {Springel} V.,
  {Rodriguez-Gomez} V.,  {Vogelsberger} M.,   {Hernquist} L.,  2020, \mn@doi
  [\mnras] {10.1093/mnras/staa1668}, \href
  {https://ui.adsabs.harvard.edu/abs/2020MNRAS.496.2673J} {496, 2673}

\bibitem[\protect\citeauthoryear{{Juri{\'c}} et~al.,}{{Juri{\'c}}
  et~al.}{2008}]{Juric2008}
{Juri{\'c}} M.,  et~al., 2008, \mn@doi [\apj] {10.1086/523619}, \href
  {https://ui.adsabs.harvard.edu/abs/2008ApJ...673..864J} {673, 864}

\bibitem[\protect\citeauthoryear{{Kannappan}, {Guie}  \& {Baker}}{{Kannappan}
  et~al.}{2009}]{Kannappan2009}
{Kannappan} S.~J.,  {Guie} J.~M.,   {Baker} A.~J.,  2009, \mn@doi [\aj]
  {10.1088/0004-6256/138/2/579}, \href
  {https://ui.adsabs.harvard.edu/abs/2009AJ....138..579K} {138, 579}

\bibitem[\protect\citeauthoryear{{Kautsch}, {Grebel}, {Barazza}  \&
  {Gallagher}}{{Kautsch} et~al.}{2006}]{Kautsch2006}
{Kautsch} S.~J.,  {Grebel} E.~K.,  {Barazza} F.~D.,   {Gallagher} J.~S. I.,
  2006, \mn@doi [\aap] {10.1051/0004-6361:20053981e}, \href
  {https://ui.adsabs.harvard.edu/abs/2006A&A...451.1171K} {451, 1171}

\bibitem[\protect\citeauthoryear{{Kelvin} et~al.,}{{Kelvin}
  et~al.}{2014}]{Kelvin2014}
{Kelvin} L.~S.,  et~al., 2014, \mn@doi [\mnras] {10.1093/mnras/stu1507}, \href
  {https://ui.adsabs.harvard.edu/abs/2014MNRAS.444.1647K} {444, 1647}

\bibitem[\protect\citeauthoryear{{Kere{\v{s}}}, {Vogelsberger}, {Sijacki},
  {Springel}  \& {Hernquist}}{{Kere{\v{s}}} et~al.}{2012}]{Keres2012}
{Kere{\v{s}}} D.,  {Vogelsberger} M.,  {Sijacki} D.,  {Springel} V.,
  {Hernquist} L.,  2012, \mn@doi [\mnras] {10.1111/j.1365-2966.2012.21548.x},
  \href {https://ui.adsabs.harvard.edu/abs/2012MNRAS.425.2027K} {425, 2027}

\bibitem[\protect\citeauthoryear{{Koppelman}, {Helmi}, {Massari},
  {Price-Whelan}  \& {Starkenburg}}{{Koppelman} et~al.}{2019}]{Koppelman2019}
{Koppelman} H.~H.,  {Helmi} A.,  {Massari} D.,  {Price-Whelan} A.~M.,
  {Starkenburg} T.~K.,  2019, \mn@doi [\aap] {10.1051/0004-6361/201936738},
  \href {https://ui.adsabs.harvard.edu/abs/2019A&A...631L...9K} {631, L9}

\bibitem[\protect\citeauthoryear{{\VAN{Kruit}{van der}{van
  der}}~Kruit}{{\VAN{Kruit}{van der}{van der}}~Kruit}{1988}]{1988vanderKruit}
{\VAN{Kruit}{van der}{van der}}~Kruit P.~C.,  1988, \aap, \href
  {https://ui.adsabs.harvard.edu/abs/1988A&A...192..117V} {192, 117}

\bibitem[\protect\citeauthoryear{{Kunder} et~al.,}{{Kunder}
  et~al.}{2017}]{Kunder2017}
{Kunder} A.,  et~al., 2017, \mn@doi [\aj] {10.3847/1538-3881/153/2/75}, \href
  {https://ui.adsabs.harvard.edu/abs/2017AJ....153...75K} {153, 75}

\bibitem[\protect\citeauthoryear{{Licquia} \& {Newman}}{{Licquia} \&
  {Newman}}{2015}]{Licquia2015}
{Licquia} T.~C.,  {Newman} J.~A.,  2015, \mn@doi [\apj]
  {10.1088/0004-637X/806/1/96}, \href
  {https://ui.adsabs.harvard.edu/abs/2015ApJ...806...96L} {806, 96}

\bibitem[\protect\citeauthoryear{{Lindegren} et~al.,}{{Lindegren}
  et~al.}{2016}]{Lindegren2016}
{Lindegren} L.,  et~al., 2016, \mn@doi [\aap] {10.1051/0004-6361/201628714},
  \href {https://ui.adsabs.harvard.edu/abs/2016A&A...595A...4L} {595, A4}

\bibitem[\protect\citeauthoryear{{Ma}, {Hopkins}, {Wetzel}, {Kirby},
  {Angl{\'e}s-Alc{\'a}zar}, {Faucher-Gigu{\`e}re}, {Kere{\v{s}}}  \&
  {Quataert}}{{Ma} et~al.}{2017}]{2017Ma}
{Ma} X.,  {Hopkins} P.~F.,  {Wetzel} A.~R.,  {Kirby} E.~N.,
  {Angl{\'e}s-Alc{\'a}zar} D.,  {Faucher-Gigu{\`e}re} C.-A.,  {Kere{\v{s}}} D.,
    {Quataert} E.,  2017, \mn@doi [\mnras] {10.1093/mnras/stx273}, \href
  {https://ui.adsabs.harvard.edu/abs/2017MNRAS.467.2430M} {467, 2430}

\bibitem[\protect\citeauthoryear{{Majewski} et~al.,}{{Majewski}
  et~al.}{2017}]{Majewski2017}
{Majewski} S.~R.,  et~al., 2017, \mn@doi [\aj] {10.3847/1538-3881/aa784d},
  \href {https://ui.adsabs.harvard.edu/abs/2017AJ....154...94M} {154, 94}

\bibitem[\protect\citeauthoryear{{Marinacci}, {Pakmor}  \&
  {Springel}}{{Marinacci} et~al.}{2014}]{Marinacci2014}
{Marinacci} F.,  {Pakmor} R.,   {Springel} V.,  2014, \mn@doi [\mnras]
  {10.1093/mnras/stt2003}, \href
  {https://ui.adsabs.harvard.edu/abs/2014MNRAS.437.1750M} {437, 1750}

\bibitem[\protect\citeauthoryear{{Marinacci} et~al.,}{{Marinacci}
  et~al.}{2018}]{Marinacci2018}
{Marinacci} F.,  et~al., 2018, \mn@doi [\mnras] {10.1093/mnras/sty2206}, \href
  {https://ui.adsabs.harvard.edu/abs/2018MNRAS.480.5113M} {480, 5113}

\bibitem[\protect\citeauthoryear{{Martig}, {Bournaud}, {Croton}, {Dekel}  \&
  {Teyssier}}{{Martig} et~al.}{2012}]{Martig2012}
{Martig} M.,  {Bournaud} F.,  {Croton} D.~J.,  {Dekel} A.,   {Teyssier} R.,
  2012, \mn@doi [\apj] {10.1088/0004-637X/756/1/26}, \href
  {https://ui.adsabs.harvard.edu/abs/2012ApJ...756...26M} {756, 26}

\bibitem[\protect\citeauthoryear{{Martin}, {Kaviraj}, {Devriendt}, {Dubois}  \&
  {Pichon}}{{Martin} et~al.}{2018}]{Martin2018}
{Martin} G.,  {Kaviraj} S.,  {Devriendt} J.~E.~G.,  {Dubois} Y.,   {Pichon} C.,
   2018, \mn@doi [\mnras] {10.1093/mnras/sty1936}, \href
  {https://ui.adsabs.harvard.edu/abs/2018MNRAS.480.2266M} {480, 2266}

\bibitem[\protect\citeauthoryear{{McConnachie}, {Irwin}, {Ferguson}, {Ibata},
  {Lewis}  \& {Tanvir}}{{McConnachie} et~al.}{2005}]{McConnachie2005}
{McConnachie} A.~W.,  {Irwin} M.~J.,  {Ferguson} A.~M.~N.,  {Ibata} R.~A.,
  {Lewis} G.~F.,   {Tanvir} N.,  2005, \mn@doi [\mnras]
  {10.1111/j.1365-2966.2004.08514.x}, \href
  {https://ui.adsabs.harvard.edu/abs/2005MNRAS.356..979M} {356, 979}

\bibitem[\protect\citeauthoryear{{McDermid} et~al.,}{{McDermid}
  et~al.}{2006}]{McDermid2006}
{McDermid} R.~M.,  et~al., 2006, \mn@doi [\mnras]
  {10.1111/j.1365-2966.2006.11065.x}, \href
  {https://ui.adsabs.harvard.edu/abs/2006MNRAS.373..906M} {373, 906}

\bibitem[\protect\citeauthoryear{{\VAN{Merel}{van der}{van
  der}}~Marel}{{\VAN{Merel}{van der}{van der}}~Marel}{2006}]{vanderMarel2006}
{\VAN{Merel}{van der}{van der}}~Marel R.~P.,  2006, The Large Magellanic Cloud:
  Structure and kinematics.
Cambridge University Press, p. 47–71, \mn@doi{10.1017/CBO9780511734908.005}

\bibitem[\protect\citeauthoryear{{Myeong}, {Evans}, {Belokurov}, {Sanders}  \&
  {Koposov}}{{Myeong} et~al.}{2018}]{Myeong2018}
{Myeong} G.~C.,  {Evans} N.~W.,  {Belokurov} V.,  {Sanders} J.~L.,   {Koposov}
  S.~E.,  2018, \mn@doi [\apjl] {10.3847/2041-8213/aad7f7}, \href
  {https://ui.adsabs.harvard.edu/abs/2018ApJ...863L..28M} {863, L28}

\bibitem[\protect\citeauthoryear{{Myeong}, {Vasiliev}, {Iorio}, {Evans}  \&
  {Belokurov}}{{Myeong} et~al.}{2019}]{Myeong2019}
{Myeong} G.~C.,  {Vasiliev} E.,  {Iorio} G.,  {Evans} N.~W.,   {Belokurov} V.,
  2019, \mn@doi [\mnras] {10.1093/mnras/stz1770}, \href
  {https://ui.adsabs.harvard.edu/abs/2019MNRAS.488.1235M} {488, 1235}

\bibitem[\protect\citeauthoryear{{Naab}, {Jesseit}  \& {Burkert}}{{Naab}
  et~al.}{2006}]{Naab2006}
{Naab} T.,  {Jesseit} R.,   {Burkert} A.,  2006, \mn@doi [\mnras]
  {10.1111/j.1365-2966.2006.10902.x}, \href
  {https://ui.adsabs.harvard.edu/abs/2006MNRAS.372..839N} {372, 839}

\bibitem[\protect\citeauthoryear{{Naidu} et~al.,}{{Naidu}
  et~al.}{2021}]{Naidu2021}
{Naidu} R.~P.,  et~al., 2021, arXiv e-prints, \href
  {https://ui.adsabs.harvard.edu/abs/2021arXiv210303251N} {p. arXiv:2103.03251}

\bibitem[\protect\citeauthoryear{{Naiman} et~al.,}{{Naiman}
  et~al.}{2018}]{Naiman2018}
{Naiman} J.~P.,  et~al., 2018, \mn@doi [\mnras] {10.1093/mnras/sty618}, \href
  {https://ui.adsabs.harvard.edu/abs/2018MNRAS.477.1206N} {477, 1206}

\bibitem[\protect\citeauthoryear{{Nelson} et~al.,}{{Nelson}
  et~al.}{2015}]{Nelson2015}
{Nelson} D.,  et~al., 2015, \mn@doi [Astronomy and Computing]
  {10.1016/j.ascom.2015.09.003}, \href
  {https://ui.adsabs.harvard.edu/abs/2015A&C....13...12N} {13, 12}

\bibitem[\protect\citeauthoryear{{Nelson} et~al.,}{{Nelson}
  et~al.}{2018}]{Nelson2018}
{Nelson} D.,  et~al., 2018, \mn@doi [\mnras] {10.1093/mnras/stx3040}, \href
  {https://ui.adsabs.harvard.edu/abs/2018MNRAS.475..624N} {475, 624}

\bibitem[\protect\citeauthoryear{{Nelson} et~al.,}{{Nelson}
  et~al.}{2019a}]{Nelson2019a}
{Nelson} D.,  et~al., 2019a, \mn@doi [Computational Astrophysics and Cosmology]
  {10.1186/s40668-019-0028-x}, \href
  {https://ui.adsabs.harvard.edu/abs/2019ComAC...6....2N} {6, 2}

\bibitem[\protect\citeauthoryear{{Nelson} et~al.,}{{Nelson}
  et~al.}{2019b}]{Nelson2019b}
{Nelson} D.,  et~al., 2019b, \mn@doi [\mnras] {10.1093/mnras/stz2306}, \href
  {https://ui.adsabs.harvard.edu/abs/2019MNRAS.490.3234N} {490, 3234}

\bibitem[\protect\citeauthoryear{{Ojha}}{{Ojha}}{2001}]{Ohja2001}
{Ojha} D.~K.,  2001, \mn@doi [\mnras] {10.1046/j.1365-8711.2001.04155.x}, \href
  {https://ui.adsabs.harvard.edu/abs/2001MNRAS.322..426O} {322, 426}

\bibitem[\protect\citeauthoryear{{Okamoto}, {Eke}, {Frenk}  \&
  {Jenkins}}{{Okamoto} et~al.}{2005}]{Okamoto2005}
{Okamoto} T.,  {Eke} V.~R.,  {Frenk} C.~S.,   {Jenkins} A.,  2005, \mn@doi
  [\mnras] {10.1111/j.1365-2966.2005.09525.x}, \href
  {https://ui.adsabs.harvard.edu/abs/2005MNRAS.363.1299O} {363, 1299}

\bibitem[\protect\citeauthoryear{{Park}, {Choi}, {Vogeley}, {Gott}, {Blanton}
  \& {SDSS Collaboration}}{{Park} et~al.}{2007}]{Park2007}
{Park} C.,  {Choi} Y.-Y.,  {Vogeley} M.~S.,  {Gott} J.~Richard I.,  {Blanton}
  M.~R.,   {SDSS Collaboration} 2007, \mn@doi [\apj] {10.1086/511059}, \href
  {https://ui.adsabs.harvard.edu/abs/2007ApJ...658..898P} {658, 898}

\bibitem[\protect\citeauthoryear{{Park} et~al.,}{{Park}
  et~al.}{2019}]{Park2019}
{Park} M.-J.,  et~al., 2019, \mn@doi [\apj] {10.3847/1538-4357/ab3afe}, \href
  {https://ui.adsabs.harvard.edu/abs/2019ApJ...883...25P} {883, 25}

\bibitem[\protect\citeauthoryear{{Park} et~al.,}{{Park}
  et~al.}{2021}]{2021Park}
{Park} M.~J.,  et~al., 2021, \mn@doi [\apjs] {10.3847/1538-4365/abe937}, \href
  {https://ui.adsabs.harvard.edu/abs/2021ApJS..254....2P} {254, 2}

\bibitem[\protect\citeauthoryear{{Peschken}, {{\L}okas}  \&
  {Athanassoula}}{{Peschken} et~al.}{2020}]{Peschken2020}
{Peschken} N.,  {{\L}okas} E.~L.,   {Athanassoula} E.,  2020, \mn@doi [\mnras]
  {10.1093/mnras/staa299}, \href
  {https://ui.adsabs.harvard.edu/abs/2020MNRAS.493.1375P} {493, 1375}

\bibitem[\protect\citeauthoryear{{Aumer}, {White}, {Naab}  \&
  {Scannapieco}}{Pil}{}]{PillepichInPrep}


\bibitem[\protect\citeauthoryear{{Pillepich} et~al.,}{{Pillepich}
  et~al.}{2014}]{Pillepich2014}
{Pillepich} A.,  et~al., 2014, \mn@doi [\mnras] {10.1093/mnras/stu1408}, \href
  {https://ui.adsabs.harvard.edu/abs/2014MNRAS.444..237P} {444, 237}

\bibitem[\protect\citeauthoryear{{Pillepich}, {Madau}  \& {Mayer}}{{Pillepich}
  et~al.}{2015}]{Pillepich2015}
{Pillepich} A.,  {Madau} P.,   {Mayer} L.,  2015, \mn@doi [\apj]
  {10.1088/0004-637X/799/2/184}, \href
  {https://ui.adsabs.harvard.edu/abs/2015ApJ...799..184P} {799, 184}

\bibitem[\protect\citeauthoryear{{Pillepich} et~al.,}{{Pillepich}
  et~al.}{2018a}]{Pillepich2018}
{Pillepich} A.,  et~al., 2018a, \mn@doi [\mnras] {10.1093/mnras/stx2656}, \href
  {https://ui.adsabs.harvard.edu/abs/2018MNRAS.473.4077P} {473, 4077}

\bibitem[\protect\citeauthoryear{{Pillepich} et~al.,}{{Pillepich}
  et~al.}{2018b}]{Pillepich2018b}
{Pillepich} A.,  et~al., 2018b, \mn@doi [\mnras] {10.1093/mnras/stx3112}, \href
  {https://ui.adsabs.harvard.edu/abs/2018MNRAS.475..648P} {475, 648}

\bibitem[\protect\citeauthoryear{{Pillepich} et~al.,}{{Pillepich}
  et~al.}{2019}]{Pillepich2019}
{Pillepich} A.,  et~al., 2019, \mn@doi [\mnras] {10.1093/mnras/stz2338}, \href
  {https://ui.adsabs.harvard.edu/abs/2019MNRAS.490.3196P} {490, 3196}

\bibitem[\protect\citeauthoryear{{Pillepich}, {Nelson}, {Truong}, {Weinberger},
  {Martin-Navarro}, {Springel}, {Faber}  \& {Hernquist}}{{Pillepich}
  et~al.}{2021}]{Pillepich2021a}
{Pillepich} A.,  {Nelson} D.,  {Truong} N.,  {Weinberger} R.,  {Martin-Navarro}
  I.,  {Springel} V.,  {Faber} S.~M.,   {Hernquist} L.,  2021, \mn@doi [\mnras]
  {10.1093/mnras/stab2779}, \href
  {https://ui.adsabs.harvard.edu/abs/2021MNRAS.508.4667P} {508, 4667}

\bibitem[\protect\citeauthoryear{{Planck Collaboration} et~al.,}{{Planck
  Collaboration} et~al.}{2016}]{Planck2016}
{Planck Collaboration} et~al., 2016, \mn@doi [\aap]
  {10.1051/0004-6361/201525830}, \href
  {https://ui.adsabs.harvard.edu/abs/2016A&A...594A..13P} {594, A13}

\bibitem[\protect\citeauthoryear{{Puech}, {Hammer}, {Hopkins}, {Athanassoula},
  {Flores}, {Rodrigues}, {Wang}  \& {Yang}}{{Puech} et~al.}{2012}]{Puech2012}
{Puech} M.,  {Hammer} F.,  {Hopkins} P.~F.,  {Athanassoula} E.,  {Flores} H.,
  {Rodrigues} M.,  {Wang} J.~L.,   {Yang} Y.~B.,  2012, \mn@doi [\apj]
  {10.1088/0004-637X/753/2/128}, \href
  {https://ui.adsabs.harvard.edu/abs/2012ApJ...753..128P} {753, 128}

\bibitem[\protect\citeauthoryear{{Qu}, {Di Matteo}, {Lehnert}  \& {van
  Driel}}{{Qu} et~al.}{2011}]{Qu2011}
{Qu} Y.,  {Di Matteo} P.,  {Lehnert} M.~D.,   {van Driel} W.,  2011, \mn@doi
  [\aap] {10.1051/0004-6361/201015224}, \href
  {https://ui.adsabs.harvard.edu/abs/2011A&A...530A..10Q} {530, A10}

\bibitem[\protect\citeauthoryear{{Renaud}, {Agertz}, {Read}, {Ryde},
  {Andersson}, {Bensby}, {Rey}  \& {Feuillet}}{{Renaud}
  et~al.}{2021}]{Renaud2021}
{Renaud} F.,  {Agertz} O.,  {Read} J.~I.,  {Ryde} N.,  {Andersson} E.~P.,
  {Bensby} T.,  {Rey} M.~P.,   {Feuillet} D.~K.,  2021, \mn@doi [\mnras]
  {10.1093/mnras/stab250}, \href
  {https://ui.adsabs.harvard.edu/abs/2021MNRAS.503.5846R} {503, 5846}

\bibitem[\protect\citeauthoryear{{Riess}, {Fliri}  \& {Valls-Gabaud}}{{Riess}
  et~al.}{2012}]{Riess2012}
{Riess} A.~G.,  {Fliri} J.,   {Valls-Gabaud} D.,  2012, \mn@doi [\apj]
  {10.1088/0004-637X/745/2/156}, \href
  {https://ui.adsabs.harvard.edu/abs/2012ApJ...745..156R} {745, 156}

\bibitem[\protect\citeauthoryear{{Rix} \& {Bovy}}{{Rix} \&
  {Bovy}}{2013}]{RixBovy2013}
{Rix} H.-W.,  {Bovy} J.,  2013, \mn@doi [\aapr] {10.1007/s00159-013-0061-8},
  \href {https://ui.adsabs.harvard.edu/abs/2013A&ARv..21...61R} {21, 61}

\bibitem[\protect\citeauthoryear{{Robertson}, {Bullock}, {Cox}, {Di Matteo},
  {Hernquist}, {Springel}  \& {Yoshida}}{{Robertson}
  et~al.}{2006}]{Robertson2006}
{Robertson} B.,  {Bullock} J.~S.,  {Cox} T.~J.,  {Di Matteo} T.,  {Hernquist}
  L.,  {Springel} V.,   {Yoshida} N.,  2006, \mn@doi [\apj] {10.1086/504412},
  \href {https://ui.adsabs.harvard.edu/abs/2006ApJ...645..986R} {645, 986}

\bibitem[\protect\citeauthoryear{{Roca-F{\`a}brega}, {Valenzuela},
  {Col{\'\i}n}, {Figueras}, {Krongold}, {Vel{\'a}zquez}, {Avila-Reese}  \&
  {Ibarra-Medel}}{{Roca-F{\`a}brega} et~al.}{2016}]{RocaFabrega2016}
{Roca-F{\`a}brega} S.,  {Valenzuela} O.,  {Col{\'\i}n} P.,  {Figueras} F.,
  {Krongold} Y.,  {Vel{\'a}zquez} H.,  {Avila-Reese} V.,   {Ibarra-Medel} H.,
  2016, \mn@doi [\apj] {10.3847/0004-637X/824/2/94}, \href
  {https://ui.adsabs.harvard.edu/abs/2016ApJ...824...94R} {824, 94}

\bibitem[\protect\citeauthoryear{{Roca-F{\`a}brega} et~al.,}{{Roca-F{\`a}brega}
  et~al.}{2021}]{RocaFabrega2021}
{Roca-F{\`a}brega} S.,  et~al., 2021, arXiv e-prints, \href
  {https://ui.adsabs.harvard.edu/abs/2021arXiv210609738R} {p. arXiv:2106.09738}

\bibitem[\protect\citeauthoryear{{Rodriguez-Gomez} et~al.,}{{Rodriguez-Gomez}
  et~al.}{2015}]{RodGom2015}
{Rodriguez-Gomez} V.,  et~al., 2015, \mn@doi [\mnras] {10.1093/mnras/stv264},
  \href {https://ui.adsabs.harvard.edu/abs/2015MNRAS.449...49R} {449, 49}

\bibitem[\protect\citeauthoryear{{Rodriguez-Gomez} et~al.,}{{Rodriguez-Gomez}
  et~al.}{2016}]{RodGom2016}
{Rodriguez-Gomez} V.,  et~al., 2016, \mn@doi [\mnras] {10.1093/mnras/stw456},
  \href {https://ui.adsabs.harvard.edu/abs/2016MNRAS.458.2371R} {458, 2371}

\bibitem[\protect\citeauthoryear{{Rodriguez-Gomez} et~al.,}{{Rodriguez-Gomez}
  et~al.}{2017}]{RodGom2017}
{Rodriguez-Gomez} V.,  et~al., 2017, \mn@doi [\mnras] {10.1093/mnras/stx305},
  \href {https://ui.adsabs.harvard.edu/abs/2017MNRAS.467.3083R} {467, 3083}

\bibitem[\protect\citeauthoryear{{Rothberg} \& {Joseph}}{{Rothberg} \&
  {Joseph}}{2004}]{Rothberg2004}
{Rothberg} B.,  {Joseph} R.~D.,  2004, \mn@doi [\aj] {10.1086/425049}, \href
  {https://ui.adsabs.harvard.edu/abs/2004AJ....128.2098R} {128, 2098}

\bibitem[\protect\citeauthoryear{Scannapieco, White, Springel  \&
  Tissera}{Scannapieco et~al.}{2009}]{Scannapieco2009}
Scannapieco C.,  White S. D.~M.,  Springel V.,   Tissera P.~B.,  2009, \mn@doi
  [\mnras] {10.1111/j.1365-2966.2009.14764.x}, 396, 696

\bibitem[\protect\citeauthoryear{{Schaye} et~al.,}{{Schaye}
  et~al.}{2015}]{Schaye2015}
{Schaye} J.,  et~al., 2015, \mn@doi [\mnras] {10.1093/mnras/stu2058}, \href
  {https://ui.adsabs.harvard.edu/abs/2015MNRAS.446..521S} {446, 521}

\bibitem[\protect\citeauthoryear{{Schwarzkopf} \& {Dettmar}}{{Schwarzkopf} \&
  {Dettmar}}{2000}]{Schwarzkopf2000}
{Schwarzkopf} U.,  {Dettmar} R.~J.,  2000, \aap, \href
  {https://ui.adsabs.harvard.edu/abs/2000A&A...361..451S} {361, 451}

\bibitem[\protect\citeauthoryear{{Sesar}, {Juri{\'c}}  \& {Ivezi{\'c}}}{{Sesar}
  et~al.}{2011}]{Sesar2011}
{Sesar} B.,  {Juri{\'c}} M.,   {Ivezi{\'c}} {\v{Z}}.,  2011, \mn@doi [\apj]
  {10.1088/0004-637X/731/1/4}, \href
  {https://ui.adsabs.harvard.edu/abs/2011ApJ...731....4S} {731, 4}

\bibitem[\protect\citeauthoryear{Sick, Courteau, Cuillandre, Dalcanton, de
  Jong, McDonald, Simard  \& Brent~Tully}{Sick et~al.}{2014}]{Sick2014}
Sick J.,  Courteau S.,  Cuillandre J.-C.,  Dalcanton J.,  de Jong R.,  McDonald
  M.,  Simard D.,   Brent~Tully R.,  2014, \mn@doi [Proceedings of the
  International Astronomical Union] {10.1017/S1743921315003440}, 10, 82–85

\bibitem[\protect\citeauthoryear{{Siegel}, {Majewski}, {Reid}  \&
  {Thompson}}{{Siegel} et~al.}{2002}]{Siegel2002}
{Siegel} M.~H.,  {Majewski} S.~R.,  {Reid} I.~N.,   {Thompson} I.~B.,  2002,
  \mn@doi [\apj] {10.1086/342469}, \href
  {https://ui.adsabs.harvard.edu/abs/2002ApJ...578..151S} {578, 151}

\bibitem[\protect\citeauthoryear{{Sijacki}, {Vogelsberger}, {Kere{\v{s}}},
  {Springel}  \& {Hernquist}}{{Sijacki} et~al.}{2012}]{Sijacki2012}
{Sijacki} D.,  {Vogelsberger} M.,  {Kere{\v{s}}} D.,  {Springel} V.,
  {Hernquist} L.,  2012, \mn@doi [\mnras] {10.1111/j.1365-2966.2012.21466.x},
  \href {https://ui.adsabs.harvard.edu/abs/2012MNRAS.424.2999S} {424, 2999}

\bibitem[\protect\citeauthoryear{{Sparre} \& {Springel}}{{Sparre} \&
  {Springel}}{2016}]{Sparre2016}
{Sparre} M.,  {Springel} V.,  2016, \mn@doi [\mnras] {10.1093/mnras/stw1793},
  \href {https://ui.adsabs.harvard.edu/abs/2016MNRAS.462.2418S} {462, 2418}

\bibitem[\protect\citeauthoryear{{Sparre} \& {Springel}}{{Sparre} \&
  {Springel}}{2017}]{Sparre2017}
{Sparre} M.,  {Springel} V.,  2017, \mn@doi [\mnras] {10.1093/mnras/stx1516},
  \href {https://ui.adsabs.harvard.edu/abs/2017MNRAS.470.3946S} {470, 3946}

\bibitem[\protect\citeauthoryear{{Sparre}, {Whittingham}, {Damle}, {Hani},
  {Richter}, {Ellison}, {Pfrommer}  \& {Vogelsberger}}{{Sparre}
  et~al.}{2021}]{Sparre2021}
{Sparre} M.,  {Whittingham} J.,  {Damle} M.,  {Hani} M.~H.,  {Richter} P.,
  {Ellison} S.~L.,  {Pfrommer} C.,   {Vogelsberger} M.,  2021, arXiv e-prints,
  \href {https://ui.adsabs.harvard.edu/abs/2021arXiv211003702S} {p.
  arXiv:2110.03702}

\bibitem[\protect\citeauthoryear{{Springel}}{{Springel}}{2010}]{Springel2010}
{Springel} V.,  2010, \mn@doi [\mnras] {10.1111/j.1365-2966.2009.15715.x},
  \href {https://ui.adsabs.harvard.edu/abs/2010MNRAS.401..791S} {401, 791}

\bibitem[\protect\citeauthoryear{{Springel} \& {Hernquist}}{{Springel} \&
  {Hernquist}}{2005}]{Springel2005b}
{Springel} V.,  {Hernquist} L.,  2005, \mn@doi [\apjl] {10.1086/429486}, \href
  {https://ui.adsabs.harvard.edu/abs/2005ApJ...622L...9S} {622, L9}

\bibitem[\protect\citeauthoryear{{Springel}, {White}, {Tormen}  \&
  {Kauffmann}}{{Springel} et~al.}{2001}]{Springel2001}
{Springel} V.,  {White} S. D.~M.,  {Tormen} G.,   {Kauffmann} G.,  2001,
  \mn@doi [\mnras] {10.1046/j.1365-8711.2001.04912.x}, \href
  {https://ui.adsabs.harvard.edu/abs/2001MNRAS.328..726S} {328, 726}

\bibitem[\protect\citeauthoryear{{Springel}, {Di Matteo}  \&
  {Hernquist}}{{Springel} et~al.}{2005}]{Springel2005a}
{Springel} V.,  {Di Matteo} T.,   {Hernquist} L.,  2005, \mn@doi [\mnras]
  {10.1111/j.1365-2966.2005.09238.x}, \href
  {https://ui.adsabs.harvard.edu/abs/2005MNRAS.361..776S} {361, 776}

\bibitem[\protect\citeauthoryear{{Springel} et~al.,}{{Springel}
  et~al.}{2018}]{Springel2018}
{Springel} V.,  et~al., 2018, \mn@doi [\mnras] {10.1093/mnras/stx3304}, \href
  {https://ui.adsabs.harvard.edu/abs/2018MNRAS.475..676S} {475, 676}

\bibitem[\protect\citeauthoryear{{Stewart}, {Bullock}, {Wechsler}, {Maller}  \&
  {Zentner}}{{Stewart} et~al.}{2008}]{Stewart2008}
{Stewart} K.~R.,  {Bullock} J.~S.,  {Wechsler} R.~H.,  {Maller} A.~H.,
  {Zentner} A.~R.,  2008, \mn@doi [\apj] {10.1086/588579}, \href
  {https://ui.adsabs.harvard.edu/abs/2008ApJ...683..597S} {683, 597}

\bibitem[\protect\citeauthoryear{{Stewart}, {Bullock}, {Wechsler}  \&
  {Maller}}{{Stewart} et~al.}{2009}]{Stewart2009}
{Stewart} K.~R.,  {Bullock} J.~S.,  {Wechsler} R.~H.,   {Maller} A.~H.,  2009,
  \mn@doi [\apj] {10.1088/0004-637X/702/1/307}, \href
  {https://ui.adsabs.harvard.edu/abs/2009ApJ...702..307S} {702, 307}

\bibitem[\protect\citeauthoryear{{Stinson}, {Brook}, {Macci{\`o}}, {Wadsley},
  {Quinn}  \& {Couchman}}{{Stinson} et~al.}{2013}]{Stinson2013}
{Stinson} G.~S.,  {Brook} C.,  {Macci{\`o}} A.~V.,  {Wadsley} J.,  {Quinn}
  T.~R.,   {Couchman} H.~M.~P.,  2013, \mn@doi [\mnras] {10.1093/mnras/sts028},
  \href {https://ui.adsabs.harvard.edu/abs/2013MNRAS.428..129S} {428, 129}

\bibitem[\protect\citeauthoryear{{Tamm}, {Tempel}, {Tenjes}, {Tihhonova}  \&
  {Tuvikene}}{{Tamm} et~al.}{2012}]{Tamm2012}
{Tamm} A.,  {Tempel} E.,  {Tenjes} P.,  {Tihhonova} O.,   {Tuvikene} T.,  2012,
  \mn@doi [\aap] {10.1051/0004-6361/201220065}, \href
  {https://ui.adsabs.harvard.edu/abs/2012A&A...546A...4T} {546, A4}

\bibitem[\protect\citeauthoryear{{Toomre}}{{Toomre}}{1977}]{Toomre1977}
{Toomre} A.,  1977, in {Tinsley} B.~M.,  {Larson} Richard B.~Gehret D.~C.,
  eds, Evolution of Galaxies and Stellar Populations. p.~401

\bibitem[\protect\citeauthoryear{{Torrey} et~al.,}{{Torrey}
  et~al.}{2015a}]{Torrey2015}
{Torrey} P.,  et~al., 2015a, \mn@doi [\mnras] {10.1093/mnras/stu2592}, \href
  {https://ui.adsabs.harvard.edu/abs/2015MNRAS.447.2753T} {447, 2753}

\bibitem[\protect\citeauthoryear{{Torrey} et~al.,}{{Torrey}
  et~al.}{2015b}]{Torrey2015b}
{Torrey} P.,  et~al., 2015b, \mn@doi [\mnras] {10.1093/mnras/stv1986}, \href
  {https://ui.adsabs.harvard.edu/abs/2015MNRAS.454.2770T} {454, 2770}

\bibitem[\protect\citeauthoryear{{Toth} \& {Ostriker}}{{Toth} \&
  {Ostriker}}{1992}]{Toth1992}
{Toth} G.,  {Ostriker} J.~P.,  1992, \mn@doi [\apj] {10.1086/171185}, \href
  {https://ui.adsabs.harvard.edu/abs/1992ApJ...389....5T} {389, 5}

\bibitem[\protect\citeauthoryear{{Vasiliev}, {Belokurov}  \&
  {Erkal}}{{Vasiliev} et~al.}{2021}]{Vasiliev2021}
{Vasiliev} E.,  {Belokurov} V.,   {Erkal} D.,  2021, \mn@doi [\mnras]
  {10.1093/mnras/staa3673}, \href
  {https://ui.adsabs.harvard.edu/abs/2021MNRAS.501.2279V} {501, 2279}

\bibitem[\protect\citeauthoryear{{Vogelsberger}, {Sijacki}, {Kere{\v{s}}},
  {Springel}  \& {Hernquist}}{{Vogelsberger} et~al.}{2012}]{Vogelsberger2012}
{Vogelsberger} M.,  {Sijacki} D.,  {Kere{\v{s}}} D.,  {Springel} V.,
  {Hernquist} L.,  2012, \mn@doi [\mnras] {10.1111/j.1365-2966.2012.21590.x},
  \href {https://ui.adsabs.harvard.edu/abs/2012MNRAS.425.3024V} {425, 3024}

\bibitem[\protect\citeauthoryear{{Vogelsberger} et~al.,}{{Vogelsberger}
  et~al.}{2014a}]{Vogelsberger2014}
{Vogelsberger} M.,  et~al., 2014a, \mn@doi [\mnras] {10.1093/mnras/stu1536},
  \href {https://ui.adsabs.harvard.edu/abs/2014MNRAS.444.1518V} {444, 1518}

\bibitem[\protect\citeauthoryear{{Vogelsberger} et~al.,}{{Vogelsberger}
  et~al.}{2014b}]{Vogelsberger2014b}
{Vogelsberger} M.,  et~al., 2014b, \mn@doi [\nat] {10.1038/nature13316}, \href
  {https://ui.adsabs.harvard.edu/abs/2014Natur.509..177V} {509, 177}

\bibitem[\protect\citeauthoryear{{Vogelsberger}, {Marinacci}, {Torrey}  \&
  {Puchwein}}{{Vogelsberger} et~al.}{2020}]{Vogelsberger2020}
{Vogelsberger} M.,  {Marinacci} F.,  {Torrey} P.,   {Puchwein} E.,  2020,
  \mn@doi [Nature Reviews Physics] {10.1038/s42254-019-0127-2}, \href
  {https://ui.adsabs.harvard.edu/abs/2020NatRP...2...42V} {2, 42}

\bibitem[\protect\citeauthoryear{{Watkins} et~al.,}{{Watkins}
  et~al.}{2009}]{Watkins2009}
{Watkins} L.~L.,  et~al., 2009, \mn@doi [\mnras]
  {10.1111/j.1365-2966.2009.15242.x}, \href
  {https://ui.adsabs.harvard.edu/abs/2009MNRAS.398.1757W} {398, 1757}

\bibitem[\protect\citeauthoryear{{Weinberger} et~al.,}{{Weinberger}
  et~al.}{2017}]{Weinberger2017}
{Weinberger} R.,  et~al., 2017, \mn@doi [\mnras] {10.1093/mnras/stw2944}, \href
  {https://ui.adsabs.harvard.edu/abs/2017MNRAS.465.3291W} {465, 3291}

\bibitem[\protect\citeauthoryear{{Wetzel}, {Hopkins}, {Kim},
  {Faucher-Gigu{\`e}re}, {Kere{\v{s}}}  \& {Quataert}}{{Wetzel}
  et~al.}{2016}]{Wetzel2016}
{Wetzel} A.~R.,  {Hopkins} P.~F.,  {Kim} J.-h.,  {Faucher-Gigu{\`e}re} C.-A.,
  {Kere{\v{s}}} D.,   {Quataert} E.,  2016, \mn@doi [\apjl]
  {10.3847/2041-8205/827/2/L23}, \href
  {https://ui.adsabs.harvard.edu/abs/2016ApJ...827L..23W} {827, L23}

\bibitem[\protect\citeauthoryear{{Worthey}, {Espa{\~n}a}, {MacArthur}  \&
  {Courteau}}{{Worthey} et~al.}{2005}]{Worthey2005}
{Worthey} G.,  {Espa{\~n}a} A.,  {MacArthur} L.~A.,   {Courteau} S.,  2005,
  \mn@doi [\apj] {10.1086/432785}, \href
  {https://ui.adsabs.harvard.edu/abs/2005ApJ...631..820W} {631, 820}

\bibitem[\protect\citeauthoryear{{Wyse}}{{Wyse}}{2001}]{Wyse2001}
{Wyse} R.~F.~G.,  2001, in {Funes} J.~G.,  {Corsini} E.~M.,  eds,  Astronomical
  Society of the Pacific Conference Series Vol. 230, Galaxy Disks and Disk
  Galaxies. pp 71--80 (\mn@eprint {arXiv} {astro-ph/0012270})

\bibitem[\protect\citeauthoryear{{Xiang} \& {Rix}}{{Xiang} \&
  {Rix}}{2022}]{XiangRix2022}
{Xiang} M.,  {Rix} H.-W.,  2022, \mn@doi [\nat] {10.1038/s41586-022-04496-5},
  \href {https://ui.adsabs.harvard.edu/abs/2022Natur.603..599X} {603, 599}

\bibitem[\protect\citeauthoryear{{Yoachim} \& {Dalcanton}}{{Yoachim} \&
  {Dalcanton}}{2006}]{2006Yoachim}
{Yoachim} P.,  {Dalcanton} J.~J.,  2006, \mn@doi [\aj] {10.1086/497970}, \href
  {https://ui.adsabs.harvard.edu/abs/2006AJ....131..226Y} {131, 226}

\bibitem[\protect\citeauthoryear{{Zeng}, {Wang}  \& {Gao}}{{Zeng}
  et~al.}{2021}]{Zeng2021}
{Zeng} G.,  {Wang} L.,   {Gao} L.,  2021, arXiv e-prints, \href
  {https://ui.adsabs.harvard.edu/abs/2021arXiv210509722Z} {p. arXiv:2105.09722}

\bibitem[\protect\citeauthoryear{{Zhu} et~al.,}{{Zhu} et~al.}{2021}]{Zhu2021}
{Zhu} L.,  et~al., 2021, arXiv e-prints, \href
  {https://ui.adsabs.harvard.edu/abs/2021arXiv211013172Z} {p. arXiv:2110.13172}

\bibitem[\protect\citeauthoryear{{Zolotov}, {Hogg}  \& {Willman}}{{Zolotov}
  et~al.}{2011}]{Zolotov2011}
{Zolotov} A.,  {Hogg} D.~W.,   {Willman} B.,  2011, \mn@doi [\apjl]
  {10.1088/2041-8205/727/1/L14}, \href
  {https://ui.adsabs.harvard.edu/abs/2011ApJ...727L..14Z} {727, L14}

\makeatother
\end{thebibliography}


\appendix
\section{Fit of radial and vertical stellar density profiles}
\label{app:fits}

In this appendix, we explain how the measurements of the stellar disk lengths and the scale heights are performed in this paper.

For the disk length, for any given galaxy, we select all stellar particles with almost circular orbits (i.e. disk stars, $\epsilon > $0.7) between one and four times the half-mass radius, so excluding the bulge region. We fit an exponential profile to the radial stellar surface density distribution in face-on projection, in bins of 2 kpc:

\begin{equation}
\Sigma(R) =\Sigma_{\rm d} \, {\rm exp} \left(-\frac{R}{\rm R_d}\right),
\end{equation}

where $\Sigma_d$ is the central stellar mass surface density of the disk and $R_d$ is the disk scalelength.
For each galaxy, 100 fits starting with random initial values (around the values taken at the limits of the cylindrical shell) are performed. We take the mode of the distribution of the 100 best-fit values as the best measure of the scale length, i.e. the one we quote for each galaxy; the error is obtained by the interquartile range.

For the disk heights, for each galaxy projected edge-on, we select stars in circular orbits in the cylindrical shell between 3.5 and 4.5 times the disk length and we fit a double function profile (either linear or square hyperbolic secant) to the vertical stellar mass distribution, between 50 pc and 5 kpc in bins of 200 pc:

\begin{equation}
\label{eq:double}
\rho(z) =\rho_{\rm{thin}} ~{\rm sech}\left(\frac{z}{\rm h_{thin}}\right)+\rho_{\rm{thick}} ~{\rm sech}\left(\frac{z}{\rm h_{thick}}\right) ~;
\end{equation}

\begin{equation}
\label{eq:double2}
\rho(z) =\rho_{\rm{thin}} ~{\rm sech^2}\left(\frac{z}{2{\rm h_{thin}}}\right)+\rho_{\rm{thick}} ~{\rm sech^2}\left(\frac{z}{2{\rm h_{thick}}}\right) ~.
\end{equation}

This gives us the scale heights of both a ``thin''  and a ``thick'' disk component (geometrical). The factor 2 for the squared case allows that the scale heights of the linear and squared cases are comparable in magnitude. Both profiles are similar at high heights but significantly different close to the galactic midplane. We justify the choice of adopting two functional forms because both have been extensively used to describe vertical galaxy density profiles and yet may give in certain instances rather different results. The sech$^2$ profile has been employed by e.g. \citet{2006Yoachim, 2014Bizyaev} for observations of edge-on spiral galaxies, and \citet{Stinson2013, 2017Ma, 2021Park} for simulated galaxies. The sech profile, justified theoretically by \citet[][]{1988vanderKruit} and \citet{Banerjee2007}, has been used by \citet{Schwarzkopf2000} in galaxy surveys or by \citet{Qu2011} for simulations.
On average, the scale heights obtained from the double sech$^2$ profile are $\sim$10-20\% smaller than those obtained from the double sech profiles (in agreement with \citet[][]{Qu2011}.


As for the lengths measurements, for each galaxy and choice of functional form, we perform 100 fits starting with random initial values of the scale heights (in the ranges 20 to 1000 pc and 800 to 7000 pc for the thin and thick disks, respectively), for each functional form choice. We take the mode of the distribution of of the 100 best-fit values as the best measure of the scale height for a given parametric function. If needed, an error can be represented by the interquartile range of such distribution. In the main part of the manuscript, we opt instead to provide an estimate of the (systematic) error by quoting, for each galaxy, the best scale-height measures from both fitting functions of eqs.~\ref{eq:double} and \ref{eq:double2}.

\bsp	
\label{lastpage}
\end{document}